\documentclass[11pt]{article}

\usepackage[final]{acl}

\usepackage{times}
\usepackage{latexsym}
\usepackage{amssymb}
\usepackage[T1]{fontenc}

\usepackage[utf8]{inputenc}
\usepackage[table]{xcolor}
\usepackage{pgfplots}
\usepackage{pifont}
\usepackage{microtype}
\usepackage{multirow} 
\usepackage{amsmath}
\usepackage{makecell}
\usepackage{inconsolata}
\usepackage{booktabs, tabularx} 
\usepackage{nicematrix,xcolor,colortbl,graphicx,booktabs,tikz}
\usepackage{pgfplots}
\pgfplotsset{compat=1.18}
\usepgfplotslibrary{groupplots}
\usetikzlibrary{arrows.meta,calc}
\usepackage[table]{xcolor}   
\usepackage{pgf}
\usepackage{graphicx}
\usepackage{threeparttable}
\usepackage[most]{tcolorbox}

\usetikzlibrary{arrows.meta,positioning,calc}
\usepackage{xcolor}

\usepackage{tikz}
\usepackage{colortbl}
\usepackage{tcolorbox}\tcbuselibrary{skins}
\usepackage{pifont}

\newtcolorbox{promptbox}[1]{
  breakable,
  enhanced,
  colback=gray!5,
  colframe=gray!60,
  boxrule=0.5pt,
  arc=2pt,
  left=5pt, right=5pt, top=4pt, bottom=4pt,
  fonttitle=\small\bfseries,
  title={#1},
  before skip=6pt, after skip=6pt
}
\title{\textsc{UniCoRn}: A Benchmark for Unified Compression and Reranking}

\author{
  Grace Byun$^{1,2}$\thanks{Equal contribution.}\,\thanks{Work done during an internship at Cohere.},
  Clifton Poth$^{2}$\footnotemark[1],
  Carlos Lassance$^{2}$,
  Nils Reimers$^{2}$
  \\
  $^{1}$Emory University \quad $^{2}$Cohere
}

\begin{document}
\maketitle
\begin{abstract}
Reranking and context compression are complementary strategies for identifying relevant document context in retrieval-augmented generation.
Despite recent methods that combine both, their joint evaluation remains limited. Existing benchmarks measure end-task accuracy in clean retrieval settings, revealing little about why methods fail or how compression interacts with reranking. We introduce \textsc{UniCoRn}, a three-tier benchmark for unified evaluation of compression and reranking. The first tier, unit testing, isolates compression failure modes under controlled, retrieval-free settings. The second-tier, integration testing, evaluates reranking and compression over a fixed retrieval pool spanning 17 datasets, 10 domains, and 5 languages. The final, system testing, measures robustness inside search agents with dynamic retrieval. Further, we introduce metrics for preservation of answer-critical evidence and reranking-pruning alignment. We evaluate 11 compression methods while training models for both hard and soft compression, enabling a direct comparison between the two regimes.
\end{abstract}

\section{Introduction}

Large language models (LLMs) offer context windows of hundreds of thousands of tokens, but often fail to exploit them effectively \citep{liu2024lost,hong2025context}.
In Retrieval-Augmented Generation (RAG) \citep{lewis2020retrieval}, reranking effectively reduces the \emph{number} of documents, but not the \emph{size} of each, so large spans of irrelevant text still reach the LLM, degrading answer quality and increasing cost.
Combining reranking with context compression has been proposed as a natural solution \citep{chirkova2025provenceefficientrobustcontext}.

A wide range of context compression methods have been introduced (\citet{wang2023learning,jiang2024longllmlinguaacceleratingenhancingllms,louis2026oscar}; \textit{inter alia}).
Their evaluation, however, remains underdeveloped. Many are tested on classic RAG benchmarks built from standardized Wikipedia corpora \citep{petroni-etal-2021-kilt}, whereas production documents are far more diverse: noisy, of varying length, drawn from specialized domains, and laden with tables and formatting artifacts. Evaluations also tend to be fragmented, reporting only end-task accuracy and rarely revealing \emph{why} a method fails. Crucially, none of the benchmarks measures compression and reranking jointly: a compressor is scored on retrieval quality without inspecting which evidence it discards.

%
%


\definecolor{cgGreen}{HTML}{5B4B8A}    
\definecolor{cgGreenMut}{HTML}{A08CC0} %
\definecolor{cgRule}{HTML}{3F4B44}
\definecolor{cgCard}{HTML}{C4BCAD}
\definecolor{cgPanel}{HTML}{FBF3FC}   
\definecolor{cgCoral}{HTML}{ED7BAA}   
\definecolor{cgCoralDk}{HTML}{B03A72}  
\definecolor{cgPurple}{HTML}{3FA9C6}   
\definecolor{cgDocEdge}{HTML}{333333}
\definecolor{cgDocLine}{HTML}{8A8A8A}
\definecolor{cgDocText}{HTML}{4A443A}
\definecolor{cgInk}{HTML}{4A3F6B}      %
\definecolor{cgMut}{HTML}{5F6B65}
\definecolor{cgTitle}{HTML}{333333}
\tikzset{
  tier card/.style={
    draw=cgCard, line width=0.6pt, rounded corners=5pt, fill=white,
  },
  tier module/.style={
    draw=cgRule, line width=0.45pt, fill=white,
    font=\tiny\sffamily\bfseries, text=cgGreen, align=center,
    minimum width=2.05cm, minimum height=0.36cm, inner sep=1pt,
  },
  tier module absent/.style={
    tier module, fill=cgPanel, dash pattern=on 1.6pt off 1.2pt,
  },
  tier module agent/.style={
    tier module, draw=cgPurple, text=cgPurple,
  },
  tier module judge/.style={
    tier module, fill=cgPanel,
  },
  tier compressor/.style={
    draw=cgCoralDk, line width=0.5pt, fill=cgCoral,
    font=\tiny\sffamily\bfseries, text=white, align=center,
    minimum width=2.05cm, minimum height=0.50cm, inner sep=1pt,
  },
  artifact/.style={font=\tiny\sffamily\bfseries, text=cgInk, align=center},
  kept/.style={font=\tiny\sffamily\bfseries, text=cgCoralDk, align=center},
  poollabel/.style={
    font=\fontsize{4.6pt}{5.2pt}\selectfont\sffamily\bfseries,
    text=cgDocText, align=center,
  },
  regime/.style={
    font=\fontsize{4.8pt}{5.4pt}\selectfont\sffamily\bfseries, align=left,
  },
  tiername/.style={
    font=\fontsize{5.6pt}{6.2pt}\selectfont\sffamily\bfseries,
    text=cgTitle, align=left,
  },
  micro/.style={font=\fontsize{4.4pt}{5pt}\selectfont\sffamily},
  flow/.style={-{Latex[length=2pt, width=1.8pt]}, cgRule, line width=0.45pt},
  loop/.style={-{Latex[length=2pt, width=1.8pt]}, cgCoral, line width=0.5pt,
    dash pattern=on 1.8pt off 1.3pt},
  docedge/.style={draw=cgDocEdge, line width=0.5pt, line join=miter},
}

\providecommand{\evtdocpage}[2]{%
  \draw[docedge, fill=#2]
    ($#1+(-0.15,0.19)$) -- ($#1+(0.06,0.19)$) -- ($#1+(0.15,0.10)$)
    -- ($#1+(0.15,-0.19)$) -- ($#1+(-0.15,-0.19)$) -- cycle;
  \draw[docedge]
    ($#1+(0.06,0.19)$) -- ($#1+(0.06,0.10)$) -- ($#1+(0.15,0.10)$);
}

\providecommand{\evtdocglyph}[2]{%
  \evtdocpage{#1}{#2}
  \foreach \ly in {0.02,-0.06,-0.14} {%
    \draw[cgDocLine, line width=0.45pt]
      ($#1+(-0.09,\ly)$) -- ($#1+(0.09,\ly)$);
  }%
}

\providecommand{\evtdocstack}[1]{%
  \evtdocpage{($#1+(-0.05,0.04)$)}{cgPanel}
  \evtdocpage{($#1+(0.00,0.00)$)}{cgPanel}
  \evtdocglyph{($#1+(0.05,-0.04)$)}{white}
}

\begin{figure}[t]
\centering
\resizebox{\columnwidth}{!}{%
\begin{tikzpicture}[font=\sffamily]

\def\colA{1.30}
\def\colB{4.08}
\def\colC{6.86}

\node[micro, text=cgMut, anchor=east] (lg) at (8.20, 0.30) {module under test};
\draw[fill=cgCoral, draw=cgCoralDk, line width=0.4pt]
  ([xshift=-7pt, yshift=-2.2pt]lg.west) rectangle ([xshift=-2pt, yshift=2.2pt]lg.west);

\begin{scope}[shift={(\colA,0)}]
  \draw[tier card] (-1.34, 0.10) rectangle (1.34, -4.72);
  \node[tiername, anchor=west] at (-1.04, -0.14)
    {\textcolor{cgGreenMut}{1.} Unit test};
  \node[regime, text=cgGreenMut, anchor=west] at (-1.04, -0.38)
    {RETRIEVAL-FREE};
  \draw[cgRule, line width=0.5pt] (-1.04, -0.52) -- (1.04, -0.52);

  \node[artifact] (u-q) at (0, -0.80) {User query};
  \node[tier module absent] (u-r) at (0, -1.22) {No retrieval step};
  \node[micro, text=cgGreenMut, anchor=west] at (-1.02, -0.94) {\itshape absent};
  \evtdocglyph{(0,-1.88)}{white}
  \node[poollabel] (u-pl) at (0, -2.20) {single-doc per query};
  \node[tier compressor] (u-c) at (0, -2.77)
    {COMPRESSOR\\[-2pt]{\fontsize{4.4pt}{5pt}\selectfont\normalfont\sffamily\color{white}rerank + compress}};
  \node[kept] (u-k) at (0, -3.30) {compressed doc};
  \node[tier module] (u-g) at (0, -3.75) {Generator};
  \node[tier module judge] (u-j) at (0, -4.31) {LLM judge};

  \draw[flow] (0,-0.88) -- (0,-1.04);
  \draw[flow] (0,-1.40) -- (0,-1.69);
  \draw[flow] (0,-2.29) -- (0,-2.52);
  \draw[flow] (0,-3.02) -- (0,-3.23);
  \draw[flow] (0,-3.37) -- (0,-3.57);
  \draw[flow] (0,-3.93) -- (0,-4.13);
\end{scope}

\begin{scope}[shift={(\colB,0)}]
  \draw[tier card] (-1.34, 0.10) rectangle (1.34, -4.72);
  \node[tiername, anchor=west] at (-1.04, -0.14)
    {\textcolor{cgGreenMut}{2.} Integration test};
  \node[regime, text=cgGreen, anchor=west] at (-1.04, -0.38)
    {RETRIEVAL-FIXED};
  \draw[cgRule, line width=0.5pt] (-1.04, -0.52) -- (1.04, -0.52);

  \node[artifact] (i-q) at (0, -0.80) {User query};
  \node[tier module] (i-r) at (0, -1.22) {Dense retriever};
  \node[micro, text=cgGreen, anchor=west] at (-1.02, -0.94) {\itshape frozen};
  \evtdocstack{(0,-1.88)}
  \node[poollabel] (i-pl) at (0, -2.20) {50 fixed candidates};
  \node[tier compressor] (i-c) at (0, -2.77)
    {COMPRESSOR\\[-2pt]{\fontsize{4.4pt}{5pt}\selectfont\normalfont\sffamily\color{white}rerank + compress}};
  \node[kept] (i-k) at (0, -3.30) {top-5 compressed docs};
  \node[tier module] (i-g) at (0, -3.75) {Generator};
  \node[tier module judge] (i-j) at (0, -4.31) {LLM judge};

  \draw[flow] (0,-0.88) -- (0,-1.04);
  \draw[flow] (0,-1.40) -- (0,-1.69);
  \draw[flow] (0,-2.29) -- (0,-2.52);
  \draw[flow] (0,-3.02) -- (0,-3.23);
  \draw[flow] (0,-3.37) -- (0,-3.57);
  \draw[flow] (0,-3.93) -- (0,-4.13);
\end{scope}

\begin{scope}[shift={(\colC,0)}]
  \draw[tier card] (-1.34, 0.10) rectangle (1.34, -4.72);
  \node[tiername, anchor=west] at (-1.04, -0.14)
    {\textcolor{cgGreenMut}{3.} System test};
  \node[regime, text=cgPurple, anchor=west] at (-1.04, -0.38)
    {RETRIEVAL-DYNAMIC};
  \draw[cgRule, line width=0.5pt] (-1.04, -0.52) -- (1.04, -0.52);

  \node[artifact] (s-q) at (0, -0.80) {Agent sub-query};
  \node[tier module] (s-r) at (0, -1.22) {Dense retriever};
  \node[micro, text=cgPurple, anchor=west] at (-1.02, -0.94) {\itshape live};
  \evtdocstack{(0,-1.88)}
  \node[poollabel] (s-pl) at (0, -2.20) {50 candidates (this call)};
  \node[tier compressor] (s-c) at (0, -2.77)
    {COMPRESSOR\\[-2pt]{\fontsize{4.4pt}{5pt}\selectfont\normalfont\sffamily\color{white}rerank + compress}};
  \node[kept] (s-k) at (0, -3.30) {top-5 snippets};
  \node[tier module agent] (s-a) at (0, -3.75) {Agent};
  \node[tier module judge] (s-j) at (0, -4.31) {LLM judge};

  \draw[flow] (0,-0.88) -- (0,-1.04);
  \draw[flow] (0,-1.40) -- (0,-1.69);
  \draw[flow] (0,-2.29) -- (0,-2.52);
  \draw[flow] (0,-3.02) -- (0,-3.23);
  \draw[flow] (0,-3.37) -- (0,-3.57);
  \draw[flow] (0,-3.93) -- (0,-4.13);
  \draw[loop] (s-a.east) -- (1.08,-3.75) -- (1.08,-1.22) -- (s-r.east);
  \node[micro, text=cgCoralDk, rotate=90, anchor=north, inner sep=1pt]
    at (1.09,-2.49) {re-query next call};
\end{scope}

\end{tikzpicture}%
}
\vspace{-1.7em}
\caption{Three tiers for evaluating context compression. The compressor is held
fixed across tiers; only the retrieval regime changes, from absent, to computed
once and frozen, to produced live on every agent call.}
\vspace{-0.7em}
\label{fig:three_evals}
\end{figure}
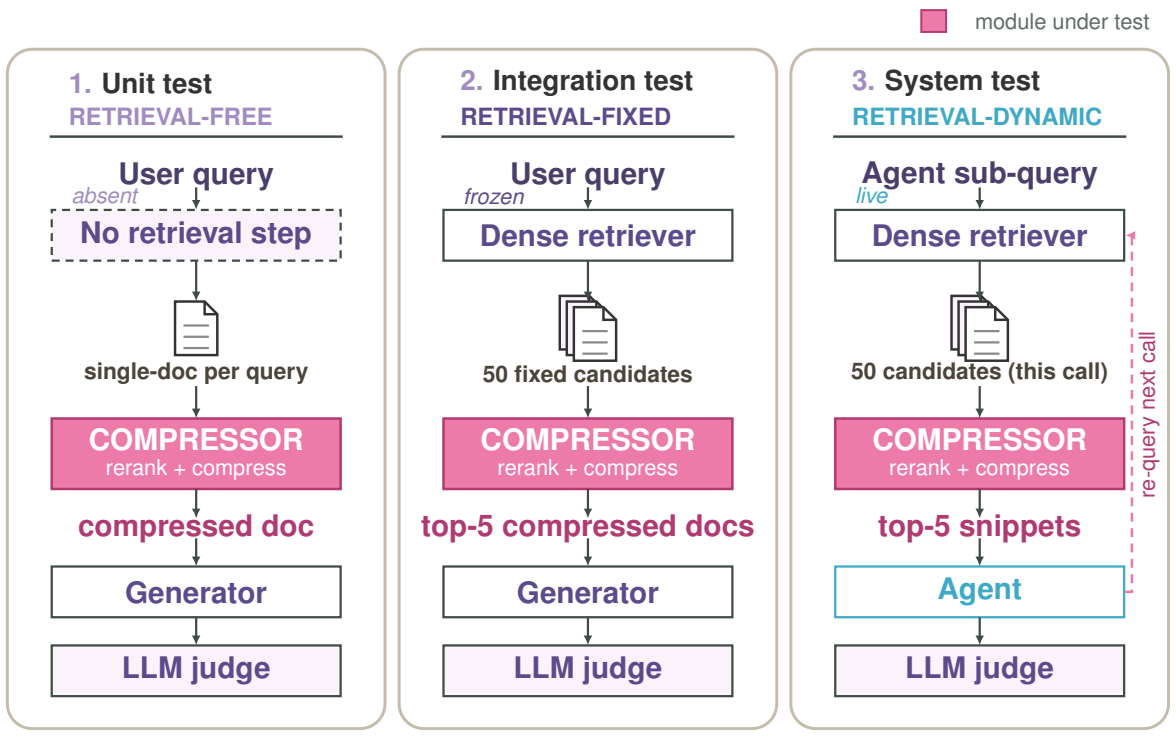

We introduce \textsc{UniCoRn}, a benchmark that evaluates compression and reranking quality across three tiers and on domains relevant to real-world enterprise settings. The \textbf{Unit Test} probes a compressor in isolation on controlled, retrieval-free inputs, exposing failure modes over evidence position, distractors, length, and language. The \textbf{Integration Test} freezes a realistic pool shared across methods, measuring rerank-and-compress quality on diverse data. The \textbf{System Test} embeds the compressor in a live search agent, where the output of the former searches shapes later queries, stressing robustness under compounding, closed-loop errors. We further propose two diagnostic metrics beyond end-task accuracy: whether the evidence needed to answer survives compression, and whether a method's reranking decisions and its compression rate agree on document relevance.

We evaluate a diverse set of publicly available hard and soft compression methods on our benchmark to obtain a comprehensive picture of their strengths and weaknesses. Because existing methods are trained with various base models and data, direct comparisons between the two compression regimes are difficult. We therefore train hard and soft baselines using matched base models and training data. With targeted improvements to the training setup, our hard compression model surpasses all prior baselines, and our soft model surpasses the prior soft baseline.


Our main contributions are as follows:

\vspace{-0.4em}
\begin{itemize}
\setlength\itemsep{-0.2em}
    \item We introduce \textsc{UniCoRn}\footnote{\url{https://huggingface.co/datasets/unicorn-rerank/unicorn-benchmark}}\footnote{\url{https://github.com/byunsj/unicorn-benchmark}} for evaluating ranking and compression quality over three tiers from unit testing to end-to-end system testing.
    \item We evaluate various compression methods on \textsc{UniCoRn}, isolating their strengths and weaknesses in realistic and noisy settings.
    \item We train and release directly comparable hard\footnote{\url{https://huggingface.co/unicorn-rerank/unicorn-comp-h}} and soft\footnote{\url{https://huggingface.co/unicorn-rerank/unicorn-comp-s}} compression baselines\ that define the current Pareto frontier on our benchmark.
\end{itemize}
\vspace{-0.5em}

\section{Background and Related Work}

\paragraph{Context Compression.}

Context compression methods can be broadly categorized into two main regimes. \textit{Hard compression} methods operate in the token space and reduce the number of tokens passed to the decoder model. Hard compression can be \textit{abstractive} \cite{xu2024recomp,yoon-etal-2024-compact}, where the compressor model generates a short summary of the document to compress, or \textit{extractive}, where the compressor selects spans to pass to the decoder on a token-level \citep{pan2024llmlingua2datadistillationefficient,jiang2024longllmlinguaacceleratingenhancingllms} or sequence-level \citep{hwang-etal-2024-dslr,mohamed2026XProvencezerocostmultilingualcontext,do2026loocompleverageleaveoneoutstrategy, xu2024recomp}. \textit{Soft compression} methods operate in the latent space of the decoder \citep{xrag, 10.1145/3701551.3703527, louis-etal-2025-pisco,louis2026oscar,li2026endtoendcontextcompressionscale,pilchen2026arcencoder} and aim to compress relevant context into vector representations.

In a RAG setting, context compression can either be \textit{query-aware}, i.e., dependent on the search query, or \textit{query-agnostic}. Query-aware compression naturally fits with the reranking stage in retrieval pipelines, which is often realized by cross-encoders \citep{nogueira2019multistagedocumentrankingbert,Lin2020PretrainedTF}.
Other work integrates reranking into the generator itself \citep{3737916.3741766}, but still passes each selected document in full.
Recent work has started to leverage the synergy between reranking and compression \citep{déjean2026efficientlistwisererankingcompressed}, and our benchmark sits exactly at this intersection.

\paragraph{RAG Benchmarks.}

Retrieval benchmarks comprehensively evaluate systems across diverse domains and languages \citep{thakur2021beir,zhang-etal-2023-miracl,muennighoff-etal-2023-mteb}.
More recent benchmarks focus on various aspects of RAG, such as challenging technical \citep{thakur2026freshstack} and enterprise \citep{loison2026vidorev3comprehensiveevaluation} documents, multi-hop \citep{tang2024multihoprag}, reasoning \citep{su2025bright}, and deep research \citep{chen2025browsecompplusfairtransparentevaluation}. Diagnostic frameworks further analyze RAG beyond end-task accuracy, either with fine-grained metrics for the retriever and generator \citep{3737916.3738608} or by studying how retriever, reranker, and reader configurations interact \citep{3780338.3781283}, but none of them treats compression as a separate component. 
Our benchmark complements these works by explicitly focusing on context compression as a new dimension to evaluate.

\section{Evaluation Settings}


\begin{table*}[ht!]
\centering
\footnotesize
\setlength{\tabcolsep}{3pt}
\renewcommand{\arraystretch}{1.15}
\begin{tabular}{@{}p{1.7cm} p{1.3cm} p{4.4cm} p{3cm} p{4.4cm}@{}}
\toprule
\textbf{Setting} & \textbf{Retrieval} & \textbf{Context Control} & \textbf{Scale} & \textbf{What it evaluates} \\
\midrule
\textbf{Unit Test} & \emph{Free} & Full; position, distractors, length, language are experimenter-set & 100 (10 Q types $\times$ 10) & Mechanistic failure modes of the compressor alone \\
\textbf{Integration Test} & \emph{Fixed} & Realistic but frozen; identical pre-retrieved pool & 12{,}775 (10 domains, 5 languages) & Generalization across domain, language, document length \\
\textbf{System Test} & \emph{Dynamic} & Live; the compressor's output shapes the agent's next query & 1,076 (2 domains, variable \#\,calls) & End-to-end robustness under compounding, closed-loop errors \\
\bottomrule
\end{tabular}
\vspace{-0.6em}
\caption{\textsc{UniCoRn} Tiers. Retrieval regime varies across tiers: a single document with no retrieval (\textbf{Unit}), a fixed pool retrieved once and shared across methods (\textbf{Integration}), or documents retrieved turn-by-turn by a live agent reacting to the compressor's own output (\textbf{System}).}
\label{tab:eval-settings}
\vspace{-0.6em}
\end{table*}

\textsc{UniCoRn} mirrors the
software-testing pyramid: \textbf{Unit Test}, \textbf{Integration Test}, and
\textbf{System Test}. 
The one technical variable that changes across tiers is how the compressor's input documents are obtained, so we tag each tier with its retrieval regime, \emph{retrieval-free}, \emph{retrieval-fixed}, or \emph{retrieval-dynamic}. See Figure \ref{fig:three_evals} and Table~\ref{tab:eval-settings}.

\subsection{Unit Test (Retrieval-Free)}
\label{sec:unit-test}
This tier removes retrieval altogether. Context is manually assembled to target one specific failure mode, so any failure is
attributable to the compressor itself rather than to what was retrieved. We build 10 queries for each of 10 diagnostic query types, for a
total of 100 queries (Table~\ref{tab:unittest-query-types}), each isolating one
mechanism that a compressor could plausibly get wrong  (e.g., evidence split across two widely separated positions, or a
same-topic adversarial distractor).
Refer to Appendix~\ref{sec:appendix_unittest_settings} and 
Table~\ref{tab:unittest-query-types-appendix} for details on data construction.

\begin{table}[t]
\centering
\small
\begin{tabularx}{\columnwidth}{@{}l >{\raggedright\arraybackslash}X@{}}
\toprule
\textbf{Query Type} & \textbf{Source} \\
\midrule
\textsc{Multi-hop Long} & HotpotQA (bridge) \\
\textsc{Numerical Tables} & FinQA, Wikipedia \\
\textsc{Scattered Evidence} & arXiv papers \\
\textsc{Information Density} & SQuAD v1.1, Wikipedia \\
\textsc{Adversarial Distractor} & HotpotQA \\
\textsc{Contextualized} & Quoref \\
\textsc{Union} & Wikipedia \\
\textsc{Summarization} & Wikipedia \\
\textsc{Ugly Formatting} & FinQA, Wikipedia \\
\textsc{Cross-lingual} & ViDoRe-FR, KorQuAD, JSQuAD \\
\bottomrule
\end{tabularx}
\vspace{-0.8em}
\caption{\textbf{Unit Test}: ten query types and source datasets.
Details and citations are in Table~\ref{tab:unittest-query-types-appendix}
(Appendix~\ref{sec:appendix_unittest_result}).}
\label{tab:unittest-query-types}
\end{table}

\subsection{Integration Test (Retrieval-Fixed)}
\label{sec:integration-test}

This tier plugs the compressor into the full
rerank\,$\to$\,compress\,$\to$\,generate\,$\to$\,judge pipeline.
Tasks are selected to represent diverse domains with a focus on noisy, real-world and enterprise data (Table~\ref{tab:benchmark}).
Retrieval is computed once per query and held
fixed. Cohere Embed-4 \citep{embed4} returns the top-50 candidates, with any missing gold document manually inserted to ensure its inclusion. All methods rerank
and compress this same 50-document pool down to their own top-5. Relative to the Unit Test, this tier trades fine-grained failure
attribution for realism such as genuine document noise, different languages, and long documents.

\begin{table*}[t]
\centering
\small
\setlength{\tabcolsep}{4pt}
\renewcommand{\arraystretch}{1.15}
\begin{tabular}{@{}p{3cm} c p{3.2cm} p{3.0cm} c p{3.2cm}@{}}
\toprule
\textbf{Dataset} & \textbf{Lang.} & \textbf{Domain} & \textbf{Query Type} & \textbf{\#Q} & \textbf{Highlighted aspect} \\
\midrule
\multirow{2}{=}{\makecell[tl]{ViDoRe V3\\ \citep{loison2026vidorev3comprehensiveevaluation}}}
 & En & Computer Science, Finance, HR, Industrial, Pharmaceuticals & Document-grounded (extractive, numerical, boolean, multi-hop) & 1{,}489 & Enterprise domains; PDF\,$\to$\,Markdown \\
\cmidrule(l){2-6}
 & Fr & Energy, Finance, Physics & Document-grounded & 930 & Multilingual \\
\midrule
\makecell[tl]{Academic Papers (New)\\ \citep{acar_arxiver2024}} & En & arXiv scientific papers & Numbers/metrics, Entities, Multi-hop logic & 372 & Long docs; synthetically generated from ArXiver \\
\midrule
\makecell[tl]{JSQuAD\\ \citep{kurihara-etal-2022-jglue}}     & Ja & Wikipedia & Extractive & 200 & Multilingual \\
\makecell[tl]{KorQuAD\\ \citep{ART002594126}}  & Ko & Wikipedia & Extractive & 100 & Multilingual; HTML\,$\to$\,MD \\
\makecell[tl]{KLUE-MRC\\ \citep{park2021klue}}   & Ko & News, Wikipedia & Extractive & 200 & Multilingual \\
\makecell[tl]{LegalQuAD\\ \citep{9723721}}  & De & Legal & Extractive & 200 & Multilingual, Legal \\
\midrule
\makecell[tl]{FinQA\\ \citep{chen-etal-2021-finqa}}     & En & Finance; earnings reports & Numerical reasoning & 1{,}138 & Tables (hard to compress) \& Calculation \\
\makecell[tl]{HotpotQA\\ \citep{yang2018hotpotqadatasetdiverseexplainable}} & En & Wikipedia & Multi-hop (2-hop) & 7{,}405 & Multi-hop reasoning  \\
\makecell[tl]{Enterprise RAG Bench\\ \citep{sun2026enterpriseragbenchragbenchmarkcompany}} & En & Simulated enterprise docs (Slack, Jira, Confluence) & Constrained, Semantic, Conflicting info, \dots & 470 & Realistic enterprise RAG \\
\makecell[tl]{ConditionalQA\\
\citep{sun-etal-2022-conditionalqa}} & En & Public policy & Extractive, boolean, conditional, multi-hop & 271 & Scenario descriptions, conditional answers \\
\midrule
\textbf{Total} & \multicolumn{3}{l}{10 domains, 5 languages} & \textbf{12{,}775} & \\
\bottomrule
\end{tabular}
\vspace{-0.6em}
\caption{\textbf{Integration Test} datasets. All examples provide a pre-retrieved candidate pool (with the gold document guaranteed within the top-50 candidates) and gold-evidence span annotations. (Academic Paper is the sole exception that uses a fixed 10-document pool; Appendix~\ref{app:academic_paper}).}
\label{tab:benchmark}
\end{table*}

\subsection{System Test (Retrieval-Dynamic)}
\label{sec:system-test}

The System Test embeds the compressor inside a live search agent that issues its
own queries over multiple turns and decides, based on what the compression model returns,
whether to answer or search again.
Retrieval is no longer fixed, so any passage the compressor processes incorrectly can steer the agent toward a worse follow-up query, letting errors compound turn over turn in ways invisible to single-shot evaluation. This tier spans two complementary settings drawn from existing agentic
benchmarks: open-domain web browsing over a large, heterogeneous corpus
(BrowseComp-Plus~\citep{chen2025browsecompplusfairtransparentevaluation}) and
structured, table- and number-heavy enterprise document QA
(OfficeQA~\citep{officeqa,opsahlong2026officeqaproenterprisebenchmark}), for a
combined 1{,}076 queries.

Following \citet{chen2025browsecompplusfairtransparentevaluation}, the baseline passes the top-5 retrieved documents to the agent without reranking or compression. We use Qwen3-Embedding-8B over the pre-built BrowseComp-Plus index for retrieval and gpt-oss-120B as the agent. OfficeQA follows the same protocol with Embed v4 as the retriever.
For all compressors we use the recommended thresholds, and conservative settings only when these compress too aggressively (Appendix~\ref{sec:appendix_browsecomp_result}). For OfficeQA we mainly report exact and numeric-tolerance accuracy following \citet{officeqa}. LLM-judge results are in Appendix~\ref{app:officeqa_judge}.
More details in Appendix~\ref{sec:appendix_browsecomp_result}.

\section{Metrics}

\textsc{UniCoRn} uses LLM judge scores as the main quality metric across tiers.
\emph{Evidence Recall} and \emph{Ranking-Compression Alignment} are introduced as additional diagnostic metrics.

\paragraph{LLM Judge.}

Across all tiers, generated answers are
scored by an LLM judge (Qwen3.5-35B-A3B, \citet{qwen3.5}) against reference answers. See Appendix~\ref{sec:appendix_prompts} for
prompting details and robustness of judge selection.

\paragraph{Evidence Recall.}
End-task accuracy alone cannot tell whether a compressor discarded the
information needed to answer. Evidence Recall measures the fraction of gold evidence that survives compression.
Three datasets (ConditionalQA, HotpotQA and LegalQuAD) provide native evidence annotations that are used directly.
For all remaining datasets, evidence spans are annotated by an ensemble of two LLMs:
a span selected by both models gets weight $w{=}2$, a span selected by one gets $w{=}1$, and the metric is
the weight of preserved spans divided by the weight of all gold spans. This
rewards consensus evidence more while still crediting single-model spans.
A human-annotated subset validates the LLM annotations
(Appendix~\ref{app:evidence}).

\paragraph{Ranking--Compression Alignment.}
Ideally, ranking and compression should agree and the top-ranked document should not be emptied entirely. We report three diagnostics
(Table~\ref{tab:alignment}). \emph{M1, top-1 empty rate}: the fraction of
queries for which compression completely removes the rank-1 document. \emph{M2, rank alignment}: per query, the Spearman agreement
between the reranker order and the order induced by token retention
$\mathrm{kept}(d)/\mathrm{orig}(d)$, averaged over queries. \emph{M3,
downstream impact}: the gap in mean judge accuracy between queries whose
rank-1 document stays nonempty and those whose rank-1 document is emptied (Appendix \ref{sec:appendix_alignment}).

\definecolor{heatLow}{RGB}{210,96,168}
\definecolor{heatMid}{RGB}{214,226,246}
\definecolor{heatHigh}{RGB}{46,110,210}
\definecolor{bestborder}{RGB}{124,72,196}

\newlength{\hcw}\setlength{\hcw}{0.95cm}  
\newlength{\havgw}\setlength{\havgw}{0.80cm}

\providecommand{\heat}{}\renewcommand{\heat}[3]{\cellcolor{#1}\textcolor{#2}{#3}}
\providecommand{\heatb}{}\renewcommand{\heatb}[3]{%
  \hspace*{-\tabcolsep}\hspace*{0.65pt}%
  \tikz[baseline=(n.base)]{%
    \node[
      fill=#1,
      draw=bestborder,
      line width=1.35pt,
      rounded corners=1pt,
      inner sep=0pt,
      outer sep=0pt,
      minimum width=\dimexpr\linewidth+2\tabcolsep-1.25pt\relax,
      minimum height=\dimexpr\arraystretch\baselineskip-1.25pt\relax,
      align=center,
      text height=0.65em,
      text depth=0.15em
    ](n){\textcolor{#2}{\textbf{#3}}};%
  }%
  \hspace*{\dimexpr-\tabcolsep+0.65pt\relax}%
}
\begin{table*}[ht!]
  \centering
  \setlength{\tabcolsep}{3.2pt}
  \renewcommand{\arraystretch}{1.32}
  \scriptsize
\begin{tabular}{l *{10}{>{\centering\arraybackslash}m{\hcw}} @{\hskip 5pt}| >{\centering\arraybackslash}m{\havgw}}
    \toprule
    \textbf{Model (threshold, comp \%)}
      & \shortstack{\textbf{Multi-hop}}
      & \shortstack{\textbf{Numeric}}
      & \shortstack{\textbf{Scattered}}
      & \shortstack{\textbf{Info-dens.}}
      & \shortstack{\textbf{Adversar.}}
      & \shortstack{\textbf{Context.}}
      & \shortstack{\textbf{Union}}
      & \shortstack{\textbf{Summariz.}}
      & \shortstack{\textbf{Ugly-fmt}}
      & \shortstack{\textbf{X-lingual}}
      & \textbf{Avg.} \\
    \midrule
    Unpruned (Oracle) & \heat{heatHigh!54!heatMid}{black}{0.77} & \heat{heatHigh!98!heatMid}{white}{0.99} & \heat{heatHigh!94!heatMid}{white}{0.97} & \heat{heatHigh!100!heatMid}{white}{1.00} & \heat{heatHigh!100!heatMid}{white}{1.00} & \heat{heatHigh!100!heatMid}{white}{1.00} & \heat{heatHigh!94!heatMid}{white}{0.97} & \heat{heatMid!70!heatLow}{black}{0.35} & \heat{heatHigh!80!heatMid}{white}{0.90} & \heat{heatHigh!92!heatMid}{white}{0.96} & \heat{heatHigh!78!heatMid}{white}{0.89} \\
    \midrule
    \textbf{Comp-H (ours)} ($\tau$0.4, 75\%) & \heat{heatHigh!4!heatMid}{black}{0.52} & \heatb{heatHigh!100!heatMid}{white}{1.00} & \heatb{heatHigh!56!heatMid}{white}{0.78} & \heatb{heatHigh!100!heatMid}{white}{1.00} & \heatb{heatHigh!100!heatMid}{white}{1.00} & \heat{heatHigh!44!heatMid}{black}{0.72} & \heatb{heatHigh!84!heatMid}{white}{0.92} & \heat{heatMid!90!heatLow}{black}{0.45} & \heat{heatHigh!74!heatMid}{white}{0.87} & \heat{heatHigh!42!heatMid}{black}{0.71} & \heatb{heatHigh!60!heatMid}{white}{0.80} \\
    \textbf{Comp-S (ours)} (94\%) & \heat{heatMid!6!heatLow}{white}{0.03} & \heat{heatHigh!42!heatMid}{black}{0.71} & \heat{heatMid!30!heatLow}{white}{0.15} & \heat{heatHigh!62!heatMid}{white}{0.81} & \heat{heatHigh!20!heatMid}{black}{0.60} & \heat{heatHigh!52!heatMid}{white}{0.76} & \heat{heatMid!46!heatLow}{white}{0.23} & \heatb{heatHigh!50!heatMid}{white}{0.75} & \heat{heatHigh!60!heatMid}{white}{0.80} & \heat{heatMid!48!heatLow}{white}{0.24} & \heat{heatHigh!2!heatMid}{black}{0.51} \\
    \midrule
    Semantic Highlight ($\tau$0.1, 76\%) & \heat{heatHigh!30!heatMid}{black}{0.65} & \heatb{heatHigh!100!heatMid}{white}{1.00} & \heat{heatHigh!48!heatMid}{white}{0.74} & \heatb{heatHigh!100!heatMid}{white}{1.00} & \heatb{heatHigh!100!heatMid}{white}{1.00} & \heat{heatMid!82!heatLow}{black}{0.41} & \heat{heatHigh!76!heatMid}{white}{0.88} & \heat{heatHigh!2!heatMid}{black}{0.51} & \heat{heatHigh!78!heatMid}{white}{0.89} & \heatb{heatHigh!66!heatMid}{white}{0.83} & \heat{heatHigh!58!heatMid}{white}{0.79} \\
    OpenProvence ($\tau$0.2, 76\%) & \heat{heatHigh!42!heatMid}{black}{0.71} & \heat{heatHigh!76!heatMid}{white}{0.88} & \heat{heatHigh!42!heatMid}{black}{0.71} & \heatb{heatHigh!100!heatMid}{white}{1.00} & \heatb{heatHigh!100!heatMid}{white}{1.00} & \heat{heatHigh!42!heatMid}{black}{0.71} & \heat{heatHigh!76!heatMid}{white}{0.88} & \heat{heatMid!72!heatLow}{black}{0.36} & \heat{heatHigh!78!heatMid}{white}{0.89} & \heat{heatHigh!42!heatMid}{black}{0.71} & \heat{heatHigh!56!heatMid}{white}{0.78} \\
    Provence ($\tau$0.2, 74\%) & \heat{heatMid!64!heatLow}{black}{0.32} & \heat{heatHigh!86!heatMid}{white}{0.93} & \heat{heatHigh!42!heatMid}{black}{0.71} & \heatb{heatHigh!100!heatMid}{white}{1.00} & \heat{heatHigh!60!heatMid}{white}{0.80} & \heat{heatHigh!24!heatMid}{black}{0.62} & \heat{heatHigh!78!heatMid}{white}{0.89} & \heat{heatMid!96!heatLow}{black}{0.48} & \heatb{heatHigh!80!heatMid}{white}{0.90} & \heat{heatHigh!40!heatMid}{black}{0.70} & \heat{heatHigh!46!heatMid}{white}{0.73} \\
    DSLR-Zerank ($\tau$0.3, 73\%) & \heat{heatHigh!2!heatMid}{black}{0.51} & \heatb{heatHigh!100!heatMid}{white}{1.00} & \heat{heatHigh!24!heatMid}{black}{0.62} & \heatb{heatHigh!100!heatMid}{white}{1.00} & \heat{heatHigh!80!heatMid}{white}{0.90} & \heat{heatMid!66!heatLow}{black}{0.33} & \heat{heatHigh!56!heatMid}{white}{0.78} & \heat{heatMid!96!heatLow}{black}{0.48} & \heat{heatHigh!78!heatMid}{white}{0.89} & \heat{heatHigh!56!heatMid}{white}{0.78} & \heat{heatHigh!46!heatMid}{white}{0.73} \\
    Sentinel ($\tau$0.3, 72\%) & \heatb{heatHigh!60!heatMid}{white}{0.80} & \heat{heatHigh!76!heatMid}{white}{0.88} & \heat{heatHigh!26!heatMid}{black}{0.63} & \heatb{heatHigh!100!heatMid}{white}{1.00} & \heat{heatHigh!80!heatMid}{white}{0.90} & \heat{heatHigh!46!heatMid}{white}{0.73} & \heat{heatMid!90!heatLow}{black}{0.45} & \heat{heatMid!78!heatLow}{black}{0.39} & \heat{heatHigh!40!heatMid}{black}{0.70} & \heat{heatHigh!34!heatMid}{black}{0.67} & \heat{heatHigh!44!heatMid}{black}{0.72} \\
    RECOMP ($\tau$0.5, 68\%) & \heat{heatMid!82!heatLow}{black}{0.41} & \heat{heatHigh!80!heatMid}{white}{0.90} & \heat{heatHigh!4!heatMid}{black}{0.52} & \heat{heatHigh!80!heatMid}{white}{0.90} & \heat{heatHigh!62!heatMid}{white}{0.81} & \heat{heatHigh!22!heatMid}{black}{0.61} & \heat{heatMid!46!heatLow}{white}{0.23} & \heat{heatMid!66!heatLow}{black}{0.33} & \heat{heatHigh!20!heatMid}{black}{0.60} & \heat{heatMid!6!heatLow}{white}{0.03} & \heat{heatHigh!6!heatMid}{black}{0.53} \\
    XProvence ($\tau$0.3, 78\%) & \heat{heatMid!6!heatLow}{white}{0.03} & \heat{heatMid!84!heatLow}{black}{0.42} & \heat{heatMid!12!heatLow}{white}{0.06} & \heat{heatHigh!82!heatMid}{white}{0.91} & \heat{heatHigh!40!heatMid}{black}{0.70} & \heat{heatHigh!28!heatMid}{black}{0.64} & \heat{heatHigh!40!heatMid}{black}{0.70} & \heat{heatHigh!20!heatMid}{black}{0.60} & \heat{heatHigh!42!heatMid}{black}{0.71} & \heat{heatHigh!2!heatMid}{black}{0.51} & \heat{heatHigh!6!heatMid}{black}{0.53} \\
    Oscar (94\%) & \heat{heatMid!4!heatLow}{white}{0.02} & \heat{heatMid!66!heatLow}{black}{0.33} & \heat{heatMid!10!heatLow}{white}{0.05} & \heat{heatHigh!74!heatMid}{white}{0.87} & \heat{heatHigh!80!heatMid}{white}{0.90} & \heatb{heatHigh!60!heatMid}{white}{0.80} & \heat{heatMid!50!heatLow}{white}{0.25} & \heat{heatMid!82!heatLow}{black}{0.41} & \heat{heatMid!84!heatLow}{black}{0.42} & \heat{heatMid!22!heatLow}{white}{0.11} & \heat{heatMid!84!heatLow}{black}{0.42} \\
    LLMLingua-2 ($\tau$0.2, 76\%) & \heat{heatMid!34!heatLow}{white}{0.17} & \heat{heatMid!64!heatLow}{black}{0.32} & \heat{heatMid!18!heatLow}{white}{0.09} & \heat{heatMid!94!heatLow}{black}{0.47} & \heat{heatMid!24!heatLow}{white}{0.12} & \heat{heatMid!82!heatLow}{black}{0.41} & \heat{heatMid!40!heatLow}{white}{0.20} & \heat{heatHigh!2!heatMid}{black}{0.51} & \heat{heatMid!86!heatLow}{black}{0.43} & \heat{heatMid!62!heatLow}{white}{0.31} & \heat{heatMid!60!heatLow}{white}{0.30} \\
    Keyword Snippet (86\%) & \heat{heatMid!6!heatLow}{white}{0.03} & \heat{heatMid!64!heatLow}{black}{0.32} & \heat{heatMid!14!heatLow}{white}{0.07} & \heat{heatMid!82!heatLow}{black}{0.41} & \heat{heatMid!42!heatLow}{white}{0.21} & \heat{heatMid!84!heatLow}{black}{0.42} & \heat{heatMid!2!heatLow}{white}{0.01} & \heat{heatMid!36!heatLow}{white}{0.18} & \heat{heatMid!24!heatLow}{white}{0.12} & \heat{heatMid!84!heatLow}{black}{0.42} & \heat{heatMid!44!heatLow}{white}{0.22} \\
    LongLLMLingua ($\tau$0.2, 65\%) & \heat{heatMid!10!heatLow}{white}{0.05} & \heat{heatMid!26!heatLow}{white}{0.13} & \heat{heatMid!8!heatLow}{white}{0.04} & \heat{heatMid!66!heatLow}{black}{0.33} & \heat{heatMid!44!heatLow}{white}{0.22} & \heat{heatMid!46!heatLow}{white}{0.23} & \heat{heatMid!12!heatLow}{white}{0.06} & \heat{heatMid!14!heatLow}{white}{0.07} & \heat{heatMid!62!heatLow}{white}{0.31} & \heat{heatMid!26!heatLow}{white}{0.13} & \heat{heatMid!30!heatLow}{white}{0.15} \\
    \bottomrule
  \end{tabular}
  \vspace{-1em}
  \caption{Per-query-type judge accuracy on the \textbf{Unit Test} at matched compression ratio (each at the threshold closest to
    75\%). Comp-S, Oscar and Keyword Snippet expose no keep-threshold and therefore appear as
single points (compression ratio in parentheses); for the two soft compressors this is
the nominal rate-16 figure. Blue outline marks the best pruning model per column, ties included.}
  \label{tab:unit-heatmap}
\end{table*}


\section{Compression Methods}

This section first presents evaluated existing methods and then our own trained baselines.

\subsection{Existing Public Baselines}
We compare different existing baselines including hard and soft compression. Provence \citep{chirkova2025provenceefficientrobustcontext}, XProvence \citep{mohamed2026XProvencezerocostmultilingualcontext}, OpenProvence \citep{yuichi-tateno-2025-open-provence}, Semantic Highlight \citep{zilliz2026semantic_highlight}, LLMLingua-2 \citep{pan2024llmlingua2datadistillationefficient}, LongLLMLingua \citep{jiang2024longllmlinguaacceleratingenhancingllms}, Sentinel \citep{zhang2026sentineldecodingcontextutilization}, RECOMP \citep{xu2024recomp}, DSLR \citep{hwang-etal-2024-dslr} \footnote{We use Zerank-2 as reranker for DSLR, following \citet{zeroentropy}.}, and OSCAR \citep{louis2026oscar}.
We additionally include Keyword Snippet, a rule-based baseline that scores spans by
query-term overlap and applies no learned model. See Table~\ref{tab:baselines} in Appendix~\ref{sec:appendix_baselines} for details.

\noindent
For Integration and System Tests, we keep the top-5 of the 50 retrieved
candidates by each method's document-level score. Semantic Highlight,
LLMLingua-2, Sentinel, RECOMP, OSCAR, and Comp-S produce no ranking signal,
so we rerank their candidates with \texttt{bge-reranker-v2-m3} \citep{chen2024bge} before
compression. For Unit and Integration Tests, all methods feed their retained text to a shared
generator, Qwen3.5-9B \citep{qwen3.5}, except OSCAR, which decodes with its own trained
model.

\subsection{Matched Hard and Soft Compression}

Existing baselines use a wide variety of base models and training corpora.
To allow for a more direct comparison between the hard and soft compression regimes, we train our own hard (\emph{Comp-H}) and soft (\emph{Comp-S}) compressor baselines using identical base model, generator and training data.

\paragraph{Compressor.} Both compression regimes share the same pre-trained base. We select BidirLM-0.6B \citep{boizard2026bidirlmtextomnimodalbidirectional}, a Qwen decoder adapted to a bidirectional encoder by continued pre-training.
For Comp-H, we follow the setup of \citet{chirkova2025provenceefficientrobustcontext} with a per-token relevance prediction head and a per-sequence reranking head.
For Comp-S, we follow \citet{louis2026oscar} in appending learnable memory tokens to the compressor input.


\paragraph{Training data.} The first pool of training data, shared with many public baselines, primarily contains queries and documents sourced from web search and Wikipedia, such as MSMarco \citep{bajaj2016ms} or Natural Questions \citep{kwiatkowski-etal-2019-natural}, extended by similar multilingual data.
We extend this data by a second pool that consists of publicly available corpora of parsed PDF documents, such as from ColPali \citep{faysse2025colpali}.
Each training sample consists of query, positive, and mined hard negative documents.
 Details on the exact training datasets used can be found in Appendix~\ref{app:train_data}.

For Comp-H, we derive sequence level relevance annotations from an LLM teacher.
To support diverse and multilingual documents, all text is segmented with the model-based method proposed in \citet{frohmann-etal-2024-segment}.
We add rule-based segmentation for splitting tables into rows.
All segmented documents are then annotated with Qwen3.5-27B \citep{qwen3.5} with reasoning enabled.
For Comp-S, we augment all training samples with answers by the generator from the non-compressed documents.
For reranking, we obtain teacher scores by scoring all samples with Cohere Rerank 4 Pro \citep{rerank4}.
As a last preparation step, we filter all training sets to remove samples with low consistency between the span relevance and the reranking teacher.
See Appendix~\ref{app:train_proc} for details on filtering and Appendix~\ref{app:train_details} for training details.

\section{Results \& Analysis}

\input{figure_latex/compression_integration_plot}

\definecolor{heatLow}{RGB}{210,96,168}
\definecolor{heatMid}{RGB}{214,226,246}
\definecolor{heatHigh}{RGB}{46,110,210}
\definecolor{bestborder}{RGB}{124,72,196}

\providecommand{\heat}{}\renewcommand{\heat}[3]{%
  \cellcolor{#1}\textcolor{#2}{#3}%
}
\ifdefined\hcw\else\newlength{\hcw}\fi
\ifdefined\hch\else\newlength{\hch}\fi
\setlength{\hcw}{17pt}
\setlength{\hch}{11.5pt}

\providecommand{\hbest}{}\renewcommand{\hbest}[3]{%
  \hspace*{-\tabcolsep}\hspace*{0.65pt}%
  \tikz[baseline=(n.base)]{%
    \node[
      fill=#1,
      draw=bestborder,
      line width=1.3pt,
      rounded corners=1.2pt,
      inner sep=0pt,
      outer sep=0pt,
      minimum width=\dimexpr\linewidth+2\tabcolsep-1.3pt\relax,
      minimum height=\dimexpr\arraystretch\baselineskip-1.3pt\relax,
      align=center,
      text height=0.7em,
      text depth=0.2em
    ](n){\textcolor{#2}{\textbf{#3}}};%
  }%
  \hspace*{\dimexpr-\tabcolsep+0.65pt\relax}%
}

\begin{table*}[t]
\centering
\resizebox{\textwidth}{!}{%
\setlength{\tabcolsep}{3pt}
\renewcommand{\arraystretch}{1.32}
\scriptsize
\begin{tabular}{l *{17}{>{\centering\arraybackslash}m{\hcw}}}
& \multicolumn{5}{c}{\textbf{EN Factoid / RAG}}
& \multicolumn{4}{c}{\textbf{Multilingual}}
& \multicolumn{5}{c}{\textbf{Vidore -- EN}}
& \multicolumn{3}{c}{\textbf{Vidore -- FR}} \\
\cmidrule(lr){2-6}
\cmidrule(lr){7-10}
\cmidrule(lr){11-15}
\cmidrule(lr){16-18}
\textbf{Model (config, ratio)}
& \textbf{Acad.} & \textbf{ConQA} & \textbf{FinQA}
& \textbf{Hotpot} & \textbf{Enter.}
& \textbf{JSQ.} & \textbf{KorQ.} & \textbf{KLUE} & \textbf{Legal}
& \textbf{CS} & \textbf{Fin.} & \textbf{HR} & \textbf{Ind.}
& \textbf{Pharm.} & \textbf{Energy} & \textbf{Fin.} & \textbf{Phys.} \\
\midrule

Gold Oracle
& \heat{heatHigh!90!heatMid}{white}{.95}
& \heat{heatHigh!52!heatMid}{white}{.76}
& \heat{heatHigh!10!heatMid}{black}{.55}
& \heat{heatHigh!74!heatMid}{white}{.87}
& \heat{heatHigh!66!heatMid}{white}{.83}
& \heat{heatHigh!92!heatMid}{white}{.96}
& \heat{heatHigh!68!heatMid}{white}{.84}
& \heat{heatHigh!76!heatMid}{white}{.88}
& \heat{heatHigh!66!heatMid}{white}{.83}
& \heat{heatHigh!38!heatMid}{black}{.69}
& \heat{heatHigh!28!heatMid}{black}{.64}
& \heat{heatHigh!22!heatMid}{black}{.61}
& \heat{heatHigh!24!heatMid}{black}{.62}
& \heat{heatHigh!16!heatMid}{black}{.58}
& \heat{heatHigh!26!heatMid}{black}{.63}
& \heat{heatHigh!16!heatMid}{black}{.58}
& \heat{heatHigh!10!heatMid}{black}{.55} \\
\midrule

\textbf{Comp-H} ($\tau$0.2, 75\%)
& \hbest{heatHigh!84!heatMid}{white}{.92}
& \heat{heatHigh!32!heatMid}{black}{.66}
& \hbest{heatMid!82!heatLow}{black}{.41}
& \heat{heatMid!92!heatLow}{black}{.46}
& \hbest{heatHigh!18!heatMid}{black}{.59}
& \heat{heatHigh!84!heatMid}{white}{.92}
& \hbest{heatHigh!50!heatMid}{white}{.75}
& \hbest{heatHigh!52!heatMid}{white}{.76}
& \heat{heatHigh!50!heatMid}{white}{.75}
& \hbest{heatHigh!32!heatMid}{black}{.66}
& \hbest{heatHigh!10!heatMid}{black}{.55}
& \hbest{heatHigh!16!heatMid}{black}{.58}
& \hbest{heatHigh!8!heatMid}{black}{.54}
& \hbest{heatHigh!4!heatMid}{black}{.52}
& \hbest{heatHigh!20!heatMid}{black}{.60}
& \hbest{heatHigh!8!heatMid}{black}{.54}
& \hbest{heatHigh!10!heatMid}{black}{.55} \\

\textbf{Comp-S} (94\%)
& \heat{heatMid!32!heatLow}{white}{.16}
& \heat{heatMid!96!heatLow}{black}{.48}
& \heat{heatMid!14!heatLow}{white}{.07}
& \heat{heatMid!84!heatLow}{black}{.42}
& \heat{heatMid!66!heatLow}{black}{.33}
& \heat{heatHigh!22!heatMid}{black}{.61}
& \heat{heatMid!36!heatLow}{white}{.18}
& \heat{heatMid!50!heatLow}{white}{.25}
& \heat{heatMid!66!heatLow}{black}{.33}
& \heat{heatHigh!24!heatMid}{black}{.62}
& \heat{heatMid!74!heatLow}{black}{.37}
& \heat{heatMid!74!heatLow}{black}{.37}
& \heat{heatMid!78!heatLow}{black}{.39}
& \heat{heatMid!66!heatLow}{black}{.33}
& \heat{heatMid!84!heatLow}{black}{.42}
& \heat{heatMid!68!heatLow}{black}{.34}
& \heat{heatMid!100!heatLow}{black}{.50} \\
\midrule

Oscar (94\%)
& \heat{heatMid!24!heatLow}{white}{.12}
& \heat{heatMid!80!heatLow}{black}{.40}
& \heat{heatMid!8!heatLow}{white}{.04}
& \hbest{heatHigh!38!heatMid}{black}{.69}
& \heat{heatMid!24!heatLow}{white}{.12}
& \heat{heatMid!98!heatLow}{black}{.49}
& \heat{heatMid!14!heatLow}{white}{.07}
& \heat{heatMid!34!heatLow}{white}{.17}
& \heat{heatMid!14!heatLow}{white}{.07}
& \heat{heatHigh!14!heatMid}{black}{.57}
& \heat{heatMid!48!heatLow}{white}{.24}
& \heat{heatMid!56!heatLow}{white}{.28}
& \heat{heatMid!68!heatLow}{black}{.34}
& \heat{heatMid!64!heatLow}{white}{.32}
& \heat{heatMid!44!heatLow}{white}{.22}
& \heat{heatMid!32!heatLow}{white}{.16}
& \heat{heatMid!52!heatLow}{white}{.26} \\

DSLR-Zerank ($\tau$0.5, 75\%)
& \heat{heatHigh!44!heatMid}{black}{.72}
& \heat{heatHigh!32!heatMid}{black}{.66}
& \heat{heatMid!54!heatLow}{white}{.27}
& \heat{heatMid!46!heatLow}{white}{.23}
& \heat{heatMid!82!heatLow}{black}{.41}
& \hbest{heatHigh!84!heatMid}{white}{.92}
& \heat{heatHigh!16!heatMid}{black}{.58}
& \heat{heatMid!98!heatLow}{black}{.49}
& \heat{heatHigh!14!heatMid}{black}{.57}
& \heat{heatHigh!2!heatMid}{black}{.51}
& \heat{heatMid!88!heatLow}{black}{.44}
& \heat{heatMid!90!heatLow}{black}{.45}
& \heat{heatMid!78!heatLow}{black}{.39}
& \heat{heatMid!92!heatLow}{black}{.46}
& \heat{heatHigh!6!heatMid}{black}{.53}
& \heat{heatMid!86!heatLow}{black}{.43}
& \heat{heatHigh!6!heatMid}{black}{.53} \\

OpenProvence ($\tau$0.2, 74\%)
& \heat{heatHigh!68!heatMid}{white}{.84}
& \hbest{heatHigh!42!heatMid}{black}{.71}
& \heat{heatMid!58!heatLow}{white}{.29}
& \heat{heatMid!88!heatLow}{black}{.44}
& \heat{heatMid!100!heatLow}{black}{.50}
& \heat{heatHigh!84!heatMid}{white}{.92}
& \heat{heatHigh!22!heatMid}{black}{.61}
& \heat{heatHigh!22!heatMid}{black}{.61}
& \heat{heatHigh!40!heatMid}{black}{.70}
& \heat{heatHigh!24!heatMid}{black}{.62}
& \heat{heatMid!90!heatLow}{black}{.45}
& \heat{heatHigh!4!heatMid}{black}{.52}
& \heat{heatMid!92!heatLow}{black}{.46}
& \heat{heatMid!90!heatLow}{black}{.45}
& \heat{heatHigh!2!heatMid}{black}{.51}
& \heat{heatMid!86!heatLow}{black}{.43}
& \heat{heatMid!90!heatLow}{black}{.45} \\

Semantic HL ($\tau$0.1, 77\%)
& \heat{heatHigh!60!heatMid}{white}{.80}
& \heat{heatHigh!26!heatMid}{black}{.63}
& \heat{heatMid!52!heatLow}{white}{.26}
& \heat{heatMid!84!heatLow}{black}{.42}
& \heat{heatHigh!10!heatMid}{black}{.55}
& \heat{heatHigh!84!heatMid}{white}{.92}
& \heat{heatHigh!44!heatMid}{black}{.72}
& \heat{heatHigh!24!heatMid}{black}{.62}
& \hbest{heatHigh!54!heatMid}{white}{.77}
& \heat{heatHigh!18!heatMid}{black}{.59}
& \heat{heatMid!94!heatLow}{black}{.47}
& \heat{heatHigh!2!heatMid}{black}{.51}
& \heat{heatMid!88!heatLow}{black}{.44}
& \heat{heatMid!100!heatLow}{black}{.50}
& \heat{heatHigh!10!heatMid}{black}{.55}
& \heat{heatMid!98!heatLow}{black}{.49}
& \heat{heatHigh!6!heatMid}{black}{.53} \\

Provence ($\tau$0.1, 74\%)
& \heat{heatHigh!56!heatMid}{white}{.78}
& \heat{heatHigh!24!heatMid}{black}{.62}
& \heat{heatMid!44!heatLow}{white}{.22}
& \heat{heatMid!70!heatLow}{black}{.35}
& \heat{heatMid!94!heatLow}{black}{.47}
& \heat{heatHigh!82!heatMid}{white}{.91}
& \heat{heatMid!94!heatLow}{black}{.47}
& \heat{heatHigh!2!heatMid}{black}{.51}
& \heat{heatHigh!2!heatMid}{black}{.51}
& \heat{heatHigh!24!heatMid}{black}{.62}
& \heat{heatMid!84!heatLow}{black}{.42}
& \heat{heatMid!94!heatLow}{black}{.47}
& \heat{heatMid!88!heatLow}{black}{.44}
& \heat{heatMid!98!heatLow}{black}{.49}
& \heat{heatMid!92!heatLow}{black}{.46}
& \heat{heatMid!80!heatLow}{black}{.40}
& \heat{heatMid!98!heatLow}{black}{.49} \\

Sentinel ($\tau$0.1, 77\%)
& \heat{heatHigh!50!heatMid}{white}{.75}
& \heat{heatHigh!8!heatMid}{black}{.54}
& \heat{heatMid!44!heatLow}{white}{.22}
& \heat{heatMid!94!heatLow}{black}{.47}
& \heat{heatMid!70!heatLow}{black}{.35}
& \heat{heatHigh!66!heatMid}{white}{.83}
& \heat{heatMid!78!heatLow}{black}{.39}
& \heat{heatHigh!4!heatMid}{black}{.52}
& \heat{heatHigh!22!heatMid}{black}{.61}
& \heat{heatMid!76!heatLow}{black}{.38}
& \heat{heatMid!76!heatLow}{black}{.38}
& \heat{heatMid!72!heatLow}{black}{.36}
& \heat{heatMid!64!heatLow}{white}{.32}
& \heat{heatMid!88!heatLow}{black}{.44}
& \heat{heatMid!84!heatLow}{black}{.42}
& \heat{heatMid!78!heatLow}{black}{.39}
& \heat{heatMid!92!heatLow}{black}{.46} \\

RECOMP ($\tau$0.6, 70\%)
& \heat{heatHigh!38!heatMid}{black}{.69}
& \heat{heatMid!90!heatLow}{black}{.45}
& \heat{heatMid!34!heatLow}{white}{.17}
& \heat{heatMid!36!heatLow}{white}{.18}
& \heat{heatMid!64!heatLow}{white}{.32}
& \heat{heatHigh!64!heatMid}{white}{.82}
& \heat{heatHigh!10!heatMid}{black}{.55}
& \heat{heatHigh!34!heatMid}{black}{.67}
& \heat{heatMid!80!heatLow}{black}{.40}
& \heat{heatHigh!18!heatMid}{black}{.59}
& \heat{heatMid!68!heatLow}{black}{.34}
& \heat{heatMid!62!heatLow}{white}{.31}
& \heat{heatMid!70!heatLow}{black}{.35}
& \heat{heatMid!70!heatLow}{black}{.35}
& \heat{heatMid!68!heatLow}{black}{.34}
& \heat{heatMid!48!heatLow}{white}{.24}
& \heat{heatMid!74!heatLow}{black}{.37} \\

XProvence ($\tau$0.3, 77\%)
& \heat{heatMid!18!heatLow}{white}{.09}
& \heat{heatMid!58!heatLow}{white}{.29}
& \heat{heatMid!12!heatLow}{white}{.06}
& \heat{heatMid!74!heatLow}{black}{.37}
& \heat{heatMid!38!heatLow}{white}{.19}
& \heat{heatHigh!78!heatMid}{white}{.89}
& \heat{heatMid!54!heatLow}{white}{.27}
& \heat{heatHigh!14!heatMid}{black}{.57}
& \heat{heatMid!46!heatLow}{white}{.23}
& \heat{heatHigh!2!heatMid}{black}{.51}
& \heat{heatMid!76!heatLow}{black}{.38}
& \heat{heatMid!78!heatLow}{black}{.39}
& \heat{heatMid!72!heatLow}{black}{.36}
& \heat{heatMid!92!heatLow}{black}{.46}
& \heat{heatMid!82!heatLow}{black}{.41}
& \heat{heatMid!70!heatLow}{black}{.35}
& \heat{heatHigh!2!heatMid}{black}{.51} \\

LLMLingua-2 ($\tau$0.2, 77\%)
& \heat{heatMid!76!heatLow}{black}{.38}
& \heat{heatMid!96!heatLow}{black}{.48}
& \heat{heatMid!10!heatLow}{white}{.05}
& \heat{heatMid!30!heatLow}{white}{.15}
& \heat{heatMid!34!heatLow}{white}{.17}
& \heat{heatMid!28!heatLow}{white}{.14}
& \heat{heatMid!64!heatLow}{white}{.32}
& \heat{heatMid!42!heatLow}{white}{.21}
& \heat{heatMid!64!heatLow}{white}{.32}
& \heat{heatMid!82!heatLow}{black}{.41}
& \heat{heatMid!48!heatLow}{white}{.24}
& \heat{heatMid!48!heatLow}{white}{.24}
& \heat{heatMid!42!heatLow}{white}{.21}
& \heat{heatMid!30!heatLow}{white}{.15}
& \heat{heatMid!58!heatLow}{white}{.29}
& \heat{heatMid!54!heatLow}{white}{.27}
& \heat{heatMid!36!heatLow}{white}{.18} \\

LongLLMLingua ($\tau$0.1, 64\%)
& \heat{heatMid!30!heatLow}{white}{.15}
& \heat{heatMid!48!heatLow}{white}{.24}
& \heat{heatMid!10!heatLow}{white}{.05}
& \heat{heatMid!84!heatLow}{black}{.42}
& \heat{heatMid!26!heatLow}{white}{.13}
& \heat{heatMid!94!heatLow}{black}{.47}
& \heat{heatMid!10!heatLow}{white}{.05}
& \heat{heatMid!12!heatLow}{white}{.06}
& \heat{heatMid!20!heatLow}{white}{.10}
& \heat{heatMid!40!heatLow}{white}{.20}
& \heat{heatMid!32!heatLow}{white}{.16}
& \heat{heatMid!30!heatLow}{white}{.15}
& \heat{heatMid!36!heatLow}{white}{.18}
& \heat{heatMid!48!heatLow}{white}{.24}
& \heat{heatMid!30!heatLow}{white}{.15}
& \heat{heatMid!30!heatLow}{white}{.15}
& \heat{heatMid!50!heatLow}{white}{.25} \\

Keyword Snippet (80\%)
& \heat{heatMid!14!heatLow}{white}{.07}
& \heat{heatMid!26!heatLow}{white}{.13}
& \heat{heatMid!10!heatLow}{white}{.05}
& \heat{heatMid!54!heatLow}{white}{.27}
& \heat{heatMid!24!heatLow}{white}{.12}
& \heat{heatHigh!84!heatMid}{white}{.92}
& \heat{heatMid!62!heatLow}{white}{.31}
& \heat{heatMid!82!heatLow}{black}{.41}
& \heat{heatMid!48!heatLow}{white}{.24}
& \heat{heatMid!34!heatLow}{white}{.17}
& \heat{heatMid!26!heatLow}{white}{.13}
& \heat{heatMid!26!heatLow}{white}{.13}
& \heat{heatMid!40!heatLow}{white}{.20}
& \heat{heatMid!48!heatLow}{white}{.24}
& \heat{heatMid!28!heatLow}{white}{.14}
& \heat{heatMid!30!heatLow}{white}{.15}
& \heat{heatMid!52!heatLow}{white}{.26} \\
\bottomrule
\end{tabular}%
}
\vspace{-0.6em}
\caption{\textbf{Integration Test} per-dataset judge accuracy at matched
compression ($\approx$75\% target); each model is shown at its configuration
closest to 75\%. Comp-H outperforms other baselines in 13 out of 17 datasets. See Appendix~\ref{sec:appendix_integrationtest_result}.}
\vspace{-5pt}
\label{tab:heatmap-integration}
\end{table*}


\subsection{Unit Test}

\paragraph{Overview.} Table~\ref{tab:unit-heatmap} reports judge accuracy on each of the ten unit-test query types at matched compression.
At the $\approx$75\% compression point, margins among the top three are narrow.
Comp-H is best or tied-best on five of the ten types and leads with 0.80, followed by Semantic Highlight (0.79) and OpenProvence (0.78). Sentinel (0.72) and RECOMP (0.53) rank below Provence and DSLR-Zerank (0.73), and all sentence-level methods stay far above LLMLingua-2, LongLLMLingua, and Keyword Snippet. Semantic Highlight matches or surpasses Comp-H at higher compression (Figure~\ref{fig:unit-accuracy-vs-threshold-comp} in Appendix~\ref{sec:appendix_unittest_result}). 

\paragraph{Compressor failure modes.}
Unit Test controls the context construction process, so failures directly reflect limitations in compression rather than retrieval errors. Table~\ref{tab:unit-heatmap} shows that compression difficulty is highly dependent on the structure of the required evidence. The types that are hard for every method break the same assumption: that the answer resides in the single most query-similar sentence, which is the signal most compressors rely on. In \textsc{contextualized} queries, the answer sentence often contains only a pronoun, while the referenced entity appears in an earlier sentence with low query similarity. Similarity-based compressors therefore discard the antecedent and break the coreference chain. \textsc{scattered} queries spread the required evidence over several spans that are all needed, so a compressor tuned to keep the top span returns a partial answer. \textsc{multi-hop Long} queries embed the two gold HotpotQA paragraphs inside a single $128$K-token document of distractor paragraphs, so removing either one breaks the reasoning chain. \textsc{cross-lingual} queries require relevance matching across languages, which is noisier and causes correct evidence to be discarded more often as compression increases. These failures are structural rather than model specific, which is why these categories remain challenging for most methods. On top of this, RECOMP drops to 0.03 on \textsc{cross-lingual}, likely because its encoder is trained only on English NQ.

\paragraph{Same average, different budgets.}
The average compression ratio hides how methods spread compression across query types. At the matched setting, every other sentence-level pruner removes over 99.6\% of the \textsc{multi-hop Long} document (fewer than 500 of 128K tokens kept), while Sentinel removes 68--74\% of every type and keeps about 38K tokens. Its top \textsc{multi-hop Long} score (0.80, above the unpruned 0.77) thus comes from keeping more context, not from more precise selection. We still match on the average ratio, since in practice one threshold is used for all inputs and the compressor decides how much to keep per document. Adapting this budget is part of what the benchmark measures.

\paragraph{Where compression is beneficial.} On \textsc{information density}, \textsc{adversarial distractor}, and \textsc{numerical tables}, the best methods stay at the uncompressed ceiling while discarding most of the context, so here compression is essentially free. On \textsc{summarization}, results look even better: the uncompressed setting reaches only 0.35 and almost every compression method passes it. These broad overview questions span multiple sub-topics and have no single answer-bearing span, and with the full context generators tend to elaborate on a few sub-topics rather than covering all of them. Compression removes redundant details, and the resulting answers cover the sub-topics more evenly. Two factors help: how much is removed --- holding the model fixed, Comp-H improves from 0.45 at $\tau$0.4 to 0.62 at $\tau$0.9 --- and the compression family, where Comp-S achieves the best performance (0.75). Soft compression summarizes the context into a dense paraphrase instead of selecting sentences, which is exactly what an overview question rewards. Detailed results and qualitative examples are in Appendix~\ref{sec:appendix_unittest_result}.

\definecolor{gnblue}{HTML}{1A5FB4}
\definecolor{lsred}{HTML}{C01C28}
\newcommand{\up}[1]{\,\textsubscript{\textcolor{gnblue}{+#1}}}
\newcommand{\dn}[1]{\,\textsubscript{\textcolor{lsred}{--#1}}}
\newcommand{\eqz}{\,\textsubscript{\textcolor{gray}{$\pm$0.0}}}
\newcommand{\cb}[1]{\makebox[11.2pt][l]{\textcolor{teal!45}{\rule[-0.2ex]{\dimexpr#1pt/9\relax}{1.4ex}}}#1}
\newcommand{\nocb}{\makebox[11.2pt][l]{}---}

\begin{table*}[t]
\centering
\small
\setlength{\tabcolsep}{4pt}
\resizebox{\textwidth}{!}{%
\begin{tabular}{ll@{\hspace{8pt}}cccc@{\hspace{8pt}\vrule width 0.8pt\hspace{8pt}}ccccccc}
\toprule
 & & \multicolumn{4}{c}{\textbf{BrowseComp-Plus}} & \multicolumn{7}{c}{\textbf{OfficeQA}} \\
\cmidrule(lr){3-6}\cmidrule(lr){7-13}
\textbf{Method} & \textbf{Mode} & \textbf{Acc.} & \textbf{Recall} & \textbf{Comp.\%} & \textbf{Search} & \textbf{$\pm5\%$} & \textbf{$\pm1\%$} & \textbf{$\pm0.1\%$} & \textbf{Exact} & \textbf{Recall} & \textbf{Comp.\%} & \textbf{Search} \\
\midrule
\multirow{2}{*}{\textbf{Baseline} \footnotemark}
& Retrieval-only &40.8 & 46.0  &0  & 9.6  & 38.3 & 31.3 & 27.2 & 22.2 & 39.7 & 0 & 5.0 \\
& Rerank-only (BGE) &45.9 & 50.7 & 0 & 9.5 & \textbf{45.0} & \textbf{39.1} & \textbf{34.9} & \textbf{29.4} & 48.5 & 0 & 3.9 \\

\midrule
\multirow{2}{*}{\textbf{Comp-H}}
 & Compr-only & 47.2 & 50.2 & \cb{98.3} & 12.5 & 30.5 & 23.2 & 20.7 & 16.3 & 40.2 & \cb{83.8} & 7.7 \\
 & Rerank+Compr. & \textbf{64.1} & \textbf{65.6} & \cb{98.8} & 10.5 & \textbf{41.9} & 32.9 & 26.8 & 22.8& \textbf{59.8} & \cb{84.6} & 6.7 \\
\midrule
\multirow{2}{*}{DSLR-Zerank}
 & Compr-only & 31.9 & 42.3 & \cb{94.6} & 12.4 & 31.4 & 24.1 & 19.2 & 14.3& 39.8 & \cb{49.5} & 6.0 \\
 & Rerank+Compr. & 44.6 & 53.7 & \cb{95.8} & 10.7 & 41.6 & \textbf{35.5} & \textbf{27.4} & \textbf{22.9} & 46.7 & \cb{37.8} & 5.0 \\
\midrule
\multirow{2}{*}{OpenProvence}
 & Compr-only & 43.3 & 47.2 & \cb{97.0} & 11.9 & 32.9 & 25.6 & 20.3 & 16.3 & 39.5 & \cb{78.3} & 7.7 \\
 & Rerank+Compr. & 56.3 & 59.0 & \cb{97.4} & 10.1 & 37.8 & 27.2 & 19.9 & 17.5& 49.9 & \cb{79.2} & 7.1 \\
\midrule
\multirow{2}{*}{Provence}
 & Compr-only & 38.6 & 46.2 & \cb{96.3} & 11.9 & 27.2 & 20.7 & 16.7 & 15.0 & 39.7 & \cb{80.1} & 8.0 \\
 & Rerank+Compr. & 52.5 & 58.1 & \cb{95.7} & 11.6 & 29.3 & 23.6 & 19.1 & 16.3 & 43.7 & \cb{81.1} & 8.0 \\
\midrule
\multirow{2}{*}{XProvence} 
 & Compr-only & 26.9 & 42.1& \cb{88.2} & 12.6 & 14.2 & 10.6 & 8.9 & 7.7 & 41.5 & \cb{94.3} & 12.2 \\
 & Rerank+Compr. & 23.1 & 37.3 & \cb{92.0} & 12.6 & 16.3 & 10.2 &7.8 & 6.9 & 51.0 & \cb{93.5} & 12.0 \\
\midrule
\multirow{2}{*}{Semantic HL}
 & Compr-only & 41.9 & 46.9 & \cb{96.4} & 12.7 & 37.0 & 30.1 & 25.2 & 19.5 & 38.8& \cb{44.4} & 5.8 \\
 & Rerank+Compr. & 48.9 & 53.1 & \cb{97.0} & 11.3 & 39.8 & 31.3 & 26.0 & 20.7 & 46.6& \cb{43.1} & 4.8 \\
\midrule
\multirow{2}{*}{LLMLingua-2}
& Compr-only & 28.9 & 35.1 & \cb{46.2} & 6.4 & 25.6 & 18.3 & 16.7 & 13.8 & 38.8 & \cb{36.8} & 6.0 \\
 & Rerank+Compr. & 23.1 & 32.5 & \cb{46.0} & 4.7 & 28.6& 20.4 & 15.9& 12.2& 46.9 & \cb{36.6} & 5.2 \\
\midrule
\multirow{2}{*}{LongLLMLingua}
 & Compr-only & 16.5 & 33.0 & \cb{48.6} & 6.9 & 4.1 & 1.2 & 1.2 & 0.8 & 37.6 & \cb{42.5} & 7.8 \\
 & Rerank+Compr. & 28.3 & 43.1 & \cb{45.7} & 7.9 & 5.7 & 3.3 & 2.0 & 1.2 & 41.9 & \cb{42.5} & 7.9 \\
\bottomrule
\end{tabular}}
\vspace{-0.6em}
\caption{\textbf{System Test} results with the gpt-oss-120B agent.  Recall is the fraction of gold documents that the agent retrieves over the whole interaction. The bar in Comp.\% is the percentage of retrieved tokens that the compressor removes. Search is the average number of search tool calls per query.} 
\vspace{-0.3em}
\label{tab:system_test_with_no_cap_result}
\end{table*}

\subsection{Integration Test}

\paragraph{Overview.}Figure \ref{fig:judge-vs-compression} shows the average judge score and evidence recall against compression ratio, aggregated over 17 datasets.
Multiple models start from similar no-compression baselines but drop differently as compression increases. Token-level compressors and keyword snippets drop fastest, reaching around 0.2 accuracy at about 75\% compression.
Comp-H defines the Pareto frontier in the accuracy--compression space. At about 75\% compression, it keeps a judge score of 0.63, close to its no-compression baseline (0.68).

\begin{figure*}[ht]
\centering
\usetikzlibrary{calc}
\definecolor{cOurs}{HTML}{1F8A65}
\definecolor{cBase}{HTML}{6A6A8C}
\definecolor{cZerank}{HTML}{4E97C8}
\definecolor{cOpenProv}{HTML}{4A5CB4}
\definecolor{cProv}{HTML}{E08A18}
\definecolor{cXProv}{HTML}{0E8EA4}
\definecolor{cSemHL}{HTML}{B02A48}
\definecolor{cLL2}{HTML}{E58FB0}
\definecolor{cLongLL}{HTML}{7A4A20}
\newcommand{\pmk}[2]{\tikz[baseline=-0.6ex]{\draw[#2] plot[only marks, mark=#1, mark size=2.6pt, line width=1pt] coordinates {(0,0)};}}
\pgfplotsset{
  base axis/.style={
    scale only axis=true, height=2.3cm,
    ymin=0, ymax=70, ytick={0,20,40,60},
    grid=both, major grid style={dashed, black!22, line width=0.4pt},
    tick align=outside, tick pos=left,
    axis line style={black!75, line width=0.6pt},
    major tick style={black!75, line width=0.6pt},
    label style={font=\scriptsize}, tick label style={font=\tiny},
    title style={font=\scriptsize\bfseries, yshift=-8pt},
  },
  rp/.style={only marks, mark=star, mark size=3.6pt, line width=1.1pt},
  po/.style={only marks, mark=*,    mark size=2.5pt},
  bl/.style={only marks, mark=o,    mark size=2.9pt, line width=1.2pt},
  bl2/.style={only marks, mark=square, mark size=2.6pt, line width=1.2pt},
  bref/.style={no marks, sharp plot, line width=0.9pt, draw=cBase, opacity=0.9},
  bret/.style={bref, dash pattern=on 1.2pt off 1.4pt},   
  brr/.style={bref,  dash pattern=on 3pt off 2pt},       
  rng/.style={error bars/.cd, y dir=plus, y explicit,
              error bar style={line width=0.9pt, opacity=0.8},
              error mark=-,
              error mark options={mark size=3.4pt, line width=1.6pt, opacity=0.8}},
}
\vspace{-4pt}
\begin{tikzpicture}
\begin{axis}[base axis, name=bcpL,
  width=2.1cm, xmin=-4, xmax=50, xtick={0,20,40},
  ylabel={Answer accuracy (\%)}, ylabel style={yshift=-8pt},
  axis x line*=bottom, axis y line*=left,
  ]
  \addplot[bret] coordinates {(-4,40.80) (50,40.80)};
  \addplot[brr]  coordinates {(-4,45.90) (50,45.90)};
  \addplot[bl,  cBase] coordinates {(0,40.80)};   
  \addplot[bl2, cBase] coordinates {(0,45.90)};   
  \addplot[po, cLL2]      coordinates {(46.21,28.92)};
  \addplot[rp, cLL2]      coordinates {(46.00,23.10)};
  \addplot[po, cLongLL]   coordinates {(48.56,16.51)};
  \addplot[rp, cLongLL]   coordinates {(45.65,28.31)};
\end{axis}
\begin{axis}[base axis, name=bcpR,
  at={($(bcpL.south east)+(0.32cm,0)$)}, anchor=south west,
  width=5.2cm, xmin=87, xmax=100, xtick={88,92,96,100},
  yticklabels={}, axis x line*=bottom, axis y line*=left,
  ]
  \addplot[bret] coordinates {(87,40.80) (100,40.80)};
  \addplot[brr]  coordinates {(87,45.90) (100,45.90)};
  \addplot[po, cOurs]     coordinates {(98.27,47.23)};
  \addplot[rp, cOurs]     coordinates {(98.82,64.10)};
  \addplot[po, cZerank]   coordinates {(94.59,31.93)};
  \addplot[rp, cZerank]   coordinates {(95.78,44.58)};
  \addplot[po, cOpenProv] coordinates {(96.99,43.25)};
  \addplot[rp, cOpenProv] coordinates {(97.35,56.27)};
  \addplot[po, cProv]     coordinates {(96.31,38.55)};
  \addplot[rp, cProv]     coordinates {(95.74,52.53)};
  \addplot[po, cXProv]    coordinates {(88.20,26.90)};
  \addplot[rp, cXProv]    coordinates {(92.00,23.10)};
  \addplot[po, cSemHL]    coordinates {(96.35,41.93)};
  \addplot[rp, cSemHL]    coordinates {(97.03,48.92)};
  \node[font=\tiny, text=cBase, anchor=south west, inner sep=1pt]
       at (axis cs:87.3,45.9) {+rerank};
  \node[font=\tiny, text=cBase, anchor=south west, inner sep=1pt]
       at (axis cs:87.3,40.8) {retrieval};
\end{axis}
\foreach \p in {bcpL.south east, bcpR.south west}{
  \draw[black!75, line width=0.7pt] ($(\p)+(-2.4pt,-3.4pt)$) -- ($(\p)+(0.4pt,3.4pt)$);
  \draw[black!75, line width=0.7pt] ($(\p)+(0.6pt,-3.4pt)$)  -- ($(\p)+(3.4pt,3.4pt)$);
}
\node[font=\scriptsize\bfseries, anchor=south]
  at ($($(bcpL.north)!0.5!(bcpR.north)$)+(15pt,0pt)$) {BrowseComp-Plus (open-web search)};
\node[font=\scriptsize, anchor=north]
  at ($($(bcpL.south)!0.5!(bcpR.south)$)+(15pt,-10pt)$) {Compression Ratio (\%)};
\begin{axis}[base axis, name=oqa,
  at={($(bcpR.south east)+(1.5cm,0)$)}, anchor=south west,
  width=6.0cm, xmin=-4, xmax=100, xtick={0,20,40,60,80,100},
  ymin=0, ymax=50, ytick={0,10,20,30,40},
  xlabel={Compression Ratio (\%)}, xlabel style={yshift=5pt},
  ylabel={Numeric accuracy (\%)}, ylabel style={yshift=-4pt},
  title={OfficeQA (enterprise docs)},
  ]
  \addplot[bret] coordinates {(-4,38.3) (100,38.3)};
  \addplot[brr]  coordinates {(-4,45.0) (100,45.0)};
  \addplot[bl,  cBase, rng] coordinates {(-0.5,22.22) +- (0,16.08)};  
  \addplot[bl2, cBase, rng] coordinates {( 0,29.41)   +- (0,15.59)};  
  \addplot[po, cOurs,     rng] coordinates {(83.77,16.26) +- (0,14.23)};
  \addplot[rp, cOurs,     rng] coordinates {(84.62,22.76) +- (0,19.11)};
  \addplot[po, cZerank,   rng] coordinates {(49.49,14.29) +- (0,17.14)};
  \addplot[rp, cZerank,   rng] coordinates {(37.80,22.86) +- (0,18.77)};
  \addplot[po, cOpenProv, rng] coordinates {(78.34,16.26) +- (0,16.67)};
  \addplot[rp, cOpenProv, rng] coordinates {(79.24,17.48) +- (0,20.32)};
  \addplot[po, cProv,     rng] coordinates {(80.08,15.04) +- (0,12.20)};
  \addplot[rp, cProv,     rng] coordinates {(81.14,16.26) +- (0,13.01)};
  \addplot[po, cXProv,    rng] coordinates {(94.30,7.70)  +- (0,6.50)};
  \addplot[rp, cXProv,    rng] coordinates {(93.50,6.90)  +- (0,9.40)};
  \addplot[po, cSemHL,    rng] coordinates {(44.42,19.51) +- (0,17.48)};
  \addplot[rp, cSemHL,    rng] coordinates {(43.14,20.73) +- (0,19.11)};
  \addplot[po, cLL2,      rng] coordinates {(36.76,13.82) +- (0,11.79)};
  \addplot[rp, cLL2,      rng] coordinates {(36.60,12.20) +- (0,16.40)};
  \addplot[po, cLongLL,   rng] coordinates {(42.45,0.81)  +- (0,3.26)};
  \addplot[rp, cLongLL,   rng] coordinates {(42.50,1.22)  +- (0,4.47)};
  \node[font=\tiny, text=cBase, anchor=south east, inner sep=1pt]
       at (axis cs:99,45.0) {+rerank};
  \node[font=\tiny, text=cBase, anchor=north east, inner sep=1pt]
       at (axis cs:99,38.3) {retrieval};
\end{axis}
\end{tikzpicture}
\vspace{-6pt}
{\scriptsize
\pmk{*}{cOurs}~Comp-H\;\,
\pmk{*}{cZerank}~DSLR\;\,
\pmk{*}{cOpenProv}~OpenProv.\;\,
\pmk{*}{cProv}~Prov.\;\,
\pmk{*}{cXProv}~XProv.\;\,
\pmk{*}{cSemHL}~SemHL\;\,
\pmk{*}{cLL2}~LLMLingua-2\;\,
\pmk{*}{cLongLL}~LongLLMLingua\;\,
\pmk{o}{cBase}~Baseline (retrieval)\;\,
\pmk{square}{cBase}~Baseline (+rerank)
}
\caption{\textbf{System Test (Left: BrowseComp-Plus, Right: OfficeQA)}: answer accuracy vs.\
compression ratio using the 120B agent. Each method is shown at both operating points
($\star$ Rerank+Compress, $\bullet$ Compress-only); $\circ$ and $\square$ are the
no-compression baselines (retrieval-only and rerank-only with BGE). On OfficeQA the marker is the exact
numeric accuracy and the whisker above it reaches the accuracy within a $\pm5\%$ tolerance;
the two dashed lines there are drawn at the $\pm5\%$ level of each baseline. Top-right is
best: our model is alone in that corner on BrowseComp-Plus. On OfficeQA, no compressor
recovers the rerank-only baseline, but Comp-H comes the closest to it and matches
DSLR-Zerank while compressing about twice as much.}
\vspace{-0.8em}
\label{fig:system_acc_vs_no_comp}
\end{figure*}
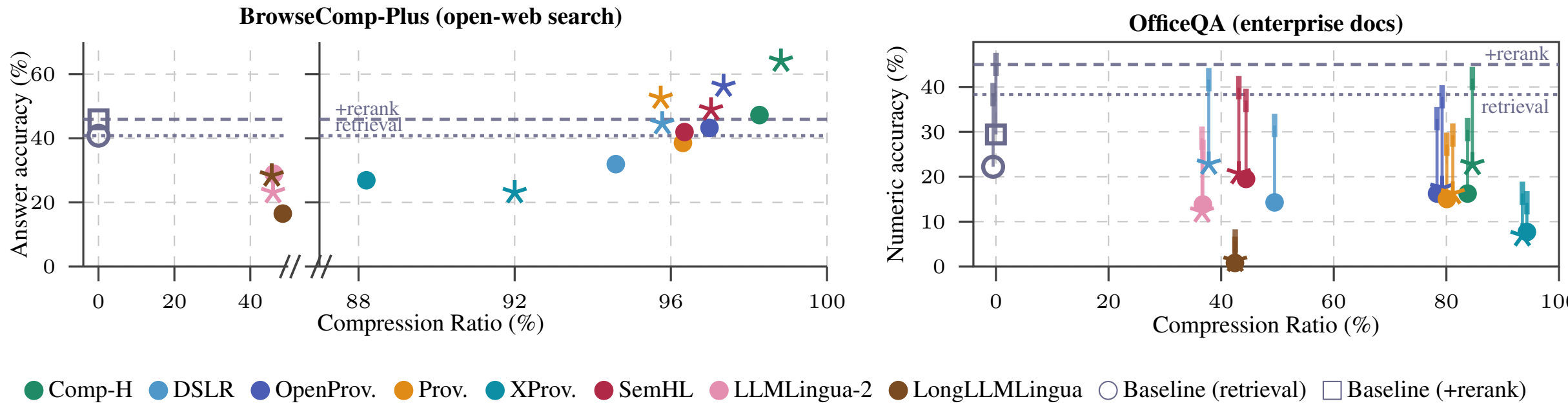

\footnotetext{For BrowseComp-Plus, we follow \citet{chen2025browsecompplusfairtransparentevaluation} and truncate retrieved documents to the first 512 tokens. See Appendix~\ref{appendix:browsecomp_token_cap} for analysis.}

\paragraph{Compression depends on answer structure.}

When answers require exact spans or structured values, fine-grained token-level compression can remove critical information. As in Table~\ref{tab:heatmap-integration}, on FinQA, every method performs below its typical level, but token-level compressors collapse (LLMLingua-2 and LongLLMLingua 0.05) compared with Comp-H (0.41). Token-level deletion removes decimal points and table delimiters, causing values to detach from row and column headers. Soft compression also fails here: replacing passages with a fixed number of memory embeddings forces the generator to regenerate rather than copy. Numeric values and proper names are easily lost under this constraint, so it struggles on numeric and entity-centric tasks (e.g., Academic Papers and FinQA).

In contrast, soft compression is preferable when answers require semantic judgment or aggregation rather than verbatim extraction. On ViDoRe Computer Science boolean queries, Comp-S (0.94) and OSCAR (0.80) match or exceed the gold-document oracle (0.82) despite 94\% compression (Appendix~\ref{sec:appendix_integration_qtype}). Since Comp-S and OSCAR only run at a fixed rate ($\approx$93.6\%), we also compare all methods at a matched ratio (Table~\ref{tab:matched_soft}, Appendix~\ref{sec:appendix_integrationtest_result}). Boolean questions often depend on the overall meaning of passages rather than a single answer-bearing sentence, causing sentence-level pruning to discard useful context. This pattern mirrors the Unit Test, where soft compression is strong for summarization or contextualized queries requiring coreference resolution. The advantage of each strategy is thus not determined by task difficulty alone, but by how evidence is distributed and answers are structured. See Appendix~\ref{sec:appendix_integration_examples} for examples.

\paragraph{Ranking and compression can disagree.}
The Integration Test reveals that producing ranking scores and the
pruning mask from the same model does not guarantee alignment between the two. Our alignment diagnostics make this clear:
XProvence ($\tau$0.3) removes all content from its rank-1 document for 33\% of queries,
achieving only 0.16 judge accuracy on these cases compared with 0.47 when the
top-ranked document retains some content (Appendix~\ref{sec:appendix_alignment_result}).
This pattern holds across methods: documents preserved by ranking can still be lost during compression, degrading answer quality.

\paragraph{Evidence recall as a proxy for final quality.}
For sentence-level hard compression methods, evidence recall correlates strongly with judge accuracy across configurations (Pearson $r{=}0.980$,
$n{=}98$; partial $r{=}0.967$ controlling for compression ratio). The association also holds per query (pooled $r{=}0.55$, positive in all 17 datasets) and transfers to held-out datasets and compressors (Appendix~\ref{sec:app_evidencerecall_judge_correlation}), indicating that performance depends on which evidence spans are retained,
not simply how much text is removed.
Evidence recall can therefore compare sentence-level compressors without running generation, but it is not a per-query substitute for end-task accuracy.

\subsection{System Test}

\paragraph{BrowseComp-Plus.}
The results exhibit a two-stage pattern. Compression alone improves over the retrieval-only baseline for three evaluated systems including Comp-H, even though both operate on the same per-search top-5 budget (Table~\ref{tab:system_test_with_no_cap_result}). This suggests that removing redundant context can help the agent exploit retrieved evidence more effectively rather than just reducing the input length. Interestingly, compression also increases interaction-level retrieval recall across many evaluated systems, suggesting that a more focused context helps the agent exploit search more effectively. Adding reranking further improves performance for most systems by improving both evidence quality and final answer accuracy. The pattern is not universal, however. For XProvence and LLMLingua-2, reranking lowers interaction-level recall. A likely reason
is that a narrow reranker and an aggressive compressor remove too much together,
leaving the agent with insufficient context to form a good next query. Such interactions are difficult to observe in static retrieval benchmarks because they depend on how the agent adapts its search behavior over the course of retrieval.

\paragraph{OfficeQA.}

OfficeQA shows a different trend. Compression alone performs worse than the retrieval-only baseline for every evaluated system, and reranking recovers only part of the loss. Reranking raises interaction-level recall for all compressors, by nearly 20 points in the best case, but only two systems exceed the retrieval-only baseline in exact match, and none reaches the rerank-only baseline. On numeric documents, therefore, compression degrades answer quality, and higher recall does not translate into higher answer accuracy. The main challenge is preserving small but critical numerical evidence, since removing a single token, row, or table cell can eliminate the exact value required for answering. Compressors that operate at the token level are the most vulnerable in this setting, which is consistent with the failures observed in our Unit and Integration Tests.

What compression still buys is cost: Comp-H matches the retrieval-only baseline (22.8 vs. 22.2 exact) while removing 84.6\% of the retrieved tokens, whereas DSLR-Zerank reaches a similar score only by keeping most of the context (37.8\% removed). 
As shown in Appendix~\ref{app:token-accounting}, this also holds when accounting for all tokens in agentic trajectories.
We additionally benchmark end-to-end latency for all methods in Appendix~\ref{app:efficiency}.


\section{Conclusion}
We introduced \textsc{UniCoRn}, a benchmark for joint evaluation of context compression and reranking across three levels, from controlled document-level tests to live search agents. Our analysis shows that compression is not simply a matter of reducing context length: the key is whether answer-critical evidence survives and whether compression decisions align with ranking. Sentence-level compressors often preserve critical evidence under high compression, while token-level methods struggle with structured evidence such as tables and exact values, and soft compression is more effective when answers require semantic aggregation rather than extraction. By training matched hard and soft compression models with the same base model and data, we enable a controlled comparison between compression regimes. We release \textsc{UniCoRn} to establish the base for a more comprehensive evaluation of compression methods on diverse and real-world data.
The failure cases revealed by \textsc{UniCoRn} simultaneously guide the way for future optimizations of compression methods in the context of RAG.

\section*{Limitations}

Our benchmark mainly relies on existing datasets, which inevitably contain a small number of queries that are not fully answerable from the available textual context alone (e.g., cases requiring non-textual evidence or lacking sufficient supporting information). Therefore, some residual noise remains. We retain these cases to preserve comparability with the original benchmarks, and because they do not systematically favor any particular compression method. Generators are instructed to abstain when the provided context is insufficient, and we additionally report gold-oracle results as a reference ceiling to contextualize the achievable performance. Removing queries flagged as unanswerable leaves the method rankings almost unchanged (overall Spearman's $\rho{=}0.993$). Further analysis and human validation are provided in Appendix~\ref{app:ev-human}.
Evidence recall matches gold spans against the compressed text, making it directly applicable to sentence-level pruning but not to soft compression.
We thus report evidence recall as a complementary diagnostic alongside end-task accuracy, which remains our primary metric.

Due to reliance on existing public data, our benchmark is also limited in the number of languages and domains covered. The majority of data is in English, with remaining data being in other high-resource languages.
Extending our benchmark to a wider range of languages and domains is a natural avenue for future work.

All reported results for trained models come from single runs due to required compute resources.



\bibliography{custom}

\appendix

\section{Baselines}
\label{sec:appendix_baselines}
Table \ref{tab:baselines} compares different baselines that we test with \textsc{UniCoRn}. 


\begin{table*}[ht!]
    \centering
\begin{threeparttable}
    \footnotesize
    \setlength{\tabcolsep}{4pt}
    \renewcommand{\arraystretch}{1.2}
    \begin{tabular}{@{}l l l c p{3.1cm} p{3cm}@{}}
    \toprule
    \textbf{Method} & \textbf{Backbone} & \textbf{Granularity} & \textbf{Q-aware} & \textbf{Doc-level ranking score} & \textbf{Notable behaviour} \\
    \midrule
    \multicolumn{6}{@{}l}{\emph{Sentence-level hard compression baselines}} \\
    Provence\tnote{a}         & DeBERTa-v3       & Sentence    & Yes & First 512-tok block only ($s_i{=}0$); later blocks discarded & 512-tok window; Internal NLTK splitter \\
    XProvence\tnote{b}        & BGE-M3           & Sentence    & Yes & First block only& 8192-tok window\\
    OpenProvence-large\tnote{c} & ModernBERT     & Sentence    & Yes & Max over 512-tok blocks (sigmoid-calibrated prob) & Sliding 512-tok blocks; multilingual splitter \\
    Semantic Highlight\tnote{d} & BGE-M3  & Sentence    & Yes & \textbf{None} internally; score by BGE-M3 & 8192-tok window; Supports English and Chinese  \\
    DSLR & Zerank-2\tnote{e} & Sentence & Yes & Max over sentence-level scores & Thresholding as described in \citet{zeroentropy} \\
    Sentinel\tnote{f} & \makecell[l]{Qwen2.5-0.5B (proxy)\\ + attention probe} & Sentence & Yes & \textbf{None} internally; score by BGE-M3 & Logistic-regression probe on proxy-LM attention; higher threshold = less compression \\
    RECOMP (extractive)\tnote{g} & Contriever-MS MARCO & Sentence & Yes & \textbf{None} internally; score by BGE-M3 & Query--sentence dual encoder trained on NQ; threshold on sentence score \\
    \midrule
    \multicolumn{6}{@{}l}{\emph{Token-level baselines}} \\
    LLMLingua-2\tnote{h}  & XLM-RoBERTa-large      & Token       & \textbf{No} & \textbf{None} internally; score by BGE-M3 &  Threshold = keep-rate (higher = less compression) \\
    LongLLMLingua\tnote{i} & Llama-2-7B      & Token       & Yes & Negated conditional perplexity & Ranks by perplexity, then compress the top-5; Threshold = keep-rate \\
    \midrule
    \multicolumn{6}{@{}l}{\emph{Soft-compression baselines}} \\
    OSCAR\tnote{j}        & \makecell[l]{Llama3.2-1B (compressor)\\ Qwen2-7B (decoder)}   & \makecell[l]{Chunk\\(128 tok)} & Yes &  \textbf{None} internally; score by BGE-M3  & Each chunk $\to$ 8 memory embeddings; answers with its \emph{own} decoder \\
    \midrule
    \multicolumn{6}{@{}l}{\emph{Heuristic (rule-based) baseline}} \\
    Keyword Snippet & None (rule-based) & \makecell[l]{Snippet\\ ($\approx$300 char)} & Yes & Fraction of unique query tokens matched (lexical); retrieval order kept otherwise & Structure-aware excerpt (prose/list/table) anchored at densest keyword match; no learned model; threshold ignored \\
    \bottomrule
    \end{tabular}
    \caption{Context-compression methods evaluated. }
    \label{tab:baselines}
\begin{tablenotes}[flushleft]\scriptsize
\item[a] \url{https://huggingface.co/naver/provence-reranker-debertav3-v1}
\item[b] \url{https://huggingface.co/naver/XProvence-reranker-bgem3-v2}
\item[c] \url{https://huggingface.co/hotchpotch/open-provence-reranker-large-v1}
\item[d] \url{https://huggingface.co/zilliz/semantic-highlight-bilingual-v1}
\item[e] \url{https://huggingface.co/zeroentropy/zerank-2-reranker}
\item[f] \url{https://huggingface.co/ReRaWo/Sentinel}
\item[g] \url{https://huggingface.co/fangyuan/nq_extractive_compressor}
\item[h] \url{https://huggingface.co/microsoft/llmlingua-2-xlm-roberta-large-meetingbank}
\item[i] \url{https://huggingface.co/NousResearch/Llama-2-7b-hf}
\item[j] \url{https://huggingface.co/naver/oscar-qwen2-7B}
\end{tablenotes}
\end{threeparttable}
    \end{table*}

\section{Academic Paper Dataset}
\label{app:academic_paper}

We select academic research papers due to their high lexical and information density, structural complexity, and the need for multi-hop reasoning over long contexts. Also, academic papers often contain hard-to-compress information such as precise numbers, experimental identifiers, and rare or domain-specific terminology. The ArXiver dataset~\citep{acar_arxiver2024} is used as a source.

\paragraph{Context construction}
Each example is a fixed 10-document pool: one target paper plus nine
retrieved hard negatives (nearest-neighbor papers from abstracts with
\texttt{BAAI/bge-base-en-v1.5}). Unlike our other datasets in Integration Test,
where documents are short enough to keep a top-50 pool, papers are
$\sim$10k tokens each, so a top-50 list would far exceed a 128k window.
We therefore use $k{=}10$ and pool under a 115k-token budget
($\approx$90\% of 128k; \texttt{cl100k\_base}), yielding $N{=}200$ with mean context length $104{,}679$ tokens (max $114{,}528$;
Table~\ref{tab:context_stats}).

\begin{table}[h]
\centering
\small
\begin{tabular}{lrp{11.3em}}
\toprule
\textbf{Metric} & \textbf{Tokens} & \textbf{Description} \\
\midrule
\multicolumn{3}{l}{\textit{Context components (average)}} \\
Target document & 9{,}745 & Ground-truth paper \\
Single distractor & 10{,}548 & One hard negative \\
Total distractors & 94{,}934 & Sum of 9 distractors \\
\midrule
\multicolumn{3}{l}{\textit{Full context (target + 9 distractor papers)}} \\
Mean length & \textbf{104{,}679} & \\
Max length & 114{,}528 &  \\
\bottomrule
\end{tabular}
\caption{Token statistics of the curated Academic Paper (\texttt{cl100k\_base}).}
\label{tab:context_stats}
\end{table}

\paragraph{Query generation}
Over each target paper, we generate QA--Evidence triples with Gemini 2.5 Flash under three query types. (Claude Opus 4.6 is used for evidence double-annotation, Appendix \ref{app:evidence-integration}.) Query types are:

\begin{itemize}
    \item \textbf{Type A (Numerical):} extract a specific metric/hyperparameter/result; the answer must contain a number attested in the evidence.
    \item \textbf{Type B (Entity):} describe a method setting or variable and ask for its name/label.
    \item \textbf{Type C (Multi-hop):} require combining facts from at least two disjoint passages (e.g., method vs.\ results).
\end{itemize}

We retain only examples that satisfy three programmatic checks: evidence spans must be exact substrings of the source paper, Type A answers must share a numeric value with the evidence, and Type C examples must cite at least two evidence spans. If the generated output does not meet the requirement, it is filtered. Because not every paper contains material for all three types (e.g., some report no extractable numerical result), the per-type counts are uneven and papers contribute a variable number of queries. The resulting dataset contains 372 queries from 197 papers, comprising 88 Type A, 144 Type B, and 140 Type C examples. See Table~\ref{tab:academic_type_examples}.

\begin{tcolorbox}[breakable, colback=gray!5, colframe=gray!50!black,
  title={Academic Paper QA-Evidence Generation}, label=box:academic-qae-prompt,
  fonttitle=\bfseries\small, fontupper=\scriptsize]
\begin{verbatim}
[System]
You are a strict 'Hard Examiner' for a PhD-level reading 
comprehension test.

Your goal is to generate "Gold Standard" Q&A pairs that are 
DIFFICULT for a keyword-based search engine (like BM25) 
to find, but solvable for a Semantic Reranker or a Domain 
Expert. You must generate Q&A pairs from the provided 
research paper and format them exactly according to the 
given template.

*** CRITICAL RULES FOR QUESTION GENERATION ***
1. **NO PAPER TITLES**: The Query MUST NOT contain the exact 
title of the paper.
2. **AVOID ABSTRACTS**: Do NOT generate questions based 
solely on the Abstract or Introduction. 
Focus on **Related Work, Methodology, Experiments, 
Analysis and Tables**.
3. **PARAPHRASE STRONGLY**: Do NOT copy-paste sentences 
from the text to form
   the question. Use synonyms and high-level descriptions.
   - Bad: "In the Unproportional Mosaicing method..." 
   (Too easy, keyword match)
   - Good: "In the data augmentation technique that 
   randomly splits images into various-sized blocks..." 
   (Harder, requires semantic understanding)
4. **STANDALONE BUT UNIQUE (FINGERPRINTING)**: The question 
must be self-contained and target ONLY this paper.
   - **Constraint:** If the paper proposes a method with a 
   very unique name (like "XYZ-Net"), do NOT use the name 
   directly. Instead, **describe its specific architectural 
   features or mechanism**.
   - **Exception:** If describing is too vague, combine the 
   model name with a specific Dataset Name or Metric 
   (e.g., "The XYZ-Net performance on CVIQD
     dataset").
5. **NO PRONOUNS**: Do NOT use terms like "in this study", 
"the proposed method", "according to the authors" 
without a qualifying noun.

Rules for Answers:
- Be concise. Give the exact number, term, or synthesized 
fact.

Rules for Evidence:
- Evidence must be verbatim, exact substrings copied from 
the provided paper.
- Ideally, pick evidence from the middle or end of the paper 
(Results, Tables, Discussion).

Output Format:
- Return EXACTLY 3 items as a JSON list (Type A, B, C).
- **Ensure strictly valid JSON syntax.** Escape all double 
quotes (") inside the question/answer strings.
- Do not wrap the output in markdown code blocks.
Just return the raw JSON string.

[User]
=====Paper Full Text=====
{paper_text}

=====[Task (Template)]=====
Generate exactly 3 Q&A pairs (One of each type A, B, C) as a 
JSON list.
Focus on the **deep details** of the paper (Tables, 
Experimental Results, Ablation Studies).

A) Numbers & Metrics (Hard):
- Ask for a specific numerical value found in a Table or 
Result section.
- **Trick:** Do not explicitly name the Table (e.g., don't 
say "In Table 2...").
  Instead, describe the experimental condition (dataset, 
  model variant, resolution) so that the question is
  standalone.
- Example: "When the ViT-B/16 model is pre-trained with 
masked patch prediction, what is the specific fine-tuning 
accuracy on ImageNet?"

B) Entity & Nomenclature (Reverse Definition):
- Describe a mechanism's function or a variable's role and 
ask for its **Symbol** or **Name**.
- **Constraint:** The description must not contain the name 
itself.
- **Constraint 2:** Focus on **Novel Terms or Specific 
Redefinitions** introduced in this paper. DO NOT ask about 
general knowledge (e.g., "What is ReLU?", "What is a 
Convolution?").
- Example: "What specific term is used to describe the 
phenomenon where the shock acts as the boundary layer of the 
corona?" (Answer: Propagating Oscillatory Shock model)

C) Multi-Hop Logic (Comparison):
- Ask to compare two distinct values, conditions, or ablation 
settings mentioned in different parts of the text.
- Example: "How do the optimal pretraining task weightings 
differ between datasets with limited unlabeled data versus 
those with extensive unlabeled data?"

=====[Evidence requirement]=====
- A & B: A LIST of ONE OR MORE exact quotes copied verbatim 
from the paper.
- C: A LIST of at least 2 exact quotes from DIFFERENT parts 
of the text (e.g., one quote for Condition A, one for 
Condition B).

=====[Output JSON schema]=====
[
  {
    "template_type": "A",
    "query": "Question text here...",
    "answer": "Answer text here...",
    "evidence": ["verbatim quote 1"]
  },
  {
    "template_type": "B",
    "query": "Question text here...",
    "answer": "Answer text here...",
    "evidence": ["verbatim quote 2"]
  },
  {
    "template_type": "C",
    "query": "Question text here...",
    "answer": "Answer text here...",
    "evidence": ["verbatim quote 3", "verbatim quote 4"]
  }
]
\end{verbatim}
\end{tcolorbox}

\begin{table*}[t]
\centering
\small
\begin{tabular}{p{0.06\textwidth}p{0.28\textwidth}p{0.18\textwidth}p{0.38\textwidth}}
\toprule
\textbf{Type} & \textbf{Query} & \textbf{Answer} & \textbf{Evidence} \\
\midrule
A ($88$) &
For ASVspoof 2019, what t-DCF and EER did the fusion of MFCC-ResNet, Spec-ResNet, and CQCC-ResNet achieve on the logical access task? &
t-DCF $0.1569$, EER $6.02$ &
(w{=}2, core)
``The fusion model reached a 0.1569 t-DCF and 6.02 EER on logical access task \ldots'' \\
\midrule
B ($144$) &
What magnetic configuration---two QSLs crossing diagonally above a flux rope with no true null---facilitates magnetic reconnection? &
Hyperbolic flux tube (HFT) &
(w{=}1, supporting)
``The analysis reveals that the crossing between the two QSLs does not possess a null point structure.''
\newline
(w{=}2, core)
``\ldots this configuration of Q-value map indicates the presence of a hyperbolic flux tube (HFT) structure, which is also a magnetic structure conducive to facilitating magnetic reconnection.'' \\
\midrule
C ($140$) &
In the triple-coverage analysis, what was the most frequent error for factual KGs, and how did it change for fictional KGs? &
Factual: \textit{hallucinated}; fictional: \textit{absent} &
(w{=}2)
``When expressing factual data, the most common error category was \textit{hallucinated}; expressing triples in a way that was incompatible with the source prompt.''
\newline
(w{=}2)
``When generating text for fictional data, the most common error was instead information missing from the generated text.'' \\
\bottomrule
\end{tabular}
\caption{Query templates and examples in the Academic Paper set ($N{=}372$).}
\label{tab:academic_type_examples}
\end{table*}

\section{Evidence}
\label{app:evidence}

\subsection{Evidence Annotation Prompts}
We annotate evidence only for Unit and Integration Tests since in System Test, the retrieval is dynamic within 100K and 83K documents.

\subsubsection{Unit Test}
We build 100 single-document items (10 types $\times$ 10). After assembling
each document, Claude Opus~4.8 \citep{anthropic2026opus48} writes a natural question, answer, and
verbatim evidence spans under type-specific prompts. Spans must match the
source as exact substrings. \textsc{Multi-hop Long} is constructed from HotpotQA gold documents and \textsc{cross-lingual}
translates only the question/answer and retains native evidence. Instructions varied by query type.

\begin{tcolorbox}[breakable, colback=gray!5, colframe=gray!50!black,
  title={QA and Evidence Generation Prompt (Unit Test)}, label=box:unit-ev-prompt,
  fonttitle=\bfseries\small, fontupper=\scriptsize]
\begin{verbatim}
[System]
You are an expert annotator building evaluation QA for a
context-compression benchmark. Every question must read 
like a NATURAL question a real user would ask about the 
document--never artificial, templated, or self-referential 
(do not say 'according to the document'). 
The answer must be derivable SOLELY from the provided 
document. You must copy every evidence span VERBATIM from 
the document: an exact character-for-character substring 
(same words, same order,
same punctuation; whitespace may differ). 
Output ONLY a single JSON object, no prose, no code fences.

[User]
{TYPE_SPECIFIC_INSTRUCTIONS}
Return JSON:
{"query": "...", "answer": "...",
 "evidence": ["<verbatim span 1>", "..."]}
DOCUMENT:
{doc_text}
\end{verbatim}
\end{tcolorbox}

\subsubsection{Integration Test}\label{app:evidence-integration}
In the Integration Test, we annotate evidence for every dataset except ConditionalQA, HotpotQA and
LegalQuAD, which already release reliable, verbatim evidence. For the remaining datasets,
Claude Opus~4.6 \citep{anthropic2026opus46card} and Gemini~2.5~Flash \citep{comanici2025gemini25pushingfrontier} generate \emph{verbatim} supporting spans given the
query, gold answer, and gold document. Both models receive the same instructions: spans
must be copied character-for-character and kept minimal but self-sufficient, with no paraphrase,
summarization, or translation. We use the same general query template for JSQuAD, KorQuAD,
KLUE, and Enterprise RAG, but vary the in-context example by dataset.

For the ViDoRe datasets, prompting the models to output evidence text directly produced
minor paraphrasing, normalization differences, and altered table formatting. Because our
pipeline verifies each span against the source at the string level, these mismatches were
caught as verification failures rather than silently accepted. To avoid them and
reduce output tokens, we instead extract gold evidence by segmenting
each document with nltk, numbering the segments, and asking the LLMs to return only the IDs
of the segments that support the gold answer. The evidence text is then reconstructed from
those IDs, with no paraphrasing.

As noted in Appendix~\ref{app:academic_paper}, the academic-paper dataset is synthetic, so
we use Gemini to create the query, gold answer, and evidence jointly, and Claude to double-annotate evidence. After human validation, the academic-paper and FinQA datasets were
re-annotated for higher evidence quality using the more detailed prompt shown below, which
was needed for tabular datasets. FinQA is a special case: its official
\texttt{gold\_inds} are paraphrased pointers rather than verbatim text, so we prompt the
models to resolve each pointer to its literal source span and to attach the table's
column-header row.

\begin{tcolorbox}[breakable, colback=gray!5, colframe=gray!50!black,
  title={Evidence Annotation Prompt (General)}, label=box:general-ev-prompt,
  fonttitle=\bfseries\small, fontupper=\scriptsize]
\begin{verbatim}
You are extracting evidence for a QA benchmark.

Task:
- Read the Query, Gold Answer, and Document.
- Return supporting evidence as exact copied spans from 
the Document.
- Spans may be clauses, phrases, or full sentences. 
They do NOT need to be sentence-aligned.

Critical constraints:
1) Verbatim only: every span must be copied exactly from 
the Document (character-level exact text).
2) Minimal but self-sufficient: keep spans as short as 
possible, but enough to justify why the Gold Answer is 
correct.
3) No paraphrase, no summarization, no translation, no 
added punctuation.
4) If no supporting text exists, return an empty list.
5) Output JSON only.

[Example]
Query: {fewshot_query}
Gold Answer: {fewshot_answer}
Document: {fewshot_doc}
Output:
{{"evidence": {fewshot_evidence_json}}}

---

Now process this:
Query: {query}
Gold Answer: {answer}
Document: {doc_text}

Return JSON only in this exact shape:
{"evidence": ["exact span 1", "exact span 2"]}
\end{verbatim}
\end{tcolorbox}

\begin{tcolorbox}[breakable, colback=gray!5, colframe=gray!50!black,
  title={Evidence Annotation Prompt (ViDoRe V3)}, label=box:vidore-ev-prompt,
  fonttitle=\bfseries\small, fontupper=\scriptsize]
\begin{verbatim}
You are a research assistant specializing in document 
analysis. The document below has LINE NUMBERS 
(e.g. "1: ...", "2: ..."). Your task: find ALL line numbers 
that contain evidence supporting the Gold Answer.

IMPORTANT: This is a difficult case where a previous
extraction attempt found NO evidence. You must try
harder. Apply these broadened rules:

1. DIRECT evidence: lines containing exact numbers, dates, 
names, or facts mentioned in the Gold Answer.
2. INDIRECT evidence: lines that provide necessary context 
to understand the answer (e.g., column headers for a table 
value, section titles, definitions).
3. PARTIAL matches: if the Gold Answer mentions a count 
(e.g. "11 diagrams"), include lines that list or 
enumerate those items even if the count itself is not 
explicitly stated.
4. TABLE handling: include the header row AND the relevant 
data row(s). For markdown tables (lines starting 
with |), always include the column-header row if you 
select any data row.
5. If the document truly contains NO relevant information at 
all, return {{"line_ids": []}}.

[Example]
[Query]
{FEWSHOT_QUERY}

[Gold Answer]
{FEWSHOT_ANSWER}

[Numbered Document]
{numbered_fewshot}

[Output]
{fewshot_output}

---

Now do the same for the following.

[Query] (the question the user asked)
{query}
[Gold Answer] (find sentences that support this)
{answer}
[Numbered Document] (select sentence numbers from here)
{numbered_doc_str}

[Instructions]
- Output: list only the sentence numbers (integers) of lines 
that contain evidence for the Gold Answer above. 
Return nothing else.
- "Evidence" = any sentence whose text is needed to support 
or explain the Gold Answer: exact numbers, dates, names, 
definitions, or key facts. When the Gold Answer mentions 
a specific figure or value (e.g. $18.841 billion),
include the line(s) in the document that contain that 
figure or value.

- For markdown tables:
  - Include the specific data row(s) that contain the facts 
  used in the Gold Answer.
  - Include the header row only if it is necessary to 
  interpret the selected values (e.g., to identify the year, 
  unit, or segment).
  - Do not include unrelated rows, repeated headers, 
  captions, or footers.
- Do not paraphrase or copy text; only return line IDs. 
Return valid JSON in this form only: 
{{"line_ids": [1, 2, 5, 6]}}."""
\end{verbatim}
\end{tcolorbox}

\begin{tcolorbox}[breakable, colback=gray!5, colframe=gray!50!black,
  title={Evidence Annotation Prompt (Academic Paper)}, label=box:academic-ev-prompt,
  fonttitle=\bfseries\small, fontupper=\scriptsize]
\begin{verbatim}
You are extracting evidence for a QA benchmark.

Task:
- Read the Query, Gold Answer, and Document.
- Return supporting evidence as exact copied spans from 
the Document.
- Spans may be clauses, phrases, or full sentences. 
They do NOT need to be sentence-aligned.

Critical constraints:
1) Verbatim only: every span must be copied exactly from 
the Document.
2) Minimal but self-sufficient: keep spans as short as 
possible, BUT they must contain ALL context needed to answer 
the Query from the spans alone. In particular:
   - For tabular data, include the table header row 
   (column names / units) and the row label together with the 
   answer cell.
   - For numbers/equations/symbols, include the surrounding 
   label or definition (variable name, units, condition, 
   dataset, ...).
   - For named entities/terms, include enough of the clause 
   to bind the entity to the concept asked about.
   - Do not include unrelated sentences or whole paragraphs.
3) No paraphrase, no summarization, no translation, no 
added punctuation.
4) If no supporting text exists, return an empty list.
5) Output JSON only.

[Example]
Query: {fewshot_query}
Gold Answer: {fewshot_answer}
Document: {fewshot_doc}
Output:
{{"evidence": {fewshot_evidence_json}}}

---

Now process this:
Query: {query}
Gold Answer: {answer}
Document: {doc_text}

Return JSON only in this exact shape:
{"evidence": ["exact span 1", "exact span 2"]}
\end{verbatim}
\end{tcolorbox}

\begin{tcolorbox}[breakable, colback=gray!5, colframe=gray!50!black,
  title={Evidence Annotation Prompt (FinQA)}, label=box:finqa-ev-prompt,
  fonttitle=\bfseries\small, fontupper=\scriptsize]
\begin{verbatim}
[system]
You are a precise text extraction assistant. 
Given a financial document and a set of evidence 
descriptions, extract verbatim spans from the document 
that the descriptions refer to, plus any spans essential 
to answer the question.

Rules:
1. Every returned text must be a character-for-character 
substring of the document.
2. The evidence descriptions are paraphrased pointers, 
not verbatim text. Use them to find the literal source span.
3. For "table_*" descriptions, extract the matching table 
row in its document format (e.g. ['row_label','value',...]). 
Include that table's column-header row once per referenced 
table when present.
4. For "text_*" descriptions, extract the matching sentence 
verbatim.
5. If other spans are required to answer (e.g. unit/scale 
qualifiers or referenced context), include them too.
6. Return only a JSON array of {"doc_id": "...", 
"text": "..."} objects.

Example:
[Document]
... amount ( in millions ) .[['', 'amount ( in millions )'], 
['2014 net revenue', '$ 5735'], ['volume/weather', '95'], 
['2015 net revenue', '$ 5829']] ...

[Evidence Descriptions]
{
  "table_1": "the 2014 net revenue of amount ( in millions ) 
  is $ 5735 ;",
  "table_8": "the 2015 net revenue of amount ( in millions ) 
  is $ 5829 ;"
}

[Gold Doc ID]
00

[Output]
[{"doc_id": "00", "text": "['', 'amount ( in millions )']"},
 {"doc_id": "00", "text": "['2014 net revenue', '$ 5735']"},
 {"doc_id": "00", "text": "['2015 net revenue', '$ 5829']"}]


[user]
[Document]
{gold_doc}

[Evidence Descriptions]
{gold_inds_json}

[Gold Doc ID]
{gold_doc_id}

Return ONLY the JSON array.


\end{verbatim}
\end{tcolorbox}

\subsection{Substring-aware Ensemble Merge}
\label{app:ev-merge}
We merge the two models' spans per document into a single weighted gold set
with the following rules:
\begin{enumerate}
  \item \textbf{Exact agreement.} If a span is produced by both models
  (identical after whitespace normalization), it is assigned weight $w{=}2$.
  \item \textbf{Containment promotion.} If one model's span is a substring of
  the other's, the \emph{shared} (shorter) span is promoted to $w{=}2$, and the
  longer span is split into its leftover fragment(s), which stay at $w{=}1$ with
  their original source.
  \item \textbf{Single-source.} Any remaining span selected by only one model
  keeps $w{=}1$.
\end{enumerate}
For ViDoRe, both models select from the same numbered sentence candidates, so their
selections are aligned by construction and exact agreement is unambiguous. The remaining
datasets---JSQuAD, KorQuAD, KLUE, Enterprise RAG, and FinQA---instead use free-form
verbatim extraction, where the two models may select the same underlying sentence but
draw slightly different span boundaries (one longer, one shorter). Containment promotion
handles this case. To avoid spurious agreement, it is applied only when the shared
span is at least four words long and occurs exactly once in the document, so that short or
repeated strings are not mistaken for consensus. Leftover fragments from splitting are kept
only if they span at least two words; punctuation-only or single-token residue (e.g.\ stray
conjunctions) is removed. Rather than discarding filtered fragments automatically, we
collected the borderline cases and reviewed them manually, retaining any that carried
genuine evidence. In practice, almost all were semantically empty connectives.

\subsection{Human Validation}
\label{app:ev-human}

\begin{figure*}[ht!]
    \centering    \includegraphics[width=0.9\textwidth]{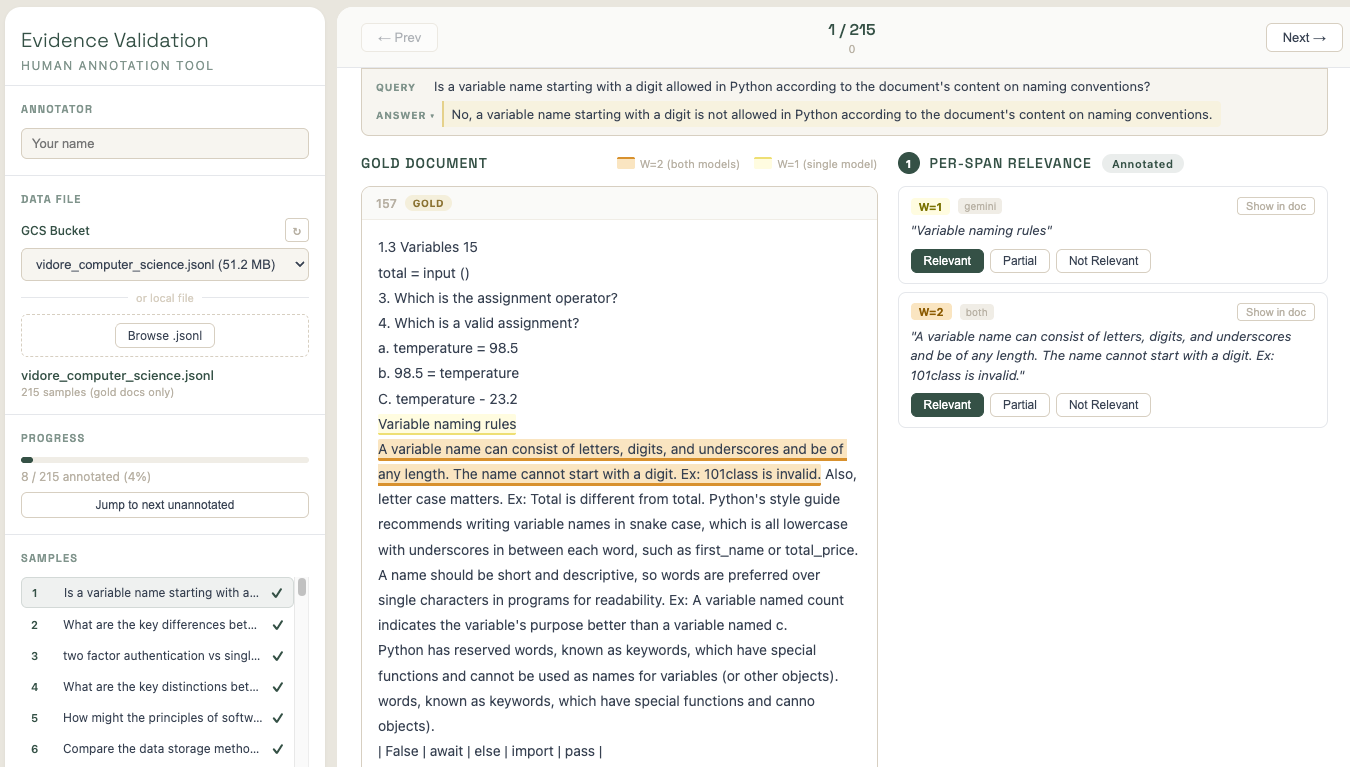}
\caption{The web interface used for human evidence validation: annotators
review highlighted evidence spans on the gold document and rate their
relevance and sufficiency. (1/2)}
    \label{fig:evidence_validation}
\end{figure*}
\begin{figure*}[ht!]
    \centering    \includegraphics[width=0.6\textwidth]{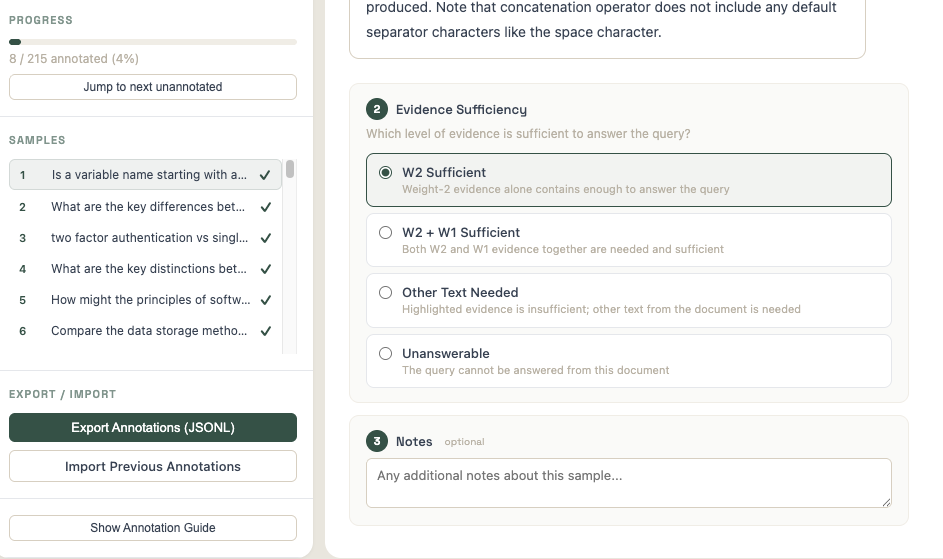}
\caption{The web interface used for human evidence validation (2/2)}
\vspace{-0.6em}
    \label{fig:evidence_validation2}
\end{figure*}

Using the interface in Figure \ref{fig:evidence_validation} and
\ref{fig:evidence_validation2}, two annotators
\footnote{co-authors of the paper} audited 110 queries,
with each query and its highlighted spans initially reviewed by one annotator. The audit covered 317
evidence-bearing query--document pairs and 1,057 highlighted spans
(453 with \(w{=}1\) and 604 with \(w{=}2\)). The initial audit identified
omitted table headers as a systematic weakness, particularly in FinQA,
whose sufficiency was 41.7\% (5/12). Header-aware re-annotation with an
updated prompt eliminated these failures (FinQA: 4/8 to 8/8; Academic
Papers: 8/10 to 10/10).

With the revised annotations, 90.0\% (99/110) of audited queries were
evidence-sufficient. Of the reviewed evidence spans, 684 (64.7\%) were fully
relevant, 260 (24.6\%) partially relevant, and 113 (10.7\%) irrelevant.
Nearly all irrelevant spans (109/113) occurred in the long,
multi-document ViDoRe examples, and most were single-model spans
(90/113; 79.6\% with \(w{=}1\)): the irrelevance rate was 19.9\% for
single-model spans, compared with 3.8\% for consensus spans (\(w{=}2\)),
so the ensemble weighting already discounts the majority of these
errors. To assess annotation reliability, we additionally double-annotated a subset of 822 spans from 65 queries across seven datasets. The two annotators achieved 81.0\% exact agreement, with Cohen's $\kappa=0.62$, weighted $\kappa=0.71$, and Gwet's AC1 $=0.75$, indicating substantial inter-annotator agreement. These results support the reliability of our human evidence-quality assessments. The correlation between evidence recall and judge accuracy in
Appendix~\ref{sec:app_evidencerecall_judge_correlation} indicates that
the annotated spans capture the information needed to answer the query
despite this residual noise.

In the human audit above, two ViDoRe queries could not be answered even from their complete sets of gold documents. This is not a missing-evidence problem in our extraction.
ViDoRe v3 annotates each qrel with a \texttt{content\_type} marking the
modality of the evidence region (\texttt{Text}, \texttt{Table},
\texttt{Image}, \texttt{Chart}, \texttt{Infographic}, or \texttt{Mixed}). Because we operate on parsed
Markdown, such regions have no textual counterpart, which is also why gold
coverage on ViDoRe is lower than elsewhere. A diagnostic check with Gemini over all
Integration Test datasets flagged 20.4\% of ViDoRe queries (493/2,419) as
unanswerable, compared with 7.9\% (795/10,085) elsewhere. We nonetheless retain all queries in the main results, since these flags come from Gemini rather than human annotators and a filtered subset would no longer be comparable to results reported on the full datasets.
To check whether this choice affects our conclusions, we re-compute the method rankings after removing the flagged queries.
The rankings are identical on 9 of the 15 affected datasets, and Spearman's $\rho \geq 0.95$ on all of them (overall $\rho{=}0.993$, Kendall's $\tau{=}0.970$).
Unanswerable queries therefore lower absolute scores but do not change the relative ordering of methods.
The Integration Test additionally reports a gold-oracle ceiling that passes all available gold documents to the generator without reranking or compression, which shows how much headroom remains.

\subsection{Evidence Statistics}

\begin{table*}[t]
\centering
\small
\setlength{\tabcolsep}{3.5pt}
\resizebox{\textwidth}{!}{%
\begin{tabular}{lrrrrrrrrr}
\toprule
Dataset & Queries & Gold pairs & Unique gold & $w{=}1$ Claude & $w{=}1$ Gemini & $w{=}2$ Both & Native $w{=}1$ & Spans / gold & Gold coverage \\
\midrule
Academic Papers          & 372   & 372    & 372    & 456   & 242  & 480  & --     & 3.17 & 100\% \\
ConditionalQA              & 271 & 271 & 59 & --    & --   & --   & 1,165  & 4.30 & 100\% \\
Enterprise RAG           & 470   & 741    & 722    & 3,876 & 377  & 1,971& --     & 8.40 & 97.2\% \\
FinQA                    & 1,138 & 1,138  & 380    & 25    & 133  & 2,711& --     & 2.52 & 100\% \\
HotpotQA                 & 7,405 & 14,810 & 13,783 & --    & --   & --   & 18,003 & 1.22 & 99.99\% \\
JSQuAD                   & 200   & 200    & 172    & 72    & 25   & 215  & --     & 1.56 & 100\% \\
KLUE-MRC                 & 200   & 200    & 199    & 149   & 20   & 235  & --     & 2.02 & 100\% \\
KorQuAD                  & 100   & 100    & 88     & 57    & 17   & 210  & --     & 2.84 & 100\% \\
LegalQuAD                & 200   & 200    & 200    & --    & --   & --   & 200    & 1.00 & 100\% \\
ViDoRe--CS & 215   & 1,049  & 589    & 876   & 533  & 1,696& --     & 2.96 & 95.6\% \\
ViDoRe--Energy           & 308   & 1,103  & 722    & 1,012 & 482  & 1,529& --     & 2.74 & 93.7\% \\
ViDoRe--Finance (EN)     & 309   & 1,461  & 785    & 1,431 & 882  & 2,357& --     & 3.20 & 89.8\% \\
ViDoRe--Finance (FR)     & 320   & 1,468  & 883    & 1,545 & 862  & 2,201& --     & 3.14 & 93.4\% \\
ViDoRe--HR               & 318   & 1,731  & 613    & 2,124 & 914  & 2,203& --     & 3.03 & 93.9\% \\
ViDoRe--Industrial       & 283   & 1,614  & 1,000  & 1,444 & 1,272& 2,800& --     & 3.42 & 92.5\% \\
ViDoRe--Pharm.  & 364   & 1,732  & 946    & 961   & 856  & 2,207& --     & 2.32 & 93.4\% \\
ViDoRe--Physics          & 302   & 2,178  & 962    & 1,121 & 408  & 2,251& --     & 1.74 & 95.6\% \\
\midrule
\textbf{Total}           & \textbf{12,775} & \textbf{30,368} & \textbf{22,475} & \textbf{15,149} & \textbf{7,023} & \textbf{23,066} & \textbf{19,368} & \textbf{2.13} & \textbf{97.3\%} \\
\bottomrule
\end{tabular}%
}
\caption{Gold-document and evidence-span statistics in \textbf{Integration Test}. Gold pairs are query--document
pairs with a positive qrel. Unique gold is the number of distinct gold documents.
Spans are extracted by two models: $w{=}1$ Claude and $w{=}1$ Gemini count spans
selected by only one model, and $w{=}2$ Both counts spans selected by both.
ConditionalQA, HotpotQA and LegalQuAD provide reliable native evidence (supporting evidence for ConditionalQA, supporting facts for HotpotQA, verbatim answer extracted from source document for LegalQuAD), which we use directly
without weighting (Native $w{=}1$). Spans / gold is averaged over all gold pairs,
including those with no span. Gold coverage is the share of gold pairs with at
least one span.}
\label{tab:evidence_statistics}
\end{table*}

Table~\ref{tab:evidence_statistics} reports gold-document and evidence-span
statistics for each Integration Test dataset. Gold coverage is at or near
100\% for all datasets except ViDoRe, where it ranges from 89.8\% to 95.6\%.
This gap comes from the source annotation rather than from our extraction.
ViDoRe v3 labels each qrel with a \texttt{content\_type} indicating the
modality of the evidence region (Text, Table,
Image, Chart, Infographic, or Mixed),
and some ViDoRe gold pairs are annotated as non-textual.
Since we operate on the parsed Markdown text, evidence in a
figure or chart region has no textual counterpart to extract, and both
extractors return an empty span for such pairs.

\section{Metrics}
\subsection{Retrieval--Compression Alignment Metrics}\label{sec:appendix_alignment}

\begin{table*}[ht!]
\centering
\small
\setlength{\tabcolsep}{5pt}
\renewcommand{\arraystretch}{1.35}
\begin{tabularx}{\textwidth}{@{}l >{\raggedright\arraybackslash}X >{\raggedright\arraybackslash}X@{}}
\toprule
\textbf{Metric} & \textbf{Definition} & \textbf{Interpretation} \\
\midrule
\textbf{M1} Top-1 empty rate
& Fraction of queries whose rank-1 document is fully compressed ($\mathrm{kept}{=}0$).
& Directly counts the sharpest self-contradiction: discarding the document ranked first. \\

\textbf{M2} Spearman alignment
& Per-query Spearman correlation between rerank order (by score) and compression order (by surviving text), averaged over queries.
& High: Compression preserves the reranker's intent; low: the two stages disagree. \\

\textbf{M3} Downstream impact
& Mean judge score ($[0,1]$) on queries whose rank-1 document was emptied vs.\ the rest.
& A large gap shows the discrepancy actually degrades answers. \\
\bottomrule
\end{tabularx}
\vspace{-0.6em}
\caption{Retrieval--Compression alignment metrics. For each query the reranker
returns its top-$K$ documents in score order, and $\mathrm{kept}$ denotes how
many tokens are preserved after compression.}
\label{tab:alignment}
\end{table*}





 See Table \ref{tab:alignment} for details on Retrieval--Compression Alignment Metrics.

\subsection{Evidence Recall}\label{sec:app_evidencerecall_judge_correlation}

\paragraph{Definition} \emph{Evidence
Recall} measures the fraction of gold evidence that survives compression. This is a metric mainly for hard sentence-level compression methods. We build an ensemble gold annotation from Claude Opus~4.6 \citep{anthropic2026opus46card} and Gemini~2.5~Flash \citep{comanici2025gemini25pushingfrontier}, merging their spans by agreement.
A span selected by both models is treated as \emph{core evidence}
(weight $w{=}2$); a span selected by only one is \emph{supporting evidence}
($w{=}1$). Let $G$ be the set of
gold evidence spans required to answer the query, and let $P \subseteq G$
be the preserved spans. With $w(s)$ the weight of span $s$, Weighted
Evidence Recall (WER) is
\[
\mathrm{WER} \;=\; \frac{\sum_{s\in P} w(s)}{\sum_{s\in G} w(s)}.
\]
\noindent
One modification is made for token-level compressors, which delete words inside sentences.
Here, a sentence counts as surviving when more than half of its content-weighted tokens remain.

\definecolor{cOurs}{HTML}{1F8A65}
\definecolor{cZerank}{HTML}{70B0D8}
\definecolor{cOpenProv}{HTML}{5A6CC0}
\definecolor{cSemHL}{HTML}{C0455F}
\definecolor{cProv}{HTML}{C06028}
\definecolor{cXProv}{HTML}{E8C030}
\definecolor{cLL2}{HTML}{C85898}
\definecolor{cLongLL}{HTML}{8A5A2E}
\definecolor{cKeyword}{HTML}{8888A8}
\definecolor{cOLS}{HTML}{555555}

\begin{figure*}[t]
  \centering
  \begin{tikzpicture}
  \begin{axis}[
    width=0.6\textwidth,
    height=0.4\textwidth,
    xlabel={Average evidence recall (\%)},
    ylabel={Average judge accuracy},
    xmin=-2, xmax=102,
    ymin=-0.02, ymax=0.95,
    xtick={0,20,40,60,80,100},
    ytick={0,0.2,0.4,0.6,0.8},
    grid=both,
    major grid style={gray!22, thin},
    minor tick num=1,
    legend style={
      at={(0.5,-0.17)},
      anchor=north,
      legend columns=-1,
      font=\scriptsize,
      draw=none,
      fill=none,
      cells={anchor=west},
      inner sep=1pt,
      /tikz/every even column/.append style={column sep=3pt},
    },
    legend cell align=left,
    tick label style={font=\small},
    label style={font=\small},
  ]

    \addplot[cOLS, dashed, line width=1.05pt, forget plot]
      coordinates { (0,0.1315) (100,0.8979) };
    
  \addplot[only marks, mark=*, mark size=2.2pt, cOurs, forget plot]
    coordinates {
      (94.60,0.8561) (92.60,0.8414) (88.87,0.8099) (86.25,0.7901)
      (84.58,0.7966) (82.05,0.7573) (77.51,0.7139) (70.16,0.6484)
      (63.22,0.5836) (49.89,0.5189)
    };
  \addplot[only marks, mark=*, mark size=2.2pt, cZerank, forget plot]
    coordinates {
      (97.47,0.8701) (86.87,0.7866) (79.87,0.7294) (72.25,0.6246)
      (62.02,0.5998) (51.50,0.5492)
    };
  \addplot[only marks, mark=*, mark size=2.2pt, cOpenProv, forget plot]
    coordinates {
      (87.53,0.8008) (80.80,0.7844) (77.37,0.7111) (74.23,0.6984)
      (69.30,0.6352) (65.37,0.5894) (60.72,0.5341) (55.31,0.4746)
      (43.46,0.3473)
    };
    \addplot[only marks, mark=*, mark size=2.2pt, cSemHL, forget plot]
      coordinates {
        (85.43,0.7923) (81.53,0.7565) (74.67,0.6799) (72.80,0.6749)
        (68.23,0.6545) (63.15,0.6246) (55.91,0.5550) (50.59,0.5037)
        (39.49,0.3982)
      };
  \addplot[only marks, mark=*, mark size=2.2pt, cProv, forget plot]
    coordinates {
      (77.47,0.7681) (75.43,0.7336) (72.63,0.6879) (69.10,0.6648)
      (67.65,0.6710) (64.53,0.6443) (59.27,0.6051) (53.68,0.5420)
      (41.38,0.4493)
    };
  \addplot[only marks, mark=*, mark size=2.2pt, cXProv, forget plot]
    coordinates {
      (79.73,0.7948) (61.15,0.6394) (45.30,0.5281) (36.58,0.4145)
      (26.56,0.2883) (21.71,0.2471) (15.28,0.1675) (9.37,0.1180)
      (4.70,0.0773)
    };
  \addplot[only marks, mark=*, mark size=2.2pt, cKeyword, forget plot]
    coordinates { (18.04,0.2192) };

  \addplot[only marks, mark=o, mark size=2.4pt, mark options={solid}, cOurs, forget plot]
    coordinates {
      (73.78,0.6521) (68.87,0.6318) (64.31,0.6166) (59.67,0.5886)
      (55.10,0.5671) (49.70,0.5400) (43.86,0.5038) (36.58,0.4580)
      (26.04,0.3736)
    };
  \addplot[only marks, mark=o, mark size=2.4pt, mark options={solid}, cZerank, forget plot]
    coordinates {
      (64.21,0.6003) (57.70,0.5680) (52.38,0.5386) (46.36,0.5054)
      (40.51,0.4797)
    };
  \addplot[only marks, mark=o, mark size=2.4pt, mark options={solid}, cOpenProv, forget plot]
    coordinates {
      (68.40,0.6154) (64.37,0.5922) (56.97,0.5594) (51.43,0.5321)
      (46.10,0.5013) (41.18,0.4783) (36.35,0.4456) (30.73,0.4059)
      (25.21,0.3583) (16.91,0.2787)
    };
  \addplot[only marks, mark=o, mark size=2.4pt, mark options={solid}, cSemHL, forget plot]
    coordinates {
      (62.69,0.5921) (57.21,0.5743) (50.06,0.5315) (44.79,0.5016)
      (36.14,0.4440) (22.34,0.3243) (14.94,0.2461)
    };
  \addplot[only marks, mark=o, mark size=2.4pt, mark options={solid}, cProv, forget plot]
    coordinates {
      (43.75,0.5070) (39.31,0.4853) (36.03,0.4613) (31.37,0.4245)
      (28.84,0.4038) (26.33,0.3804) (23.77,0.3469) (19.15,0.2956)
    };
  \addplot[only marks, mark=o, mark size=2.4pt, mark options={solid}, cXProv, forget plot]
    coordinates {
      (57.30,0.5846) (40.12,0.4790) (27.55,0.3732) (13.53,0.2198)
      (9.04,0.1654)
    };
  \addplot[only marks, mark=o, mark size=2.4pt, mark options={solid}, cKeyword, forget plot]
    coordinates { (16.85,0.2325) };

  \addplot[only marks, mark=square*, mark size=2.4pt,
    mark options={solid, fill=cLL2, draw=cLL2}, forget plot]
    coordinates {
      (0.00,0.0849) (2.15,0.3010) (8.67,0.4430) (17.39,0.6189)
      (58.50,0.8148) (83.90,0.8497)
    };
  \addplot[only marks, mark=square*, mark size=2.4pt,
    mark options={solid, fill=cLongLL, draw=cLongLL}, forget plot]
    coordinates { (16.00,0.1546) (18.63,0.2051) };
  \addplot[only marks, mark=square, mark size=2.6pt,
    mark options={solid, draw=cLL2, fill=none}, forget plot]
    coordinates {
      (0.56,0.1324) (1.83,0.2480) (4.47,0.3351) (9.32,0.4235)
      (19.58,0.4958) (38.54,0.5455) (56.70,0.5902)
    };
  \addplot[only marks, mark=square, mark size=2.6pt,
    mark options={solid, draw=cLongLL, fill=none}, forget plot]
    coordinates {
      (11.61,0.1856) (12.62,0.2147) (14.09,0.2459) (16.54,0.2836)
      (20.03,0.3122) (24.98,0.3424) (32.59,0.3758)
    };
  \addplot[only marks, mark=*, mark size=2.2pt, cOurs] coordinates {(-10,-10)};
  \addlegendentry{Ours\,$\cdot$\,Hard}
  \addplot[only marks, mark=*, mark size=2.2pt, cZerank] coordinates {(-10,-10)};
  \addlegendentry{DSLR-Zerank}
  \addplot[only marks, mark=*, mark size=2.2pt, cOpenProv] coordinates {(-10,-10)};
  \addlegendentry{OpenProv.}
  \addplot[only marks, mark=*, mark size=2.2pt, cSemHL] coordinates {(-10,-10)};
  \addlegendentry{SemanticHL}
  \addplot[only marks, mark=*, mark size=2.2pt, cProv] coordinates {(-10,-10)};
  \addlegendentry{Provence}
  \addplot[only marks, mark=*, mark size=2.2pt, cXProv] coordinates {(-10,-10)};
  \addlegendentry{XProvence}
  \addplot[only marks, mark=*, mark size=2.2pt, cKeyword] coordinates {(-10,-10)};
  \addlegendentry{Keyword}
  \addplot[only marks, mark=square, mark size=2.5pt,
    mark options={solid, draw=cLL2, fill=none}] coordinates {(-10,-10)};
  \addlegendentry{LLMLingua-2$^\ddagger$}
  \addplot[only marks, mark=square, mark size=2.5pt,
    mark options={solid, draw=cLongLL, fill=none}] coordinates {(-10,-10)};
  \addlegendentry{LongLLMLingua$^\ddagger$}

  \end{axis}
  \end{tikzpicture}
\caption{%
  Average evidence recall vs.\ average judge accuracy
  (model$\times$threshold aggregates).
  Each point is one compressor at one threshold.
  \textbf{Filled}: unit test (unweighted mean over 10 query types);
  \textbf{open}: integration test (mean over 17 datasets);
  \textbf{squares}: token-level methods$^\ddagger$.
  Dashed: linear fit on sentence-level compressors + Keyword
  (excl.\ $^\ddagger$; Pearson $r{=}0.98$, $n{=}98$).
  $^\ddagger$These methods delete words inside sentences, so their evidence recall is measured by
  majority vote: a sentence counts as recalled when more than half of its
  content-weighted tokens survive. They are plotted but excluded from the fit because
  recall predicts their accuracy far more loosely ($r{=}0.82$, $n{=}22$).
  Soft compressors emit no readable text and have no evidence recall.
}
    \label{fig:evidence-recall-vs-accuracy}
\end{figure*}
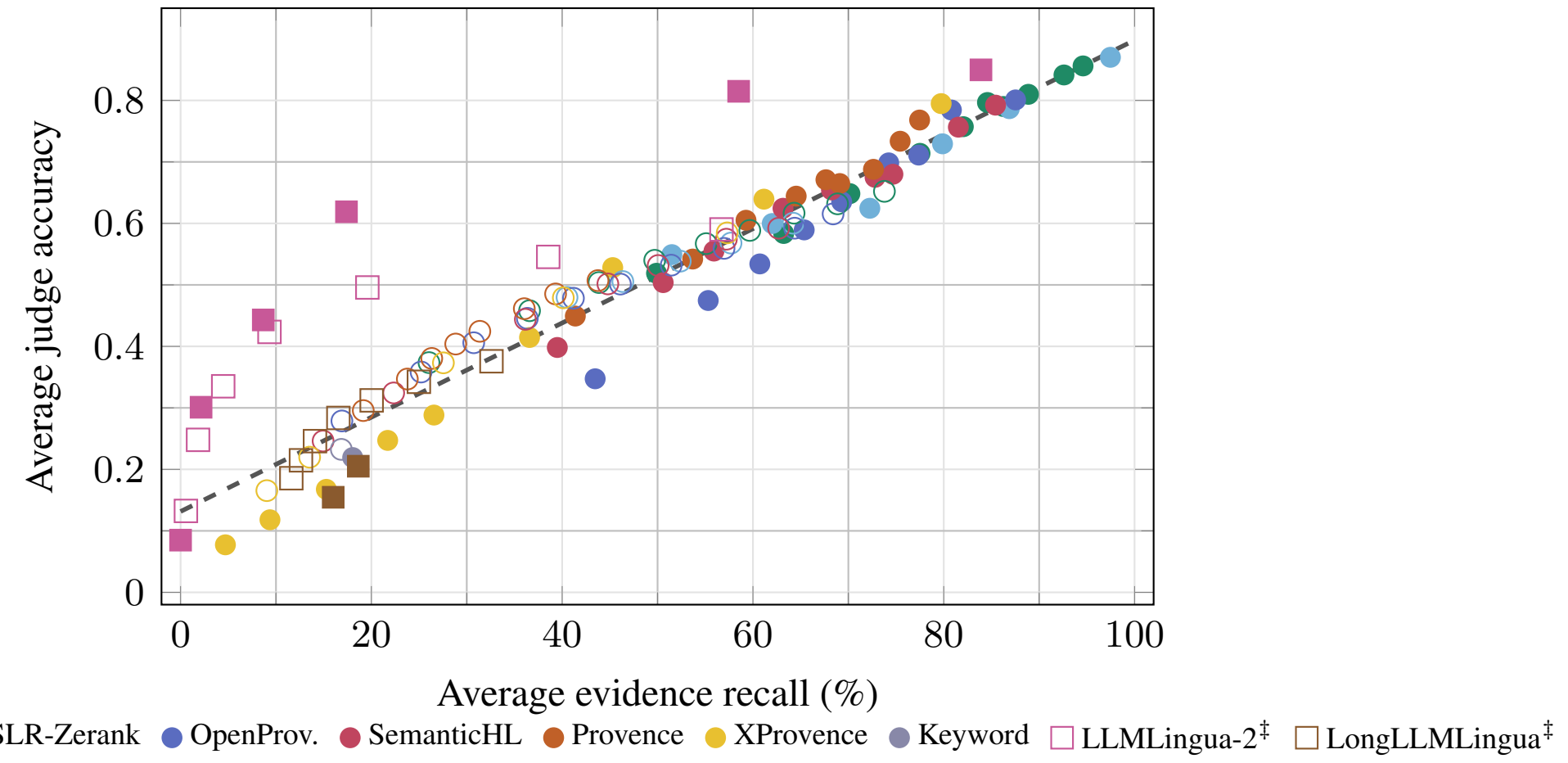


\begin{table}[ht]
\centering
\small
\begin{tabular}{lrrr}
\toprule
Set & $n$ & Pearson $r$ & Partial $r$ \\
\midrule
Unit Test & 53 & 0.987 & 0.968 \\
Integration Test & 45 & 0.978 & 0.964 \\
Pooled & 98 & 0.980 & 0.967 \\
\midrule
Token-level & 22 & 0.817 & 0.700 \\
\bottomrule
\end{tabular}
\caption{Correlation between average evidence recall and average judge
accuracy. Partial $r$: both metrics residualized on compression ratio. Token-level methods are scored with the majority-vote criterion and are excluded from the pooled row.}
\label{tab:er-partial}
\end{table}

\begin{table}[ht]
\centering
\small
\begin{tabular}{lrrr}
\toprule
Compression (\%) & $n$ & Pearson $r$ & Spearman $\rho$ \\
\midrule
$[50,75)$ & 22 & 0.979 & 0.977 \\
$[75,85)$ & 39 & 0.962 & 0.972 \\
$[85,95)$ & 28 & 0.965 & 0.973 \\
$[95,100)$ & 7 & 0.962 & 0.964 \\
\bottomrule
\end{tabular}
\caption{Evidence recall vs.\ judge accuracy within bands of similar
compression ratio (Unit + Integration pooled).
The association holds inside every band, so compression rate alone does
not explain it. Two points below $50\%$ are excluded.}
\vspace{-1.5em}
\label{tab:er-bands}
\end{table}

\paragraph{Evidence Recall and Judge Accuracy Correlation}Figure~\ref{fig:evidence-recall-vs-accuracy} relates average evidence recall to average judge accuracy at the
model$\times$threshold level.
Each point aggregates query-level scores. The Unit Test uses an
unweighted mean over ten diagnostic query types, and the Integration
Test averages over 17 datasets.
Correlations are computed across these aggregate points; we report query-level and held-out analyses separately below. Across sentence-level compressors and Keyword Snippet, the two metrics
are strongly associated
(Pearson $r{=}0.980$, Spearman $\rho{=}0.986$, $n{=}98$).
The association is not an artifact of pooling the two tiers: it holds
within each separately
(Unit $r{=}0.987$, $n{=}53$; Integration $r{=}0.978$, $n{=}45$).
Because both metrics fall as more tokens are removed, the raw
correlation might only reflect how many tokens are kept.
We check this in two ways.
First, we remove the effect of compression ratio from both metrics by
linear regression, and the correlation between the residuals is still
$r{=}0.967$ (Table~\ref{tab:er-partial}).
Second, we group the points into bands of similar compression ratio.
Within a band all methods keep about the same fraction of tokens, but
evidence recall still ranks them by accuracy (Table~\ref{tab:er-bands}).
Evidence recall therefore tracks \emph{which} spans are kept, not only
\emph{how many}.
Token-level compressors (LLMLingua-2, LongLLMLingua) are plotted but excluded
from the fit. Because they delete words inside sentences, their recall is measured
by majority vote.
Even under this criterion recall predicts
their accuracy far more loosely than it does for sentence-level pruners
($r{=}0.817$, $\rho{=}0.676$, $n{=}22$). The gap widens once compression is
partialled out ($0.700$ vs.\ $0.967$): for token-level methods the association is
driven largely by how many tokens are kept rather than by which spans survive.
Soft compressors emit no readable text, so evidence recall is undefined.
Evidence recall is therefore a useful diagnostic especially for sentence-level
compression. When accuracy drops, evidence recall tells us whether the compressor
discarded the evidence or whether the generator failed on evidence that
was still there, which is a distinction accuracy alone cannot make.

\begin{table}[t]
\centering\tiny
\setlength{\tabcolsep}{4pt}
\begin{tabular}{lcc}
\toprule
Analysis & Statistic & Value \\
\midrule
Query level & Pooled $r$ & 0.551 \\
 & Query fixed effects $r$ & 0.565 \\
 & Per-dataset $r$ (min--max) & 0.36--0.77 \\
 & Judge at ER$=0$ / ER$=1$ & 0.10 / 0.74 \\
\midrule
Leave-one-dataset-out & Median $\tau$ (MAE) & 0.91 (0.090) \\
Leave-one-compressor-out & $\tau$ (MAE), ER & 0.93 (0.022) \\
 & $\tau$ (MAE), compression & 0.44 (0.085) \\
\midrule
Compressor bootstrap & 95\% CI of $r$ & [0.971, 0.987] \\
\bottomrule
\end{tabular}
\caption{Query-level and held-out analyses of the evidence recall--judge accuracy association (Integration Test fit set of Figure~\ref{fig:evidence-recall-vs-accuracy}: 45 configurations, 17 datasets, 12{,}775 queries). $\tau$: Kendall's $\tau$; MAE in judge-accuracy units.}
\label{tab:er-robustness}
\end{table}

\paragraph{Query-level and held-out analyses.}
The aggregate points above share queries and compressors and are therefore not independent. We re-analyze the Integration Test outputs of the fit set in Figure~\ref{fig:evidence-recall-vs-accuracy}.
At the query level, evidence recall and judge accuracy correlate at $r{=}0.55$ over all configuration--query pairs. The correlation is positive in every dataset ($0.36$--$0.77$) and unchanged after removing per-query means ($r{=}0.57$). Mean judge accuracy rises monotonically with evidence recall, from $0.10$ when no gold evidence survives to $0.74$ when all of it does.
To test generalization, we fit a linear map from evidence recall to judge accuracy and apply it to unseen data. Leaving one dataset out, the predictions rank the 45 configurations on the held-out dataset with a median Kendall $\tau{=}0.91$ (MAE $0.090$). Leaving one compressor out, the pooled out-of-fold predictions reach $\tau{=}0.93$ (MAE $0.022$), versus $\tau{=}0.44$ (MAE $0.085$) when compression ratio is the predictor.
Resampling compressors with replacement gives a 95\% bootstrap interval of $[0.971, 0.987]$ for the configuration-level correlation.
Because the query-level correlation is moderate, evidence recall is best used to compare compressors and thresholds without running generation, not to predict the outcome of an individual query.

\section{Evaluation Settings}\label{sec:appendix_prompts}
\subsection{Models, Decoding, and Resources}
Table~\ref{tab:evaluation-settings} summarizes the essential settings. Generation is greedy
(temperature $0$) in every tier. Unit and Integration tests share the same generator \texttt{Qwen/Qwen3.5-9B} and
cap output at 256 tokens. Agents in the System test, \texttt{openai/gpt-oss-20b} and \texttt{120b}, run with medium reasoning effort and a much
larger output budget because they interleave reasoning with live search calls. The judge
model is fixed to \texttt{Qwen/Qwen3.5-35B-A3B} across all tiers and decodes
deterministically (temperature $0$). All inference is served with vLLM on a single node of
8$\times$NVIDIA H100 GPUs except for the System test, which used B200 GPUs.

\begin{table}[ht]
\centering\small
\setlength{\tabcolsep}{4pt}
\renewcommand{\arraystretch}{1.15}
\begin{tabular}{@{}l ccc@{}}
\toprule
& \textbf{Unit} & \textbf{Integration} & \textbf{System} \\
\midrule
\multicolumn{4}{@{}l}{\textit{Generation}} \\
\quad Model        & \multicolumn{2}{c}{Qwen3.5-9B} & \shortstack{gpt-oss-20/120b} \\
\quad Temperature  & \multicolumn{3}{c}{$0$} \\
\quad Max tokens   & $256$ & $256$ & $10{,}000$ \\
\quad Reason.\ effort & \multicolumn{2}{c}{--} & medium \\
\quad Context      & \shortstack{single doc} & \shortstack{top-5 of 50} & \shortstack{live search} \\
\addlinespace[3pt]
\multicolumn{4}{@{}l}{\textit{Judging}} \\
\quad Model        & \multicolumn{3}{c}{Qwen3.5-35B-A3B} \\
\quad Temperature  & \multicolumn{3}{c}{$0$} \\
\quad Scoring      & \multicolumn{2}{c}{yes/no prob.} & rubric \\
\bottomrule
\end{tabular}
\vspace{-0.6em}
\caption{Key evaluation settings per tier.}
\label{tab:evaluation-settings}
\end{table}

\subsection{Generator and Judge Prompts}
Prompts are shared within each evaluation tier. The Unit and Integration tests use an
identical generator prompt and an identical judge prompt following \citet{rau2024bergenbenchmarkinglibraryretrievalaugmented}. The System tests (BrowseComp
and OfficeQA) share their own generator and judge prompts (following \citet{chen2025browsecompplusfairtransparentevaluation}, which differ from the
Unit/Integration ones because generation is run by a search agent and grading uses a
structured rubric). OfficeQA additionally reports a numeric-tolerance metric on top of
the same LLM judge.

\paragraph{(A) Generator prompt — Unit and Integration Tests}
System message followed by the context and question.

\begin{tcolorbox}[breakable, colback=gray!5, colframe=gray!50!black,
  title={Generator (Unit \& Integration Tests)},
  fonttitle=\bfseries\small, fontupper=\scriptsize]
\begin{verbatim}
[system] 
Answer the Question, using only information provided in the 
Context. If no useful information is provided, you must 
output `No answer'.

[user]
Context:
{context}

Question:
{query}
\end{verbatim}
\end{tcolorbox}

\paragraph{(B) Generator prompt — System Tests}
The agent answers by issuing search-tool calls.

\begin{tcolorbox}[breakable, colback=gray!5, colframe=gray!50!black,
  title={Generator (System Test)},
  fonttitle=\bfseries\small, fontupper=\scriptsize]
\begin{verbatim}
You are a deep research agent. You need to answer the given 
question by interacting with a search engine, using the 
search tool provided. Please perform reasoning and use 
the tool step by step, in an interleaved manner. 
You may use the search tool multiple times.

Question: {Question}

Your response should be in the following format:

Explanation: {your explanation for your final answer. 
For this explanation section only, you should cite 
your evidence documents inline by enclosing their docids 
in square brackets [] at the end of sentences. 
For example, [20].}

Exact Answer: {your succinct, final answer}

Confidence: {your confidence score between 0% and 100% for 
your answer}
\end{verbatim}
\end{tcolorbox}

\paragraph{(C) Judge prompt — Unit and Integration Tests}
We follow the evaluation method of \citet{rau2024bergenbenchmarkinglibraryretrievalaugmented}. The judge answers Yes/No; the score is the softmax weight on the `Yes' first token, giving a value in $[0,1]$.

\begin{tcolorbox}[breakable, colback=gray!5, colframe=gray!50!black,
  title={Judge (Unit \& Integration Tests)},
  fonttitle=\bfseries\small, fontupper=\scriptsize]
\begin{verbatim}
[system]
You are an evaluation tool. Answer with one of Yes, No.

[user] 
Here is a question, a golden answer and an AI-generated 
answer. Can you judge whether the AI-generated answer 
is correct according to the question and golden answer, 
simply answer with one of Yes, No.

Question: {question}

Golden answer: {answer}

Generated answer: {prediction}
\end{verbatim}
\end{tcolorbox}

\paragraph{(D) Judge Prompt — System Tests}
We minimally modify the BrowseComp-Plus judge prompt to ensure Qwen consistently outputs the required \texttt{correct: yes/no} field, preventing parse failures that would otherwise reduce accuracy and calibration coverage.

\begin{tcolorbox}[breakable, colback=gray!5, colframe=gray!50!black,
  title={Judge (System Test)},
  fonttitle=\bfseries\small, fontupper=\scriptsize]
\begin{verbatim}
Judge whether the following [response] to [question] is 
correct or not based on the precise and unambiguous 
[correct_answer] below.

[question]: {question}

[response]: {response}

[correct_answer]: {correct_answer}

Your judgement must be in the format and criteria 
specified below:

extracted_final_answer: The final exact answer extracted 
from the [response].

[correct_answer]: Repeat the [correct_answer] given above.

reasoning: Explain why the extracted_final_answer is 
correct or incorrect based on [correct_answer], in the 
context of this [question]. You should judge whether the 
extracted_final_answer is semantically equivalent to 
[correct_answer], allowing the extracted_final_answer to 
be string variations of [correct_answer]. You should also 
allow the extracted_final_answer to be more precise or 
verbose than [correct_answer], as long as its additional 
details are correct. Do not comment on any background to the 
problem, do not attempt to solve the problem, do not argue 
for any answer different than [correct_answer], focus only 
on whether the answers are semantically equivalent.

correct: Answer 'yes' if extracted_final_answer matches the 
[correct_answer] given above, or is within a small margin of 
error for numerical problems. Answer 'no' otherwise, 
i.e. if there is any inconsistency, ambiguity, 
non-equivalency, or if the extracted answer is incorrect.

confidence: The extracted confidence score between 0% 
and 100% from [response]. Put 100 if there is no 
confidence score available.
\end{verbatim}
\end{tcolorbox}

\subsection{Choice of Judge Model}

\begin{figure}[h!]
    \centering    \includegraphics[width=\linewidth]{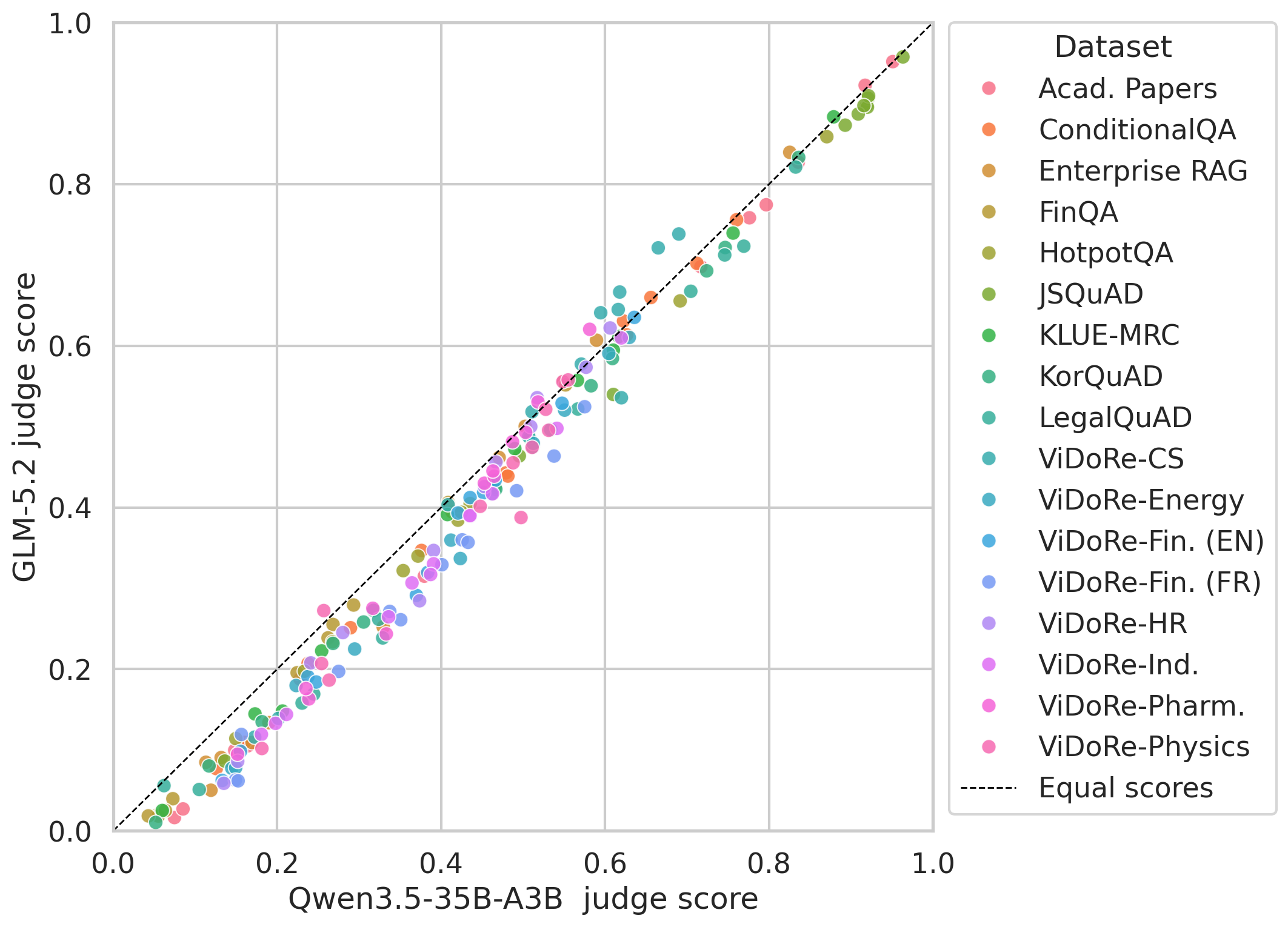}
\caption{Integration Test scores per dataset of Qwen3.5-35B-A3B judge vs GLM-5.2 judge. We plot all model configs shown in Table~\ref{tab:heatmap-integration}.}
\vspace{-0.7em}
\label{fig:judge-comparison}
\end{figure}

\begin{table}[t]
\centering
\small
\setlength{\tabcolsep}{3pt}
\begin{tabular}{@{}lrr@{}}
\toprule
Dataset & Pearson & Spearman \\
\midrule
Academic Papers & 0.9675 & 1.0000 \\
ConditionalQA & 0.9261 & 0.9860 \\
Enterprise RAG & 0.9344 & 0.9930 \\
FinQA & 0.9605 & 0.9441 \\
HotpotQA & 0.9692 & 0.9930 \\
JSQuAD & 0.9435 & 0.9790 \\
KLUE-MRC & 0.9751 & 1.0000 \\
KorQuAD & 0.9450 & 1.0000 \\
LegalQuAD & 0.9069 & 0.9860 \\
ViDoRe-CS & 0.8867 & 0.9301 \\
ViDoRe-Energy & 0.9016 & 0.9930 \\
ViDoRe-Finance (EN) & 0.9133 & 0.9930 \\
ViDoRe-Finance (FR) & 0.9141 & 0.9790 \\
ViDoRe-HR & 0.8971 & 1.0000 \\
ViDoRe-Industrial & 0.9079 & 0.9930 \\
ViDoRe-Pharmaceuticals & 0.8839 & 0.9790 \\
ViDoRe-Physics & 0.8419 & 0.9510 \\
\midrule
Total & 0.9532 & 1.0000 \\
\bottomrule
\end{tabular}
\caption{Per-answer Pearson correlation and per-method Spearman rank correlation between Qwen3.5-35B-A3B as judge and GLM-5.2 as judge.}
\label{tab:judge-method-ranking-correlation}
\end{table}

We ablate whether the choice of judge model significantly influences the results of our evaluations.
For this purpose, we take evaluation runs shown in Table~\ref{tab:heatmap-integration} and re-judge them using GLM-5.2 \citep{glm5team2026glm5vibecodingagentic}.
The per-dataset judge scores are plotted against each other in Figure~\ref{fig:judge-comparison}.
In Table~\ref{tab:judge-method-ranking-correlation}, we additionally show Pearson's correlation between raw per-answer judge scores and Spearman's rank correlation between orderings of models, averaged over datasets.
Both confirm that judge scores are highly correlated across datasets and that our evaluation framework is robust to the choice of judge model.

\section{Unit Test}\label{sec:appendix_unittest_result}
\subsection{Settings}\label{sec:appendix_unittest_settings}
\paragraph{Data} Table~\ref{tab:unittest-query-types-appendix} shows how the Unit Test is constructed. All queries were reviewed and modified by a human annotator for quality control. One query in \textsc{information-density} and three queries in \textsc{contextualized} were modified and four queries were replaced with different query--document pairs. Three queries in \textsc{union} were modified.

\begin{table*}[t]
\centering
\small
\begin{tabularx}{\textwidth}{p{3.3cm} p{3.3cm} >{\raggedright\arraybackslash}X@{}}
\toprule
\textbf{Query Type} & \textbf{Source} & \textbf{Document Construction} \\
\midrule
\textsc{Multi-hop Long} & HotpotQA \citep{yang2018hotpotqadatasetdiverseexplainable} (bridge) & Two gold bridge paragraphs embedded at ${\sim}25\%$ and ${\sim}70\%$ depth in ${\sim}128$K tokens of HotpotQA distractor paragraphs. \\
\textsc{Numerical Tables} & FinQA \citep{chen-etal-2021-finqa} (6), Wikipedia (4) & Documents dense in multi-digit figures, identifiers, and rare named entities. \\
\textsc{Scattered Evidence} & arXiv papers \citep{acar_arxiver2024} & Full paper body (8--26K chars) with relevant facts spread across sections. \\
\textsc{Information Density} & SQuAD v1.1 \citep{rajpurkar-etal-2016-squad}, Wikipedia & Dense (5): tight SQuAD paragraph; sparse (5): SQuAD paragraph buried among ${\sim}6$ unrelated Wikipedia paragraphs. \\
\textsc{Adversarial Distractor} & HotpotQA & One gold + three same-topic, highest-retrieval-score non-gold paragraphs, shuffled; each record carries a same-type wrong-answer span. \\
\textsc{Contextualized} & Quoref (dev) \citep{dasigi2019quorefreadingcomprehensiondataset} & Wikipedia paragraph where the answer requires coreference resolution; the answer name appears only in the antecedent sentence. \\
\textsc{Union} & Wikipedia  & Article whose set members (e.g., filmographies, band lineups) are scattered across ${\ge}3$ sentences, never listed in a single sentence. \\
\textsc{Summarization} & Wikipedia  & Overview sections of a broad topic with no single answer span. \\
\textsc{Ugly Formatting} & FinQA (5), Wikipedia (5) & Natural (serialized tables) and synthetic noise (markup, OCR headers/footers, mojibake, hyphenation); evidence offsets remapped to remain verbatim. \\
\textsc{Cross-lingual} & ViDoRe-FR \citep{loison2026vidorev3comprehensiveevaluation} (3), KorQuAD \citep{ART002594126} (4), JSQuAD \citep{kurihara-etal-2022-jglue}(3) & English query over a native FR/KO/JA document; only the query and answer are translated. \\
\bottomrule
\end{tabularx}
\caption{Unit Test: the ten query types, their source datasets, and document construction procedures.}
\label{tab:unittest-query-types-appendix}
\end{table*}

\input{figure_latex/unit_test_acc_compression}

\subsection{Results}\label{sec:appendix_unittest_results}

Figure~\ref{fig:unit-accuracy-vs-threshold-comp} reports the unweighted mean judge
accuracy (top row) and evidence recall (bottom row) across the ten unit-test query
types.

Since each unit-test instance contains a single document, reranking is not involved and
the task isolates compression quality alone. Under this setting, Comp-H and
Semantic Highlight are the two strongest methods, both staying close to the uncompressed
ceiling over a wide range of compression ratios. The two are nearly tied at the matched $\sim$75\% point (0.80 vs.\ 0.79).
In the mid-to-high compression range ($\sim$78--87\%), Semantic Highlight
sits slightly above Comp-H.

\subsection{Qualitative Examples}
\definecolor{good}{RGB}{27,138,101}
\definecolor{bad}{RGB}{176,58,58}
\newtcolorbox{qexample}[1]{enhanced, colback=black!2, colframe=black!30,
  boxrule=0.4pt, arc=2pt, left=5pt, right=5pt, top=4pt, bottom=4pt,
  fonttitle=\bfseries\small, title=#1}
\newcommand{\uscore}[1]{\textcolor{bad}{\textbf{#1}}}
\newcommand{\gscore}[1]{\textcolor{good}{\textbf{#1}}}


\begin{figure}[t]
\begin{qexample}{\texttt{contextualized}: a pruned antecedent leaves an unresolvable pronoun}
\textbf{Q.} How many kilograms does the Rosetta Stone weigh?\quad\textbf{Gold.} 760\\[3pt]
\begin{minipage}[t]{0.47\linewidth}
\textbf{Unpruned} (\gscore{0.99})\\[1pt]
\footnotesize The Rosetta Stone weighs approximately 760 kilograms.
\end{minipage}\hfill
\begin{minipage}[t]{0.47\linewidth}
\textbf{Comp-H, t0.7} (\uscore{0.07})\\[1pt]
\footnotesize Keeps only ``It weighs approximately 760 kg.'' With no referent for
``It'', the model refuses to answer.
\end{minipage}\\[4pt]
\textbf{Analysis.} The number 760 is literally present, but pruning dropped the
low-overlap subject sentence naming the Rosetta Stone. Similarity-based pruning keeps
the answer sentence and discards its antecedent, breaking the coreference.
\end{qexample}
\vspace{-1.2em}
\caption{\textbf{Unit Test Example (Compression Failure)}: Coreference broken by aggressive pruning.}
\label{fig:qual-contextualized}
\end{figure}

\begin{figure}[t]
\begin{qexample}{\texttt{cross\_lingual}: an English query never lexically matches the Japanese evidence}
\textbf{Q.} Is summer in the Kinki region longer or shorter than in Hokuriku? (EN query,
JA document)\quad\textbf{Gold.} Longer\\[3pt]
\begin{minipage}[t]{0.47\linewidth}
\textbf{Unpruned} (\gscore{1.00})\\[1pt]
\footnotesize Summer in the Kinki region is longer than in the Hokuriku region.
\end{minipage}\hfill
\begin{minipage}[t]{0.47\linewidth}
\textbf{Comp-H, t0.7} (\uscore{0.02})\\[1pt]
\footnotesize The gold Japanese sentence is scored as low relevance and cut, so the
model refuses.
\end{minipage}\\[4pt]
\textbf{Analysis.} Relevance must be scored across languages, which is noisy, so the
one Japanese sentence the English query needs is pruned. At \texttt{t0.1} the same
query stays correct (1.00), so this is a cost of aggressive compression, not of
compression itself.
\end{qexample}
\vspace{-1.2em}
\caption{\textbf{Unit Test Example (Compression Failure)}: Cross-lingual evidence lost under aggressive pruning.}
\label{fig:qual-crosslingual}
\end{figure}


\begin{figure}[t]
\begin{qexample}{\texttt{summarization}: pruning frees budget to cover the whole question}
\textbf{Q.} Can you give an overview of quantum computing and where it stands today?\\
\textbf{Gold.} Uses superposition and entanglement, qubit is the basic unit, not yet
practical due to decoherence, heavy government investment, quantum advantage claimed on
narrow tasks.\\[3pt]
\begin{minipage}[t]{0.47\linewidth}
\textbf{Unpruned} (\uscore{0.27})\\[1pt]
\footnotesize A long mechanistic deep dive on qubits, superposition, and measurement
that reaches the output length limit before it ever covers ``where it stands today''.
\end{minipage}\hfill
\begin{minipage}[t]{0.47\linewidth}
\textbf{Comp-H, t0.7} (\gscore{0.89})\\[1pt]
\footnotesize Overview plus current status: experimental, not yet practical,
governments investing in scalable qubits. Covers both halves of the question.
\end{minipage}\\[4pt]
\textbf{Analysis.} Same source facts, but the full context is so rich that the generator
front-loads one sub-topic and runs out of output budget. Compressing to the essential
spans lets the answer address every part of the question.
\end{qexample}
\vspace{-1.2em}
\caption{\textbf{Unit Test Example (Success)}: Compression beats the full context on broad overview questions.}
\label{fig:qual-summarization}
\end{figure}

\begin{figure}[t]
\begin{qexample}{\texttt{multi\_hop}: removing a distractor enables the bridge}
\textbf{Q.} Who was known by the stage name Aladin and worked as a consultant helping
organizations improve performance?\quad\textbf{Gold.} Eenasul Fateh\\[3pt]
\begin{minipage}[t]{0.47\linewidth}
\textbf{Unpruned} (\uscore{0.20})\\[1pt]
\footnotesize ``There is no single individual who fits both descriptions.'' Distracted
by a separate management-consulting paragraph, the model refuses to bridge.
\end{minipage}\hfill
\begin{minipage}[t]{0.47\linewidth}
\textbf{Comp-H, t0.7} (\gscore{1.00})\\[1pt]
\footnotesize Eenasul Fateh was known by the stage name Aladin and was a former
international management consultant.
\end{minipage}\\[4pt]
\textbf{Analysis.} The distractor paragraph blocks the two-hop connection. Removing it
links Aladin to Eenasul Fateh to consultant, and the model answers correctly.
\end{qexample}
\vspace{-1.2em}
\caption{\textbf{Unit Test Example (Success)}: Compression removes a distractor that blocked multi-hop reasoning.}
\label{fig:qual-multihop}
\end{figure}
All answers in Figures~\ref{fig:qual-contextualized}--\ref{fig:qual-multihop} are real
generations on the Unit Test. Figures~\ref{fig:qual-contextualized}
and~\ref{fig:qual-crosslingual} are failure cases and
Figures~\ref{fig:qual-summarization} and~\ref{fig:qual-multihop} are cases where
compression matches or beats the full context. The failure cases use the aggressive
\texttt{t0.7} setting to expose where hard compression removes needed evidence. At the lower threshold, the same queries are
preserved and answered correctly.

\clearpage

\section{Integration Test}\label{sec:appendix_integrationtest_result}

\subsection{Results}
\subsubsection{General Results}

\begin{table}[ht]
\centering
\small
\setlength{\tabcolsep}{3.5pt}
\definecolor{npgray}{gray}{0.92}
\newcommand{\sh}{\cellcolor{npgray}}
\resizebox{\columnwidth}{!}{%
\begin{tabular}{llcccc}
\toprule
\textbf{Model} & \textbf{Config} & \textbf{Ratio} & \textbf{Judge} & \textbf{Evi.Recall} & \textbf{N@5} \\
\midrule
\textbf{Comp-H}
 & \sh\emph{no compr.} & \sh-- & \sh 0.68 & \sh 81.4 & \sh 79.3 \\
 & $t{=}0.2$ & 74.6 & \textbf{0.63} & \textbf{68.9} & \textbf{79.3} \\
\addlinespace[2pt]

Semantic HL
 & \sh\emph{no compr.} & \sh-- & \sh 0.67 & \sh 77.1 & \sh 73.0 \\
 & $t{=}0.1$ & 77.3 & 0.57 & 57.2 & 73.0 \\
\addlinespace[2pt]

OpenProvence
 & \sh\emph{no compr.} & \sh-- & \sh 0.66 & \sh 76.6 & \sh 74.0 \\
 & $t{=}0.2$ & 74.4 & 0.56 & 57.0 & 73.3 \\
\addlinespace[2pt]

DSLR-Zerank
 & \sh\emph{no compr.} & \sh-- & \sh 0.65 & \sh 75.4 & \sh 70.2 \\
 & $t{=}0.5$ & 75.1 & 0.51 & 46.4 & 70.2 \\
\addlinespace[2pt]

Provence
 & \sh\emph{no compr.} & \sh-- & \sh 0.62 & \sh 68.5 & \sh 64.7 \\
 & $t{=}0.1$ & 73.6 & 0.51 & 43.8 & 64.7 \\
\addlinespace[2pt]

XProvence
 & \sh\emph{no compr.} & \sh-- & \sh 0.63 & \sh 71.1 & \sh 65.0 \\
 & $t{=}0.3$ & 77.4 & 0.37 & 27.6 & 65.0 \\
\addlinespace[2pt]

LLMLingua-2
 & \sh\emph{no compr.} & \sh-- & \sh 0.67 & \sh 77.1 & \sh 73.0 \\
 & $t{=}0.2$ & 76.8 & 0.25 & 1.8 & 73.0 \\
\addlinespace[2pt]

Sentinel
 & \sh\emph{no compr.} & \sh-- & \sh 0.67 & \sh 77.1 & \sh 73.0 \\
 & $t{=}0.1$ & 77.2 & 0.46 & 40.2 & 73.0 \\
\addlinespace[2pt]

RECOMP
 & \sh\emph{no compr.} & \sh-- & \sh 0.67 & \sh 77.1 & \sh 73.0 \\
 & $t{=}0.6$ & 70.0 & 0.42 & 35.5 & 73.0 \\
\addlinespace[2pt]

Keyword Snippet
 & \sh\emph{no compr.} & \sh-- & \sh 0.56 & \sh 59.8 & \sh 52.1 \\
 & -- & 80.5 & 0.23 & 16.9 & 52.1 \\
\addlinespace[2pt]

LongLLMLingua
 & \sh\emph{no compr.} & \sh-- & \sh 0.62 & \sh 68.0 & \sh 64.3 \\
 & $t{=}0.1$ & 63.6 & 0.19 & 11.6 & 64.3 \\
\addlinespace[2pt]

\textbf{Comp-S}$^{*}$
 & fixed & 93.6 & 0.36 & n/a & 73.0 \\
\addlinespace[2pt]

OSCAR$^{*}$
 & fixed & 93.6 & 0.27 & n/a & 73.0 \\
\bottomrule
\end{tabular}}
\caption{\textbf{Integration Test.} Comparison at matched compression
($\approx$75\% target) over 17 datasets, with the no-compression upper
reference (shaded; full reranked top-5, 0\% compression). Ratio: compression
(\%); Evi.Recall: evidence recall (\%); N@5: NDCG@5. N@5 matches across the two
rows by construction except OpenProvence, whose
\texttt{zero\_score\_when\_empty} default zeroes the score of emptied
documents. Soft compression models, Sentinel, RECOMP, LLMLingua-2 and Semantic Highlight use BGE-M3 for reranking. $^{*}$Fixed-rate soft
compressors ($\approx$93.6\%) cannot reach 75\%. See Table~\ref{tab:matched_soft} for a comparison at matched $\approx$93.6\%.}
\label{tab:appendix-matched-compression}
\end{table}

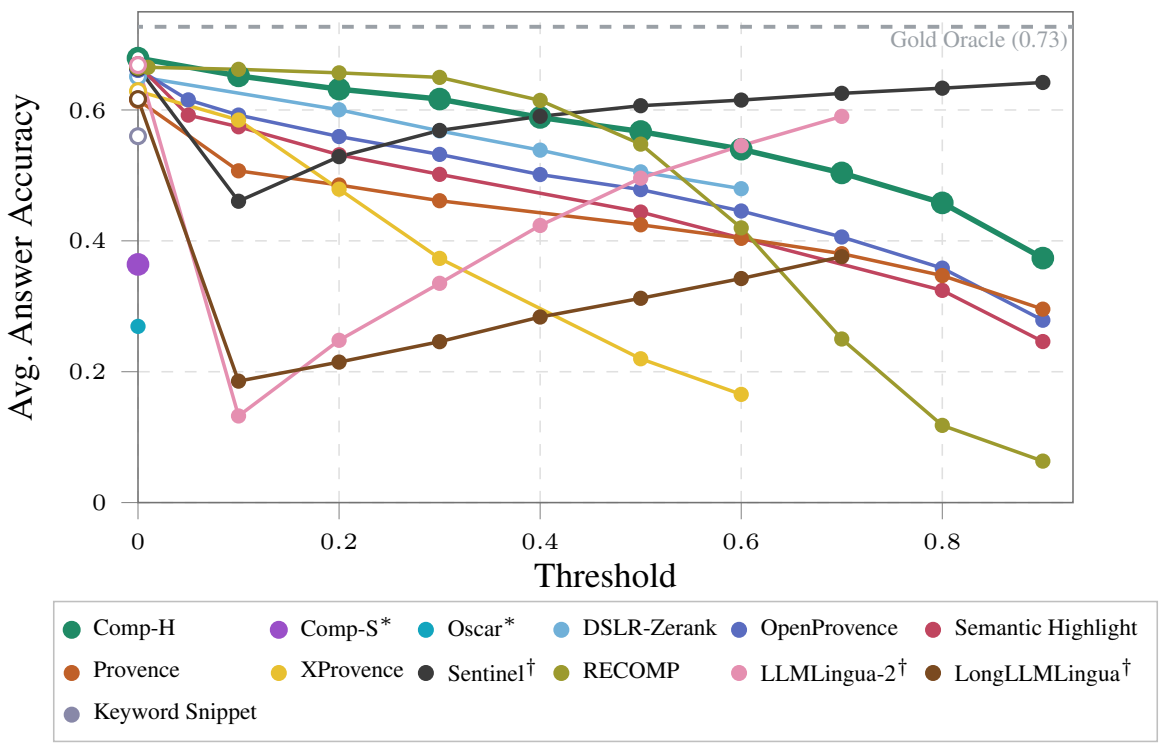
\begin{figure}[h!]
  \centering
  \definecolor{cOursExt}{HTML}{1F8A65}
  \definecolor{cOursAbs}{HTML}{9A4FC8}
  \definecolor{cOscar}{HTML}{12A5BE}
  \definecolor{cZerank}{HTML}{70B0D8}
  \definecolor{cOpenProv}{HTML}{5A6CC0}
  \definecolor{cZilliz}{HTML}{C0455F}
  \definecolor{cProv}{HTML}{C06028}
  \definecolor{cXProv}{HTML}{E8C030}
  \definecolor{cLL2}{HTML}{E58FB0}
  \definecolor{cLongLL}{HTML}{7A4A20}
  \definecolor{cKeyword}{HTML}{8888A8}
  \definecolor{cSentinel}{HTML}{3B3B3B}
  \definecolor{cRecomp}{HTML}{9C9A2E}
  \definecolor{cGold}{HTML}{9AA0A6}
  \pgfplotsset{
    ours/.style={line width=1.4pt, mark=*, mark size=2.2pt},
    base/.style={line width=0.9pt, mark=*, mark size=1.5pt},
    baselinemark/.style={only marks, mark=*, mark size=1.9pt,
                         mark options={fill=white, line width=0.8pt}, forget plot},
  }
  \resizebox{\columnwidth}{!}{%
  \begin{tikzpicture}
  \begin{axis}[
      width=10cm, height=6cm,
      xmin=0, xmax=0.93, ymin=0, ymax=0.75,
      xtick={0,0.2,0.4,0.6,0.8},
      ytick={0,0.2,0.4,0.6},
      xlabel={Threshold},
      ylabel={Avg.\ Answer Accuracy},
      grid=both,
      major grid style={dashed, black!12},
      tick align=outside, tick pos=left,
      axis line style={black!55},
      label style={font=\footnotesize}, 
      xlabel style={yshift=4pt},
      tick label style={font=\tiny},
      legend columns=6,
      legend cell align=left,
      legend style={
        at={(0.5,-0.2)}, anchor=north, draw=black!25,
        font=\tiny, column sep=2pt,
        /tikz/every even column/.append style={column sep=2pt},
      },
      legend image code/.code={
        \draw[#1, only marks, mark=*, mark size=1.6pt]
          plot coordinates {(0.08cm,0cm)};
      },
    ]

    \addplot[cGold, dashed, line width=1pt, forget plot]
      coordinates {(-0.03,0.7271) (0.95,0.7271)};
    \node[anchor=east, cGold, font=\tiny] at (axis cs:0.94,0.705)
      {Gold Oracle (0.73)};

    \addplot[cOursExt, ours] coordinates {
      (0,0.6792) (0.1,0.6521) (0.2,0.6318) (0.3,0.6166) (0.4,0.5886)
      (0.5,0.5671) (0.6,0.5400) (0.7,0.5038) (0.8,0.4580) (0.9,0.3736)};
    \addlegendentry{Comp-H}
    \addplot[cOursExt, baselinemark] coordinates {(0,0.6792)};

    \addplot[cOursAbs, ours] coordinates {(0,0.3640)};
    \addlegendentry{Comp-S$^\ast$}

    \addplot[cOscar, base] coordinates {(0,0.2693)};
    \addlegendentry{Oscar$^\ast$}

    \addplot[cZerank, base] coordinates {
      (0,0.6511) (0.2,0.6003) (0.3,0.5680) (0.4,0.5386) (0.5,0.5054)
      (0.6,0.4797)};
    \addlegendentry{DSLR-Zerank}
    \addplot[cZerank, baselinemark] coordinates {(0,0.6511)};

    \addplot[cOpenProv, base] coordinates {
      (0,0.6625) (0.05,0.6154) (0.1,0.5922) (0.2,0.5594) (0.3,0.5321)
      (0.4,0.5013) (0.5,0.4783) (0.6,0.4456) (0.7,0.4059) (0.8,0.3583)
      (0.9,0.2787)};
    \addlegendentry{OpenProvence}
    \addplot[cOpenProv, baselinemark] coordinates {(0,0.6625)};

    \addplot[cZilliz, base] coordinates {
      (0,0.6686) (0.05,0.5921) (0.1,0.5743) (0.2,0.5315) (0.3,0.5016)
      (0.5,0.4440) (0.8,0.3243) (0.9,0.2461)};
    \addlegendentry{Semantic Highlight}
    \addplot[cZilliz, baselinemark] coordinates {(0,0.6686)};

    \addplot[cProv, base] coordinates {
      (0,0.6155) (0.1,0.5070) (0.2,0.4853) (0.3,0.4613) (0.5,0.4245)
      (0.6,0.4038) (0.7,0.3804) (0.8,0.3469) (0.9,0.2956)};
    \addlegendentry{Provence}
    \addplot[cProv, baselinemark] coordinates {(0,0.6155)};

    \addplot[cXProv, base] coordinates {
      (0,0.6288) (0.1,0.5846) (0.2,0.4790) (0.3,0.3732) (0.5,0.2198)
      (0.6,0.1654)};
    \addlegendentry{XProvence}
    \addplot[cXProv, baselinemark] coordinates {(0,0.6288)};

    \addplot[cSentinel, base] coordinates {
      (0,0.6646) (0.1,0.4605) (0.2,0.5287) (0.3,0.5687) (0.4,0.5903)
      (0.5,0.6065) (0.6,0.6151) (0.7,0.6254) (0.8,0.6333) (0.9,0.6420)};
    \addlegendentry{Sentinel$^\dagger$}
    \addplot[cSentinel, baselinemark] coordinates {(0,0.6646)};

    \addplot[cRecomp, base] coordinates {
      (0,0.6663) (0.01,0.6652) (0.1,0.6621) (0.2,0.6567) (0.3,0.6498)
      (0.4,0.6146) (0.5,0.5478) (0.6,0.4198) (0.7,0.2500) (0.8,0.1180)
      (0.9,0.0634)};
    \addlegendentry{RECOMP}
    \addplot[cRecomp, baselinemark] coordinates {(0,0.6663)};

    \addplot[cLL2, base] coordinates {
      (0,0.6686) (0.1,0.1324) (0.2,0.2480) (0.3,0.3351) (0.4,0.4235)
      (0.5,0.4958) (0.6,0.5455) (0.7,0.5902)};
    \addlegendentry{LLMLingua-2$^\dagger$}
    \addplot[cLL2, baselinemark] coordinates {(0,0.6686)};

    \addplot[cLongLL, base] coordinates {
      (0,0.6167) (0.1,0.1856) (0.2,0.2147) (0.3,0.2459) (0.4,0.2836)
      (0.5,0.3122) (0.6,0.3424) (0.7,0.3758)};
    \addlegendentry{LongLLMLingua$^\dagger$}
    \addplot[cLongLL, baselinemark] coordinates {(0,0.6167)};

    \addplot[cKeyword, base] coordinates {(0,0.5599)};
    \addlegendentry{Keyword Snippet}
    \addplot[cKeyword, baselinemark] coordinates {(0,0.5599)};

  \end{axis}
  \end{tikzpicture}%
  }
  \vspace{-1.6em}
  \caption{Average accuracy vs.\ pruning threshold, aggregated over 17 datasets
    in \textbf{Integration Test}.  Open circles at threshold $0$ are each
    reranker's no-pruning top-5 baseline. $^\dagger$For the token-level
    compressors and Sentinel, the threshold controls how much context is kept
    rather than acting as a pruning cutoff, so their curves run in the opposite direction. $^\ast$The soft
    compressors and Keyword Snippet have no threshold;
    they are drawn as single markers.}
  \label{fig:judge-vs-threshold}
\end{figure}


\noindent Table~\ref{tab:appendix-matched-compression} compares judge score, evidence recall, and NDCG@5 at matched compression.
Overall, Comp-H shows the strongest results across all three metrics. Figure~\ref{fig:judge-vs-threshold} shows how accuracy varies with the compression threshold. 

\paragraph{Matched comparison with soft compression.}
Soft compressors (Comp-S, OSCAR) have no threshold and run at a fixed $\approx$93.6\% ratio, so they cannot be compared at the 75\% target in Table~\ref{tab:heatmap-integration}.
For a controlled hard-vs-soft comparison, Table~\ref{tab:matched_soft} reports each hard compressor at the threshold closest to 93.6\%.
Comp-H remains the best on average (0.46), followed by Semantic Highlight (0.44), and both clearly outperform Comp-S (0.36) and OSCAR (0.27).
Comp-S is best on ViDoRe CS and Physics, which is consistent with our finding that soft compression is strong when answers require semantic aggregation.
Overall, the matched comparison does not change our main findings.

\begin{table*}[t]
\centering
\small
\setlength{\tabcolsep}{3pt}
\resizebox{\textwidth}{!}{%
\begin{tabular}{lccccccccccccccccccc c}
\toprule
& & & \multicolumn{5}{c}{EN Factoid / RAG} & \multicolumn{4}{c}{Multilingual} & \multicolumn{5}{c}{ViDoRe -- EN} & \multicolumn{3}{c}{ViDoRe -- FR} & \\
\cmidrule(lr){4-8}\cmidrule(lr){9-12}\cmidrule(lr){13-17}\cmidrule(lr){18-20}
Model & Config & Comp.\% & Acad. & ConQA & FinQA & Hotpot & Enter. & JSQ. & KorQ. & KLUE & Legal & CS & Fin. & HR & Ind. & Pharm. & Energy & Fin. & Phys. & Avg. \\
\midrule
Gold Oracle & -- & -- & .95 & .76 & .55 & .87 & .83 & .96 & .84 & .88 & .83 & .69 & .64 & .61 & .62 & .58 & .63 & .58 & .55 & .73 \\
\midrule
Comp-H (ours) & $\tau$ 0.8 & 93.3 & \textbf{.62} & .37 & \textbf{.19} & .22 & .35 & \textbf{.90} & \textbf{.57} & .44 & .74 & .49 & .36 & \textbf{.44} & \textbf{.41} & .39 & \textbf{.46} & \textbf{.40} & .46 & \textbf{.46} \\
Comp-S (ours) & -- & 93.6 & .16 & .48 & .07 & .42 & .33 & .61 & .18 & .25 & .33 & \textbf{.62} & \textbf{.37} & .37 & .39 & .33 & .42 & .34 & \textbf{.50} & .36 \\
OSCAR & -- & 93.6 & .12 & .40 & .04 & \textbf{.69} & .12 & .49 & .07 & .17 & .07 & .57 & .24 & .28 & .34 & .32 & .22 & .16 & .26 & .27 \\
OpenProvence & $\tau$ 0.8 & 94.7 & .53 & \textbf{.53} & .13 & .22 & .29 & .89 & .34 & .11 & .54 & .38 & .31 & .42 & .30 & .31 & .35 & .23 & .22 & .36 \\
Semantic HL & $\tau$ 0.5 & 91.8 & .57 & .41 & .14 & .27 & \textbf{.40} & .88 & .45 & \textbf{.45} & \textbf{.76} & .47 & \textbf{.37} & .39 & .36 & \textbf{.40} & .45 & .38 & .38 & .44 \\
XProvence & $\tau$ 0.6 & 94.5 & .06 & .12 & .04 & .25 & .09 & .46 & .04 & .21 & .09 & .18 & .14 & .15 & .18 & .26 & .15 & .12 & .28 & .17 \\
Provence & $\tau$ 0.9 & 92.6 & .45 & .31 & .09 & .18 & .27 & .74 & .28 & .25 & .22 & .23 & .27 & .24 & .27 & .32 & .32 & .25 & .34 & .30 \\
RECOMP & $\tau$ 0.8 & 97.1 & .11 & .07 & .05 & .04 & .08 & .34 & .21 & .25 & .12 & .17 & .08 & .07 & .12 & .10 & .07 & .07 & .07 & .12 \\
\midrule
\multicolumn{21}{c}{\textit{Methods that cannot reach $\approx$94\% compression (shown at their maximum)}} \\
\midrule
DSLR-Zerank & $\tau$ 0.6 & 80.2 & .67 & .65 & .23 & .19 & .40 & .93 & .59 & .46 & .58 & .44 & .40 & .43 & .35 & .45 & .49 & .40 & .50 & .48 \\
LLMLingua-2 & $\tau$ 0.1 & 85.7 & .19 & .28 & .04 & .06 & .09 & .13 & .18 & .09 & .19 & .15 & .14 & .13 & .11 & .07 & .15 & .15 & .09 & .13 \\
Keyword Snippet & -- & 80.5 & .07 & .13 & .05 & .27 & .12 & .92 & .31 & .41 & .24 & .17 & .13 & .13 & .20 & .24 & .14 & .15 & .26 & .23 \\
Sentinel & $\tau$ 0.1 & 77.2 & .75 & .54 & .22 & .47 & .35 & .83 & .39 & .52 & .61 & .38 & .38 & .36 & .32 & .44 & .42 & .39 & .46 & .46 \\
\bottomrule
\end{tabular}}
\caption{Integration Test per-dataset judge accuracy at matched $\approx$93.6\% compression. Each hard compressor is shown at the threshold whose mean compression over the 17 datasets is closest to the 93.6\% rate of the soft compressors. Bold marks the best among matched methods per column (ties included); methods below the separator run at lower compression and are not compared.}
\label{tab:matched_soft}
\end{table*}

\subsubsection{Retrieval--Compression Alignment}\label{sec:appendix_alignment_result}

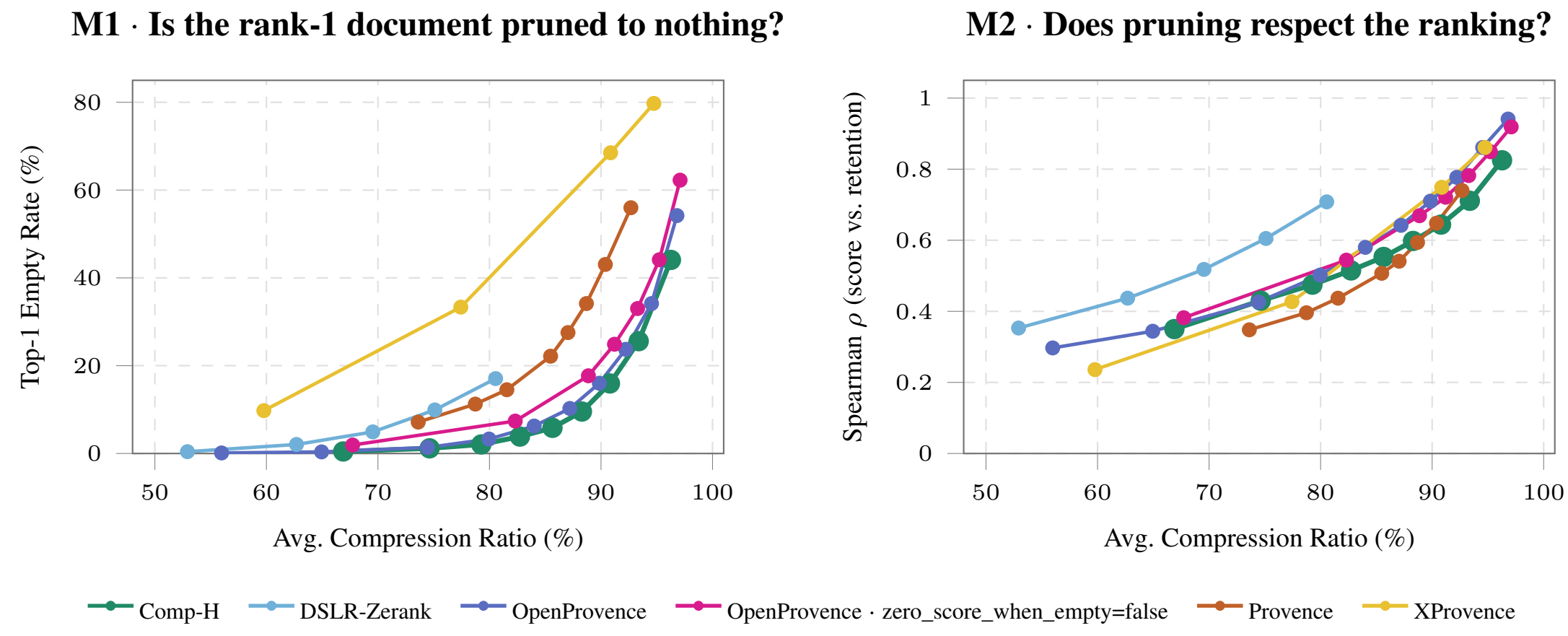
\begin{figure*}[t]
\centering
\definecolor{cOursExt}{HTML}{1F8A65}
\definecolor{cZerank}{HTML}{70B0D8}
\definecolor{cOpenProv}{HTML}{5A6CC0}
\definecolor{cOpenProvZero}{HTML}{D81B8C}
\definecolor{cProv}{HTML}{C06028}
\definecolor{cXProv}{HTML}{E8C030}
\pgfplotsset{
  ours/.style={line width=1.3pt, mark=*, mark size=2pt},
  base/.style={line width=0.9pt, mark=*, mark size=1.5pt},
  disc axis/.style={
    width=7cm, height=5cm, xmin=48, xmax=101,
    xtick={50,60,70,80,90,100},
    xlabel={Avg.\ Compression Ratio (\%)},
    grid=both, major grid style={dashed, black!12},
    tick align=outside, tick pos=left, axis line style={black!55},
    label style={font=\scriptsize}, tick label style={font=\tiny},
    title style={font=\small\bfseries, align=center},
  },
}
\newcommand{\lgd}[2]{%
  \mbox{\raisebox{2pt}{\tikz{\draw[#1, line width=1pt] (0,0) -- (0.42,0);
    \fill[#1] (0.21,0) circle (1.3pt);}}\,#2}\hspace{7pt}}
\begin{tikzpicture}
\begin{groupplot}[
    group style={group size=2 by 1, horizontal sep=2.2cm},
    disc axis,
  ]
  \nextgroupplot[ymin=0, ymax=85, ytick={0,20,40,60,80},
    ylabel={Top-1 Empty Rate (\%)},
    title={M1 $\cdot$ Is the rank-1 document pruned to nothing?}]
  \addplot[cOursExt, ours] coordinates {
    (66.89,0.46)(74.63,1.17)(79.27,2.06)(82.73,3.85)(85.65,5.84)
    (88.29,9.55)(90.8,15.95)(93.38,25.58)(96.27,44.09)};
  \addplot[cZerank, base] coordinates {
    (52.92,0.41)(62.7,2.05)(69.54,4.92)(75.1,9.92)(80.55,17.06)};
  \addplot[cOpenProv, base] coordinates {
    (55.98,0.12)(64.95,0.36)(74.44,1.37)(79.96,3.29)(84.01,6.25)
    (87.21,10.26)(89.85,16)(92.21,23.72)(94.53,34.15)(96.81,54.19)};
  \addplot[cOpenProvZero, base] coordinates {
    (67.73,1.93)(82.31,7.38)(88.87,17.7)(91.21,24.88)(93.28,33.02)
    (95.23,44.13)(97.09,62.26)};
  \addplot[cProv, base] coordinates {
    (73.61,7.15)(78.73,11.25)(81.56,14.52)(85.48,22.16)
    (87.04,27.54)(88.69,34.15)(90.39,43.06)(92.68,55.98)};
  \addplot[cXProv, base] coordinates {
    (59.77,9.76)(77.43,33.34)(90.86,68.5)(94.74,79.74)};
  \nextgroupplot[ymin=0, ymax=1.05, ytick={0,0.2,0.4,0.6,0.8,1.0},
    ylabel={Spearman $\rho$ (score vs.\ retention)},
    title={M2 $\cdot$ Does pruning respect the ranking?}]
  \addplot[cOursExt, ours] coordinates {
    (66.89,0.35)(74.63,0.43)(79.27,0.475)(82.73,0.515)(85.65,0.553)
    (88.29,0.598)(90.8,0.644)(93.38,0.711)(96.27,0.825)};
  \addplot[cZerank, base] coordinates {
    (52.92,0.353)(62.7,0.437)(69.54,0.518)(75.1,0.605)(80.55,0.708)};
  \addplot[cOpenProv, base] coordinates {
    (55.98,0.297)(64.95,0.344)(74.44,0.426)(79.96,0.502)(84.01,0.58)
    (87.21,0.642)(89.85,0.71)(92.21,0.777)(94.53,0.861)(96.81,0.941)};
  \addplot[cOpenProvZero, base] coordinates {
    (67.73,0.382)(82.31,0.544)(88.87,0.669)(91.21,0.721)(93.28,0.782)
    (95.23,0.849)(97.09,0.919)};
  \addplot[cProv, base] coordinates {
    (73.61,0.348)(78.73,0.396)(81.56,0.437)(85.48,0.507)
    (87.04,0.541)(88.69,0.594)(90.39,0.648)(92.68,0.74)};
  \addplot[cXProv, base] coordinates {
    (59.77,0.236)(77.43,0.427)(90.86,0.749)(94.74,0.861)};
\end{groupplot}
\end{tikzpicture}
\vspace{-1pt}
\sbox0{\tiny
  \lgd{cOursExt}{Comp-H}%
  \lgd{cZerank}{DSLR-Zerank}%
  \lgd{cOpenProv}{OpenProvence}%
  \lgd{cOpenProvZero}{OpenProvence $\cdot$ zero\_score\_when\_empty=false}%
  \lgd{cProv}{Provence}%
  \lgd{cXProv}{XProvence}\unskip}
\noindent\makebox[\textwidth][c]{%
  \ifdim\wd0>\textwidth\resizebox{\textwidth}{!}{\usebox0}\else\usebox0\fi}
\caption{\textbf{Integration Test $\cdot$ M1 and M2 against average compression ratio.} \textbf{M1}: Top-1 empty rate. \textbf{M2}: mean per-query Spearman rank correlation (over the reranked top-$K$
documents) between the reranker's ranking and the per-document token retention
rate (kept$\div$original tokens), which is unaffected by document length. Comp-H keeps the rank-1 document intact at
compression ratios where other baselines already discard it.}
\label{fig:integration_M1_M2_discrepancy}
\end{figure*}
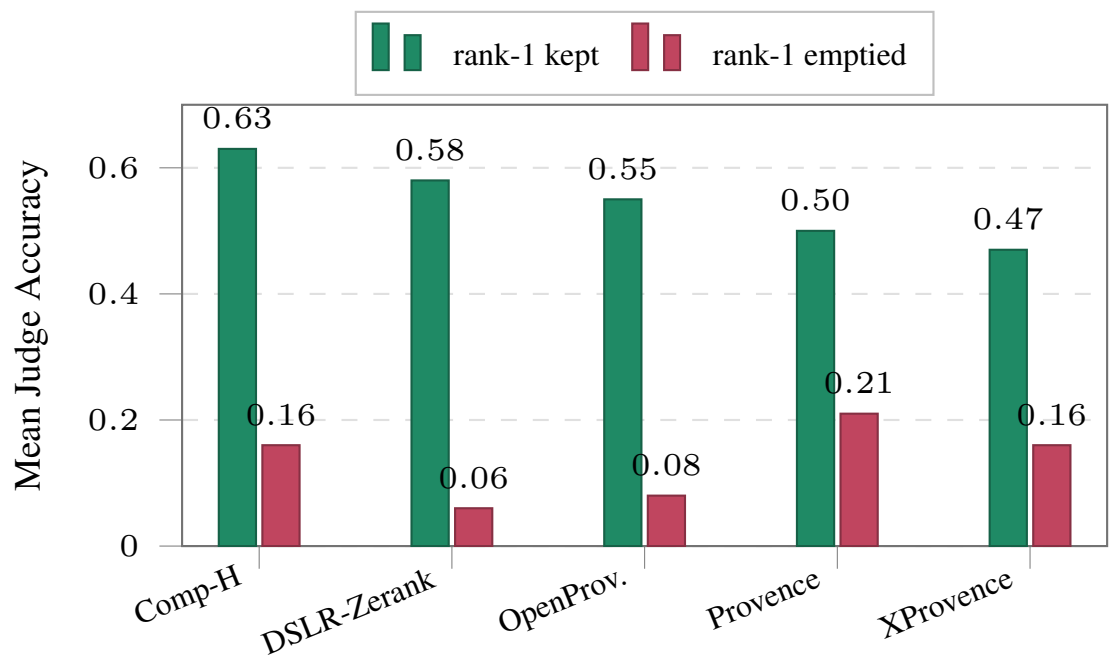
\begin{figure}[t]
\centering
\definecolor{cKept}{HTML}{1F8A65}
\definecolor{cEmpty}{HTML}{C0455F}
\begin{tikzpicture}
\begin{axis}[
    width=\columnwidth, height=4.5cm,
    ybar=1.2pt, bar width=7pt,
    ymin=0, ymax=0.7, ytick={0,0.2,0.4,0.6},
    ylabel={Mean Judge Accuracy},
    symbolic x coords={Ours, zerank, OpenProv, Provence, xProv},
    xtick=data,
    xticklabels={Comp-H, DSLR-Zerank, OpenProv., Provence, XProvence},
    x tick label style={font=\tiny, rotate=22, anchor=east},
    tick label style={font=\tiny}, label style={font=\scriptsize},
    ymajorgrids, major grid style={dashed, black!12},
    axis line style={black!55}, tick align=outside, tick pos=left,
    nodes near coords, /pgf/number format/precision=2,
    every node near coord/.append style={
      font=\tiny, /pgf/number format/fixed, /pgf/number format/fixed zerofill},
    legend style={at={(0.5,1.02)}, anchor=south, legend columns=2,
      font=\tiny, draw=black!25, column sep=4pt},
  ]
  \addplot[fill=cKept, draw=cKept!70!black] coordinates {
    (Ours,0.63)(zerank,0.58)(OpenProv,0.55)(Provence,0.5)(xProv,0.47)};
  \addlegendentry{rank-1 kept}
  \addplot[fill=cEmpty, draw=cEmpty!70!black] coordinates {
    (Ours,0.16)(zerank,0.06)(OpenProv,0.08)(Provence,0.21)(xProv,0.16)};
  \addlegendentry{rank-1 emptied}
\end{axis}
\end{tikzpicture}
\vspace{-1em}
\caption{\textbf{Integration Test $\cdot$ M3 $\cdot$ downstream impact.} Mean judge accuracy on queries
where the rank-1 document is \emph{kept} vs.\ \emph{emptied}, at threshold $0.3$.}
\label{fig:appendix_m3_result}
\end{figure}

As defined in Section~\ref{sec:appendix_alignment}, a faithful compressor should compress in agreement with its own reranker. Figure~\ref{fig:integration_M1_M2_discrepancy} probes this with two of the diagnostics, M1 and M2. Our focus here is the internal inconsistency between reranking and compression, so we exclude models that do not rerank internally.

\paragraph{M1 (top-1 empty rate)} Comp-H almost never destroys its
own rank-1 document: it stays $0.46\%$ at ${\sim}67\%$ compression and
${\sim}2\%$ at ${\sim}79\%$, becoming non-trivial only under extreme
(${>}90\%$) compression. Every other reranker baseline empties the rank-1 document far
more often at comparable compression (e.g.\ Provence ($7.2\%$), DSLR-Zerank ($9.9\%$), and XProvence ($33\%$)). OpenProvence appears
deceptively low ($1.4\%$ at $74\%$) because its default
\texttt{zero\_score\_when\_empty=True} sets the reranking score of a
fully-compressed document to zero, demoting it out of the top rank so it is no
longer counted as an `empty rank-1'. To isolate this effect we also ran the
ablation with the flag disabled (\emph{zero=false}). Here the top-1 empty rate rises---at thresholds $0.1/0.3/0.5$ it grows from $0.4/3.3/10.3\%$
(default) to $1.9/7.4/17.7\%$.

\paragraph{M2 (rank alignment)} M2 checks whether the compressor keeps more text in the
paragraph its own reranker scores higher. However, higher M2 is not
by itself better. At $\approx$75\% compression, DSLR-Zerank's alignment score reaches 0.61 against Comp-H
0.43, but empties the rank-1 document in 9.9\% of queries versus our 1.2\%, and
scores 0.51 judge accuracy against Comp-H 0.63. M2 diagnoses whether compression follows the ranking, not whether the surviving context is useful, and must be read together with M1.

\paragraph{M3 (downstream impact)} M1 and M2 characterize the discrepancy, but do
not show whether it hurts. M3 tests this directly by splitting queries by whether
the rank-1 document was emptied and comparing mean judge accuracy. Figure \ref{fig:appendix_m3_result} shows the M3 scores at threshold $0.3$.
Across every model, emptying the rank-1 document sharply degrades answers. Comp-H drops from $0.63$ (rank-1 kept) to $0.16$ (rank-1 emptied), and the
baselines collapse just as steeply: DSLR-Zerank $0.58{\to}0.06$, OpenProvence
$0.55{\to}0.08$, Provence $0.50{\to}0.21$, and
XProvence $0.47{\to}0.16$. Whenever a compressor prunes away the document it ranked first, answer quality drops sharply. This confirms that ranking–compression misalignment is a real failure.

\subsubsection{Performance by Query Type} \label{sec:appendix_integration_qtype}

\definecolor{cellgreen}{HTML}{1F8A65}
\begin{table*}[ht]
\centering
\scriptsize
\setlength{\tabcolsep}{4pt}
\renewcommand{\arraystretch}{1.05}
\begin{minipage}[t]{0.46\textwidth}
\begin{tabular}{lccc}
\multicolumn{4}{l}{\scriptsize\textbf{Academic Papers --- EN synthetic factoid \textperiodcentered 372 q}}\\[2pt]
& \makecell{Num.\\\&Metrics (88)} & \makecell{Entity\\\&Nom. (144)} & \makecell{Multi-Hop (140)}\\[-1pt]
\midrule
\textit{Gold} & \cellcolor{cellgreen!90}\textcolor{white}{\textit{.95}} & \cellcolor{cellgreen!90}\textcolor{white}{\textit{.98}} & \cellcolor{cellgreen!90}\textcolor{white}{\textit{.92}}\\
\textbf{Comp-H} & \cellcolor{cellgreen!89}\textcolor{white}{\textbf{.89}} & \cellcolor{cellgreen!90}\textcolor{white}{\textbf{.97}} & \cellcolor{cellgreen!87}\textcolor{white}{\textbf{.87}}\\
\textbf{Comp-S} & \cellcolor{cellgreen!12}\textcolor{black}{.15} & \cellcolor{cellgreen!15}\textcolor{black}{.19} & \cellcolor{cellgreen!11}\textcolor{black}{.15}\\
Oscar & \cellcolor{cellgreen!15}\textcolor{black}{.18} & \cellcolor{cellgreen!32}\textcolor{black}{.35} & \cellcolor{cellgreen!9}\textcolor{black}{.13}\\
Zerank-2 & \cellcolor{cellgreen!75}\textcolor{white}{.76} & \cellcolor{cellgreen!90}\textcolor{white}{.90} & \cellcolor{cellgreen!48}\textcolor{black}{.50}\\
OpenProv. & \cellcolor{cellgreen!72}\textcolor{white}{.73} & \cellcolor{cellgreen!90}\textcolor{white}{.93} & \cellcolor{cellgreen!80}\textcolor{white}{.81}\\
Provence & \cellcolor{cellgreen!71}\textcolor{white}{.72} & \cellcolor{cellgreen!89}\textcolor{white}{.90} & \cellcolor{cellgreen!68}\textcolor{white}{.69}\\
xProvence & \cellcolor{cellgreen!7}\textcolor{black}{.11} & \cellcolor{cellgreen!6}\textcolor{black}{.07} & \cellcolor{cellgreen!6}\textcolor{black}{.08}\\
SemHL+BGE & \cellcolor{cellgreen!67}\textcolor{white}{.68} & \cellcolor{cellgreen!90}\textcolor{white}{.92} & \cellcolor{cellgreen!73}\textcolor{white}{.74}\\
LL2+BGE & \cellcolor{cellgreen!19}\textcolor{black}{.22} & \cellcolor{cellgreen!54}\textcolor{white}{.55} & \cellcolor{cellgreen!27}\textcolor{black}{.30}\\
\end{tabular}
\end{minipage}%
\hfill
\begin{minipage}[t]{0.46\textwidth}
\begin{tabular}{lcc}
\multicolumn{3}{l}{\scriptsize\textbf{HotpotQA --- EN multi-hop \textperiodcentered 7{,}405 q}}\\[2pt]
& \makecell{Bridge\,2-hop (5918)} & \makecell{Comparison (1487)}\\[-1pt]
\midrule
\textit{Gold} & \cellcolor{cellgreen!87}\textcolor{white}{\textit{.88}} & \cellcolor{cellgreen!84}\textcolor{white}{\textit{.85}}\\
\textbf{Comp-H} & \cellcolor{cellgreen!43}\textcolor{black}{.45} & \cellcolor{cellgreen!50}\textcolor{black}{.52}\\
\textbf{Comp-S} & \cellcolor{cellgreen!35}\textcolor{black}{.38} & \cellcolor{cellgreen!58}\textcolor{white}{.60}\\
Oscar & \cellcolor{cellgreen!67}\textcolor{white}{\textbf{.68}} & \cellcolor{cellgreen!72}\textcolor{white}{\textbf{.73}}\\
Zerank-2 & \cellcolor{cellgreen!23}\textcolor{black}{.26} & \cellcolor{cellgreen!10}\textcolor{black}{.14}\\
OpenProv. & \cellcolor{cellgreen!40}\textcolor{black}{.42} & \cellcolor{cellgreen!47}\textcolor{black}{.49}\\
Provence & \cellcolor{cellgreen!33}\textcolor{black}{.36} & \cellcolor{cellgreen!32}\textcolor{black}{.34}\\
xProvence & \cellcolor{cellgreen!38}\textcolor{black}{.40} & \cellcolor{cellgreen!21}\textcolor{black}{.25}\\
SemHL+BGE & \cellcolor{cellgreen!45}\textcolor{black}{.47} & \cellcolor{cellgreen!20}\textcolor{black}{.23}\\
LL2+BGE & \cellcolor{cellgreen!10}\textcolor{black}{.14} & \cellcolor{cellgreen!15}\textcolor{black}{.18}\\
\end{tabular}
\end{minipage}
\vspace{4pt}
\begin{center}\begin{tabular}{lcccccccc}
\multicolumn{9}{l}{\scriptsize\textbf{Enterprise RAG Bench --- EN internal docs \textperiodcentered 470 q}}\\[2pt]
& \makecell{Basic (175)} & \makecell{Semantic (125)} & \makecell{Intra-doc\,Reason. (40)} & \makecell{Project-\,Rel. (40)} & \makecell{Constrained (30)} & \makecell{Conflicting (20)} & \makecell{Compl. (20)} & \makecell{Misc (20)}\\[-1pt]
\midrule
\textit{Gold} & \cellcolor{cellgreen!90}\textcolor{white}{\textit{.95}} & \cellcolor{cellgreen!89}\textcolor{white}{\textit{.89}} & \cellcolor{cellgreen!90}\textcolor{white}{\textit{.98}} & \cellcolor{cellgreen!18}\textcolor{black}{\textit{.22}} & \cellcolor{cellgreen!57}\textcolor{white}{\textit{.59}} & \cellcolor{cellgreen!90}\textcolor{white}{\textit{.99}} & \cellcolor{cellgreen!29}\textcolor{black}{\textit{.32}} & \cellcolor{cellgreen!90}\textcolor{white}{\textit{.96}}\\
\textbf{Comp-H} & \cellcolor{cellgreen!68}\textcolor{white}{\textbf{.69}} & \cellcolor{cellgreen!58}\textcolor{white}{\textbf{.59}} & \cellcolor{cellgreen!70}\textcolor{white}{\textbf{.71}} & \cellcolor{cellgreen!11}\textcolor{black}{.15} & \cellcolor{cellgreen!48}\textcolor{black}{\textbf{.50}} & \cellcolor{cellgreen!65}\textcolor{white}{.67} & \cellcolor{cellgreen!20}\textcolor{black}{.23} & \cellcolor{cellgreen!77}\textcolor{white}{\textbf{.78}}\\
\textbf{Comp-S} & \cellcolor{cellgreen!51}\textcolor{black}{.53} & \cellcolor{cellgreen!18}\textcolor{black}{.21} & \cellcolor{cellgreen!14}\textcolor{black}{.17} & \cellcolor{cellgreen!6}\textcolor{black}{.07} & \cellcolor{cellgreen!8}\textcolor{black}{.12} & \cellcolor{cellgreen!33}\textcolor{black}{.36} & \cellcolor{cellgreen!10}\textcolor{black}{.14} & \cellcolor{cellgreen!58}\textcolor{white}{.59}\\
Oscar & \cellcolor{cellgreen!11}\textcolor{black}{.15} & \cellcolor{cellgreen!6}\textcolor{black}{.07} & \cellcolor{cellgreen!6}\textcolor{black}{.10} & \cellcolor{cellgreen!6}\textcolor{black}{.01} & \cellcolor{cellgreen!6}\textcolor{black}{.04} & \cellcolor{cellgreen!21}\textcolor{black}{.24} & \cellcolor{cellgreen!6}\textcolor{black}{.10} & \cellcolor{cellgreen!26}\textcolor{black}{.29}\\
Zerank-2 & \cellcolor{cellgreen!52}\textcolor{black}{.54} & \cellcolor{cellgreen!43}\textcolor{black}{.45} & \cellcolor{cellgreen!19}\textcolor{black}{.22} & \cellcolor{cellgreen!6}\textcolor{black}{.09} & \cellcolor{cellgreen!22}\textcolor{black}{.25} & \cellcolor{cellgreen!52}\textcolor{white}{.54} & \cellcolor{cellgreen!13}\textcolor{black}{.17} & \cellcolor{cellgreen!48}\textcolor{black}{.50}\\
OpenProv. & \cellcolor{cellgreen!61}\textcolor{white}{.62} & \cellcolor{cellgreen!45}\textcolor{black}{.48} & \cellcolor{cellgreen!59}\textcolor{white}{.60} & \cellcolor{cellgreen!9}\textcolor{black}{.13} & \cellcolor{cellgreen!37}\textcolor{black}{.40} & \cellcolor{cellgreen!70}\textcolor{white}{.71} & \cellcolor{cellgreen!6}\textcolor{black}{.10} & \cellcolor{cellgreen!50}\textcolor{black}{.52}\\
Provence & \cellcolor{cellgreen!59}\textcolor{white}{.60} & \cellcolor{cellgreen!36}\textcolor{black}{.39} & \cellcolor{cellgreen!59}\textcolor{white}{.60} & \cellcolor{cellgreen!11}\textcolor{black}{.15} & \cellcolor{cellgreen!24}\textcolor{black}{.27} & \cellcolor{cellgreen!56}\textcolor{white}{.57} & \cellcolor{cellgreen!18}\textcolor{black}{.22} & \cellcolor{cellgreen!62}\textcolor{white}{.64}\\
xProvence & \cellcolor{cellgreen!22}\textcolor{black}{.25} & \cellcolor{cellgreen!10}\textcolor{black}{.14} & \cellcolor{cellgreen!8}\textcolor{black}{.12} & \cellcolor{cellgreen!6}\textcolor{black}{.09} & \cellcolor{cellgreen!7}\textcolor{black}{.11} & \cellcolor{cellgreen!28}\textcolor{black}{.31} & \cellcolor{cellgreen!8}\textcolor{black}{.12} & \cellcolor{cellgreen!33}\textcolor{black}{.36}\\
SemHL+BGE & \cellcolor{cellgreen!65}\textcolor{white}{.67} & \cellcolor{cellgreen!47}\textcolor{black}{.49} & \cellcolor{cellgreen!70}\textcolor{white}{.71} & \cellcolor{cellgreen!13}\textcolor{black}{\textbf{.16}} & \cellcolor{cellgreen!35}\textcolor{black}{.38} & \cellcolor{cellgreen!79}\textcolor{white}{\textbf{.80}} & \cellcolor{cellgreen!20}\textcolor{black}{\textbf{.24}} & \cellcolor{cellgreen!64}\textcolor{white}{.66}\\
LL2+BGE & \cellcolor{cellgreen!17}\textcolor{black}{.20} & \cellcolor{cellgreen!14}\textcolor{black}{.18} & \cellcolor{cellgreen!7}\textcolor{black}{.11} & \cellcolor{cellgreen!6}\textcolor{black}{.05} & \cellcolor{cellgreen!9}\textcolor{black}{.13} & \cellcolor{cellgreen!24}\textcolor{black}{.28} & \cellcolor{cellgreen!6}\textcolor{black}{.05} & \cellcolor{cellgreen!23}\textcolor{black}{.27}\\
\end{tabular}\end{center}
\vspace{2pt}
\begin{center}\begin{tabular}{lcccccc}
\multicolumn{7}{l}{\scriptsize\textbf{ViDoRe \textperiodcentered Computer Science --- EN \textperiodcentered 215 q}}\\[2pt]
& \makecell{Extr. (74)} & \makecell{Open-\,ended (67)} & \makecell{Compare-\,contrast (59)} & \makecell{Enum. (42)} & \makecell{Multi-\,hop (23)} & \makecell{Bool. (20)}\\[-1pt]
\midrule
\textit{Gold} & \cellcolor{cellgreen!72}\textcolor{white}{\textit{.73}} & \cellcolor{cellgreen!60}\textcolor{white}{\textit{.62}} & \cellcolor{cellgreen!70}\textcolor{white}{\textit{.71}} & \cellcolor{cellgreen!53}\textcolor{white}{\textit{.55}} & \cellcolor{cellgreen!61}\textcolor{white}{\textit{.63}} & \cellcolor{cellgreen!81}\textcolor{white}{\textit{.82}}\\
\textbf{Comp-H} & \cellcolor{cellgreen!71}\textcolor{white}{\textbf{.72}} & \cellcolor{cellgreen!60}\textcolor{white}{.62} & \cellcolor{cellgreen!60}\textcolor{white}{.61} & \cellcolor{cellgreen!51}\textcolor{black}{\textbf{.53}} & \cellcolor{cellgreen!50}\textcolor{black}{\textbf{.52}} & \cellcolor{cellgreen!79}\textcolor{white}{.80}\\
\textbf{Comp-S} & \cellcolor{cellgreen!55}\textcolor{white}{.57} & \cellcolor{cellgreen!55}\textcolor{white}{.57} & \cellcolor{cellgreen!68}\textcolor{white}{\textbf{.69}} & \cellcolor{cellgreen!31}\textcolor{black}{.34} & \cellcolor{cellgreen!47}\textcolor{black}{.49} & \cellcolor{cellgreen!90}\textcolor{white}{\textbf{.94}}\\
Oscar & \cellcolor{cellgreen!61}\textcolor{white}{.62} & \cellcolor{cellgreen!49}\textcolor{black}{.51} & \cellcolor{cellgreen!40}\textcolor{black}{.42} & \cellcolor{cellgreen!33}\textcolor{black}{.36} & \cellcolor{cellgreen!48}\textcolor{black}{.50} & \cellcolor{cellgreen!79}\textcolor{white}{.80}\\
Zerank-2 & \cellcolor{cellgreen!54}\textcolor{white}{.56} & \cellcolor{cellgreen!48}\textcolor{black}{.50} & \cellcolor{cellgreen!42}\textcolor{black}{.45} & \cellcolor{cellgreen!30}\textcolor{black}{.33} & \cellcolor{cellgreen!23}\textcolor{black}{.26} & \cellcolor{cellgreen!63}\textcolor{white}{.64}\\
OpenProv. & \cellcolor{cellgreen!62}\textcolor{white}{.63} & \cellcolor{cellgreen!61}\textcolor{white}{\textbf{.62}} & \cellcolor{cellgreen!60}\textcolor{white}{.62} & \cellcolor{cellgreen!43}\textcolor{black}{.46} & \cellcolor{cellgreen!50}\textcolor{black}{.52} & \cellcolor{cellgreen!72}\textcolor{white}{.73}\\
Provence & \cellcolor{cellgreen!64}\textcolor{white}{.65} & \cellcolor{cellgreen!59}\textcolor{white}{.61} & \cellcolor{cellgreen!55}\textcolor{white}{.56} & \cellcolor{cellgreen!44}\textcolor{black}{.46} & \cellcolor{cellgreen!45}\textcolor{black}{.47} & \cellcolor{cellgreen!71}\textcolor{white}{.72}\\
xProvence & \cellcolor{cellgreen!53}\textcolor{white}{.54} & \cellcolor{cellgreen!52}\textcolor{black}{.54} & \cellcolor{cellgreen!57}\textcolor{white}{.59} & \cellcolor{cellgreen!38}\textcolor{black}{.40} & \cellcolor{cellgreen!26}\textcolor{black}{.29} & \cellcolor{cellgreen!48}\textcolor{black}{.50}\\
SemHL+BGE & \cellcolor{cellgreen!66}\textcolor{white}{.67} & \cellcolor{cellgreen!56}\textcolor{white}{.57} & \cellcolor{cellgreen!59}\textcolor{white}{.60} & \cellcolor{cellgreen!41}\textcolor{black}{.43} & \cellcolor{cellgreen!34}\textcolor{black}{.37} & \cellcolor{cellgreen!72}\textcolor{white}{.73}\\
LL2+BGE & \cellcolor{cellgreen!37}\textcolor{black}{.40} & \cellcolor{cellgreen!33}\textcolor{black}{.36} & \cellcolor{cellgreen!35}\textcolor{black}{.37} & \cellcolor{cellgreen!15}\textcolor{black}{.19} & \cellcolor{cellgreen!21}\textcolor{black}{.24} & \cellcolor{cellgreen!75}\textcolor{white}{.76}\\
\end{tabular}\end{center}
\vspace{2pt}
\begin{center}\begin{tabular}{lccccccc}
\multicolumn{8}{l}{\scriptsize\textbf{ViDoRe \textperiodcentered Energy --- FR \textperiodcentered 308 q}}\\[2pt]
& \makecell{Extr. (112)} & \makecell{Num. (99)} & \makecell{Open-\,ended (74)} & \makecell{Enum. (57)} & \makecell{Compare-\,contrast (51)} & \makecell{Bool. (40)} & \makecell{Multi-\,hop (27)}\\[-1pt]
\midrule
\textit{Gold} & \cellcolor{cellgreen!70}\textcolor{white}{\textit{.71}} & \cellcolor{cellgreen!67}\textcolor{white}{\textit{.68}} & \cellcolor{cellgreen!51}\textcolor{black}{\textit{.53}} & \cellcolor{cellgreen!57}\textcolor{white}{\textit{.58}} & \cellcolor{cellgreen!49}\textcolor{black}{\textit{.51}} & \cellcolor{cellgreen!80}\textcolor{white}{\textit{.80}} & \cellcolor{cellgreen!49}\textcolor{black}{\textit{.51}}\\
\textbf{Comp-H} & \cellcolor{cellgreen!65}\textcolor{white}{\textbf{.66}} & \cellcolor{cellgreen!66}\textcolor{white}{\textbf{.67}} & \cellcolor{cellgreen!44}\textcolor{black}{\textbf{.46}} & \cellcolor{cellgreen!49}\textcolor{black}{.51} & \cellcolor{cellgreen!54}\textcolor{white}{\textbf{.56}} & \cellcolor{cellgreen!78}\textcolor{white}{\textbf{.79}} & \cellcolor{cellgreen!53}\textcolor{white}{\textbf{.55}}\\
\textbf{Comp-S} & \cellcolor{cellgreen!44}\textcolor{black}{.46} & \cellcolor{cellgreen!43}\textcolor{black}{.46} & \cellcolor{cellgreen!25}\textcolor{black}{.28} & \cellcolor{cellgreen!19}\textcolor{black}{.23} & \cellcolor{cellgreen!31}\textcolor{black}{.34} & \cellcolor{cellgreen!76}\textcolor{white}{.76} & \cellcolor{cellgreen!38}\textcolor{black}{.41}\\
Oscar & \cellcolor{cellgreen!20}\textcolor{black}{.23} & \cellcolor{cellgreen!22}\textcolor{black}{.25} & \cellcolor{cellgreen!14}\textcolor{black}{.17} & \cellcolor{cellgreen!6}\textcolor{black}{.09} & \cellcolor{cellgreen!7}\textcolor{black}{.11} & \cellcolor{cellgreen!52}\textcolor{black}{.54} & \cellcolor{cellgreen!6}\textcolor{black}{.07}\\
Zerank-2 & \cellcolor{cellgreen!60}\textcolor{white}{.61} & \cellcolor{cellgreen!60}\textcolor{white}{.61} & \cellcolor{cellgreen!41}\textcolor{black}{.43} & \cellcolor{cellgreen!38}\textcolor{black}{.41} & \cellcolor{cellgreen!43}\textcolor{black}{.45} & \cellcolor{cellgreen!68}\textcolor{white}{.69} & \cellcolor{cellgreen!36}\textcolor{black}{.38}\\
OpenProv. & \cellcolor{cellgreen!58}\textcolor{white}{.59} & \cellcolor{cellgreen!50}\textcolor{black}{.52} & \cellcolor{cellgreen!37}\textcolor{black}{.40} & \cellcolor{cellgreen!40}\textcolor{black}{.43} & \cellcolor{cellgreen!43}\textcolor{black}{.45} & \cellcolor{cellgreen!68}\textcolor{white}{.69} & \cellcolor{cellgreen!34}\textcolor{black}{.37}\\
Provence & \cellcolor{cellgreen!48}\textcolor{black}{.50} & \cellcolor{cellgreen!39}\textcolor{black}{.41} & \cellcolor{cellgreen!34}\textcolor{black}{.37} & \cellcolor{cellgreen!42}\textcolor{black}{.44} & \cellcolor{cellgreen!41}\textcolor{black}{.43} & \cellcolor{cellgreen!65}\textcolor{white}{.66} & \cellcolor{cellgreen!37}\textcolor{black}{.40}\\
xProvence & \cellcolor{cellgreen!43}\textcolor{black}{.46} & \cellcolor{cellgreen!39}\textcolor{black}{.42} & \cellcolor{cellgreen!33}\textcolor{black}{.36} & \cellcolor{cellgreen!32}\textcolor{black}{.34} & \cellcolor{cellgreen!38}\textcolor{black}{.40} & \cellcolor{cellgreen!55}\textcolor{white}{.57} & \cellcolor{cellgreen!35}\textcolor{black}{.37}\\
SemHL+BGE & \cellcolor{cellgreen!63}\textcolor{white}{.64} & \cellcolor{cellgreen!57}\textcolor{white}{.58} & \cellcolor{cellgreen!41}\textcolor{black}{.43} & \cellcolor{cellgreen!53}\textcolor{white}{\textbf{.55}} & \cellcolor{cellgreen!43}\textcolor{black}{.45} & \cellcolor{cellgreen!68}\textcolor{white}{.69} & \cellcolor{cellgreen!45}\textcolor{black}{.47}\\
LL2+BGE & \cellcolor{cellgreen!28}\textcolor{black}{.30} & \cellcolor{cellgreen!27}\textcolor{black}{.30} & \cellcolor{cellgreen!25}\textcolor{black}{.28} & \cellcolor{cellgreen!20}\textcolor{black}{.23} & \cellcolor{cellgreen!20}\textcolor{black}{.23} & \cellcolor{cellgreen!40}\textcolor{black}{.43} & \cellcolor{cellgreen!25}\textcolor{black}{.28}\\
\end{tabular}\end{center}
\caption{\textbf{Integration Test} per-query-type judge accuracy (1/2). Each sub-table is one dataset; columns are query subtypes (counts in parentheses). Models are shown at its matched-compression config (\textasciitilde75\%; soft compressors at fixed 94\%). \textbf{Bold} = best non-gold model in that column; \textit{italics} = gold-document oracle.}
\label{tab:integration-qtype}
\vspace{2pt}
\end{table*} 
\definecolor{cellgreen}{HTML}{1F8A65}
\begin{table*}[ht]
\centering
\scriptsize
\setlength{\tabcolsep}{4pt}
\renewcommand{\arraystretch}{1.05}

\begin{center}\begin{tabular}{lccccccc}
\multicolumn{8}{l}{\scriptsize\textbf{ViDoRe \textperiodcentered Finance-EN --- EN \textperiodcentered 309 q}}\\[2pt]
& \makecell{Extr. (136)} & \makecell{Num. (101)} & \makecell{Open-\,ended (54)} & \makecell{Compare-\,contrast (52)} & \makecell{Bool. (48)} & \makecell{Enum. (33)} & \makecell{Multi-\,hop (24)}\\[-1pt]
\midrule
\textit{Gold} & \cellcolor{cellgreen!68}\textcolor{white}{\textit{.69}} & \cellcolor{cellgreen!67}\textcolor{white}{\textit{.68}} & \cellcolor{cellgreen!54}\textcolor{white}{\textit{.56}} & \cellcolor{cellgreen!39}\textcolor{black}{\textit{.42}} & \cellcolor{cellgreen!83}\textcolor{white}{\textit{.83}} & \cellcolor{cellgreen!59}\textcolor{white}{\textit{.61}} & \cellcolor{cellgreen!31}\textcolor{black}{\textit{.34}}\\
\textbf{Comp-H} & \cellcolor{cellgreen!59}\textcolor{white}{\textbf{.61}} & \cellcolor{cellgreen!52}\textcolor{black}{\textbf{.54}} & \cellcolor{cellgreen!49}\textcolor{black}{\textbf{.51}} & \cellcolor{cellgreen!31}\textcolor{black}{\textbf{.34}} & \cellcolor{cellgreen!64}\textcolor{white}{\textbf{.65}} & \cellcolor{cellgreen!54}\textcolor{white}{\textbf{.56}} & \cellcolor{cellgreen!27}\textcolor{black}{\textbf{.30}}\\
\textbf{Comp-S} & \cellcolor{cellgreen!43}\textcolor{black}{.46} & \cellcolor{cellgreen!31}\textcolor{black}{.34} & \cellcolor{cellgreen!28}\textcolor{black}{.31} & \cellcolor{cellgreen!15}\textcolor{black}{.18} & \cellcolor{cellgreen!55}\textcolor{white}{.56} & \cellcolor{cellgreen!29}\textcolor{black}{.32} & \cellcolor{cellgreen!11}\textcolor{black}{.14}\\
Oscar & \cellcolor{cellgreen!22}\textcolor{black}{.25} & \cellcolor{cellgreen!14}\textcolor{black}{.17} & \cellcolor{cellgreen!25}\textcolor{black}{.28} & \cellcolor{cellgreen!10}\textcolor{black}{.14} & \cellcolor{cellgreen!38}\textcolor{black}{.41} & \cellcolor{cellgreen!16}\textcolor{black}{.19} & \cellcolor{cellgreen!14}\textcolor{black}{.18}\\
Zerank-2 & \cellcolor{cellgreen!47}\textcolor{black}{.49} & \cellcolor{cellgreen!42}\textcolor{black}{.45} & \cellcolor{cellgreen!37}\textcolor{black}{.40} & \cellcolor{cellgreen!23}\textcolor{black}{.26} & \cellcolor{cellgreen!58}\textcolor{white}{.60} & \cellcolor{cellgreen!40}\textcolor{black}{.43} & \cellcolor{cellgreen!11}\textcolor{black}{.15}\\
OpenProv. & \cellcolor{cellgreen!46}\textcolor{black}{.48} & \cellcolor{cellgreen!35}\textcolor{black}{.38} & \cellcolor{cellgreen!43}\textcolor{black}{.45} & \cellcolor{cellgreen!21}\textcolor{black}{.24} & \cellcolor{cellgreen!61}\textcolor{white}{.62} & \cellcolor{cellgreen!47}\textcolor{black}{.49} & \cellcolor{cellgreen!22}\textcolor{black}{.25}\\
Provence & \cellcolor{cellgreen!45}\textcolor{black}{.47} & \cellcolor{cellgreen!36}\textcolor{black}{.39} & \cellcolor{cellgreen!43}\textcolor{black}{.46} & \cellcolor{cellgreen!21}\textcolor{black}{.24} & \cellcolor{cellgreen!46}\textcolor{black}{.48} & \cellcolor{cellgreen!48}\textcolor{black}{.50} & \cellcolor{cellgreen!17}\textcolor{black}{.20}\\
xProvence & \cellcolor{cellgreen!42}\textcolor{black}{.45} & \cellcolor{cellgreen!29}\textcolor{black}{.32} & \cellcolor{cellgreen!34}\textcolor{black}{.37} & \cellcolor{cellgreen!24}\textcolor{black}{.27} & \cellcolor{cellgreen!45}\textcolor{black}{.47} & \cellcolor{cellgreen!43}\textcolor{black}{.45} & \cellcolor{cellgreen!14}\textcolor{black}{.18}\\
SemHL+BGE & \cellcolor{cellgreen!51}\textcolor{black}{.53} & \cellcolor{cellgreen!45}\textcolor{black}{.47} & \cellcolor{cellgreen!36}\textcolor{black}{.39} & \cellcolor{cellgreen!27}\textcolor{black}{.30} & \cellcolor{cellgreen!52}\textcolor{white}{.54} & \cellcolor{cellgreen!52}\textcolor{black}{.54} & \cellcolor{cellgreen!27}\textcolor{black}{.30}\\
LL2+BGE & \cellcolor{cellgreen!24}\textcolor{black}{.27} & \cellcolor{cellgreen!19}\textcolor{black}{.22} & \cellcolor{cellgreen!20}\textcolor{black}{.23} & \cellcolor{cellgreen!6}\textcolor{black}{.10} & \cellcolor{cellgreen!35}\textcolor{black}{.37} & \cellcolor{cellgreen!17}\textcolor{black}{.20} & \cellcolor{cellgreen!8}\textcolor{black}{.12}\\
\end{tabular}\end{center}
\vspace{1pt}
\begin{center}\begin{tabular}{lccccccc}
\multicolumn{8}{l}{\scriptsize\textbf{ViDoRe \textperiodcentered Finance-FR --- FR \textperiodcentered 320 q}}\\[2pt]
& \makecell{Extr. (151)} & \makecell{Num. (105)} & \makecell{Open-\,ended (90)} & \makecell{Compare-\,contrast (56)} & \makecell{Enum. (31)} & \makecell{Bool. (16)} & \makecell{Multi-\,hop (14)}\\[-1pt]
\midrule
\textit{Gold} & \cellcolor{cellgreen!71}\textcolor{white}{\textit{.72}} & \cellcolor{cellgreen!71}\textcolor{white}{\textit{.72}} & \cellcolor{cellgreen!37}\textcolor{black}{\textit{.40}} & \cellcolor{cellgreen!47}\textcolor{black}{\textit{.49}} & \cellcolor{cellgreen!45}\textcolor{black}{\textit{.47}} & \cellcolor{cellgreen!76}\textcolor{white}{\textit{.77}} & \cellcolor{cellgreen!27}\textcolor{black}{\textit{.30}}\\
\textbf{Comp-H} & \cellcolor{cellgreen!64}\textcolor{white}{\textbf{.66}} & \cellcolor{cellgreen!61}\textcolor{white}{\textbf{.63}} & \cellcolor{cellgreen!37}\textcolor{black}{\textbf{.39}} & \cellcolor{cellgreen!46}\textcolor{black}{\textbf{.48}} & \cellcolor{cellgreen!35}\textcolor{black}{.38} & \cellcolor{cellgreen!73}\textcolor{white}{\textbf{.73}} & \cellcolor{cellgreen!25}\textcolor{black}{.28}\\
\textbf{Comp-S} & \cellcolor{cellgreen!41}\textcolor{black}{.44} & \cellcolor{cellgreen!39}\textcolor{black}{.41} & \cellcolor{cellgreen!19}\textcolor{black}{.22} & \cellcolor{cellgreen!14}\textcolor{black}{.17} & \cellcolor{cellgreen!17}\textcolor{black}{.20} & \cellcolor{cellgreen!67}\textcolor{white}{.68} & \cellcolor{cellgreen!22}\textcolor{black}{.25}\\
Oscar & \cellcolor{cellgreen!19}\textcolor{black}{.23} & \cellcolor{cellgreen!14}\textcolor{black}{.17} & \cellcolor{cellgreen!6}\textcolor{black}{.09} & \cellcolor{cellgreen!6}\textcolor{black}{.05} & \cellcolor{cellgreen!6}\textcolor{black}{.09} & \cellcolor{cellgreen!30}\textcolor{black}{.33} & \cellcolor{cellgreen!6}\textcolor{black}{.09}\\
Zerank-2 & \cellcolor{cellgreen!52}\textcolor{black}{.54} & \cellcolor{cellgreen!50}\textcolor{black}{.52} & \cellcolor{cellgreen!31}\textcolor{black}{.33} & \cellcolor{cellgreen!29}\textcolor{black}{.32} & \cellcolor{cellgreen!23}\textcolor{black}{.27} & \cellcolor{cellgreen!68}\textcolor{white}{.69} & \cellcolor{cellgreen!17}\textcolor{black}{.21}\\
OpenProv. & \cellcolor{cellgreen!48}\textcolor{black}{.50} & \cellcolor{cellgreen!45}\textcolor{black}{.48} & \cellcolor{cellgreen!32}\textcolor{black}{.35} & \cellcolor{cellgreen!35}\textcolor{black}{.38} & \cellcolor{cellgreen!29}\textcolor{black}{.32} & \cellcolor{cellgreen!54}\textcolor{white}{.56} & \cellcolor{cellgreen!20}\textcolor{black}{.23}\\
Provence & \cellcolor{cellgreen!48}\textcolor{black}{.50} & \cellcolor{cellgreen!44}\textcolor{black}{.46} & \cellcolor{cellgreen!26}\textcolor{black}{.29} & \cellcolor{cellgreen!29}\textcolor{black}{.32} & \cellcolor{cellgreen!33}\textcolor{black}{.36} & \cellcolor{cellgreen!58}\textcolor{white}{.59} & \cellcolor{cellgreen!14}\textcolor{black}{.18}\\
xProvence & \cellcolor{cellgreen!41}\textcolor{black}{.44} & \cellcolor{cellgreen!29}\textcolor{black}{.32} & \cellcolor{cellgreen!26}\textcolor{black}{.29} & \cellcolor{cellgreen!18}\textcolor{black}{.22} & \cellcolor{cellgreen!32}\textcolor{black}{.35} & \cellcolor{cellgreen!48}\textcolor{black}{.50} & \cellcolor{cellgreen!16}\textcolor{black}{.19}\\
SemHL+BGE & \cellcolor{cellgreen!59}\textcolor{white}{.61} & \cellcolor{cellgreen!53}\textcolor{white}{.54} & \cellcolor{cellgreen!35}\textcolor{black}{.37} & \cellcolor{cellgreen!39}\textcolor{black}{.41} & \cellcolor{cellgreen!37}\textcolor{black}{\textbf{.39}} & \cellcolor{cellgreen!56}\textcolor{white}{.57} & \cellcolor{cellgreen!27}\textcolor{black}{\textbf{.30}}\\
LL2+BGE & \cellcolor{cellgreen!30}\textcolor{black}{.33} & \cellcolor{cellgreen!27}\textcolor{black}{.30} & \cellcolor{cellgreen!22}\textcolor{black}{.25} & \cellcolor{cellgreen!11}\textcolor{black}{.14} & \cellcolor{cellgreen!18}\textcolor{black}{.21} & \cellcolor{cellgreen!39}\textcolor{black}{.41} & \cellcolor{cellgreen!23}\textcolor{black}{.26}\\
\end{tabular}\end{center}
\vspace{1pt}
\begin{center}\begin{tabular}{lccccccc}
\multicolumn{8}{l}{\scriptsize\textbf{ViDoRe \textperiodcentered HR --- EN \textperiodcentered 318 q}}\\[2pt]
& \makecell{Open-\,ended (88)} & \makecell{Extr. (87)} & \makecell{Num. (84)} & \makecell{Compare-\,contrast (61)} & \makecell{Enum. (61)} & \makecell{Bool. (41)} & \makecell{Multi-\,hop (30)}\\[-1pt]
\midrule
\textit{Gold} & \cellcolor{cellgreen!49}\textcolor{black}{\textit{.51}} & \cellcolor{cellgreen!74}\textcolor{white}{\textit{.74}} & \cellcolor{cellgreen!70}\textcolor{white}{\textit{.71}} & \cellcolor{cellgreen!46}\textcolor{black}{\textit{.48}} & \cellcolor{cellgreen!51}\textcolor{black}{\textit{.53}} & \cellcolor{cellgreen!70}\textcolor{white}{\textit{.70}} & \cellcolor{cellgreen!53}\textcolor{white}{\textit{.55}}\\
\textbf{Comp-H} & \cellcolor{cellgreen!46}\textcolor{black}{\textbf{.48}} & \cellcolor{cellgreen!67}\textcolor{white}{\textbf{.68}} & \cellcolor{cellgreen!62}\textcolor{white}{\textbf{.63}} & \cellcolor{cellgreen!36}\textcolor{black}{\textbf{.39}} & \cellcolor{cellgreen!48}\textcolor{black}{\textbf{.50}} & \cellcolor{cellgreen!82}\textcolor{white}{\textbf{.82}} & \cellcolor{cellgreen!43}\textcolor{black}{.46}\\
\textbf{Comp-S} & \cellcolor{cellgreen!32}\textcolor{black}{.35} & \cellcolor{cellgreen!41}\textcolor{black}{.43} & \cellcolor{cellgreen!34}\textcolor{black}{.36} & \cellcolor{cellgreen!22}\textcolor{black}{.25} & \cellcolor{cellgreen!23}\textcolor{black}{.26} & \cellcolor{cellgreen!51}\textcolor{black}{.53} & \cellcolor{cellgreen!19}\textcolor{black}{.22}\\
Oscar & \cellcolor{cellgreen!24}\textcolor{black}{.27} & \cellcolor{cellgreen!32}\textcolor{black}{.35} & \cellcolor{cellgreen!23}\textcolor{black}{.26} & \cellcolor{cellgreen!14}\textcolor{black}{.17} & \cellcolor{cellgreen!15}\textcolor{black}{.19} & \cellcolor{cellgreen!40}\textcolor{black}{.42} & \cellcolor{cellgreen!12}\textcolor{black}{.15}\\
Zerank-2 & \cellcolor{cellgreen!42}\textcolor{black}{.44} & \cellcolor{cellgreen!52}\textcolor{black}{.54} & \cellcolor{cellgreen!51}\textcolor{black}{.53} & \cellcolor{cellgreen!27}\textcolor{black}{.30} & \cellcolor{cellgreen!28}\textcolor{black}{.31} & \cellcolor{cellgreen!57}\textcolor{white}{.59} & \cellcolor{cellgreen!21}\textcolor{black}{.25}\\
OpenProv. & \cellcolor{cellgreen!43}\textcolor{black}{.45} & \cellcolor{cellgreen!61}\textcolor{white}{.62} & \cellcolor{cellgreen!51}\textcolor{black}{.53} & \cellcolor{cellgreen!36}\textcolor{black}{.38} & \cellcolor{cellgreen!37}\textcolor{black}{.39} & \cellcolor{cellgreen!66}\textcolor{white}{.67} & \cellcolor{cellgreen!44}\textcolor{black}{\textbf{.46}}\\
Provence & \cellcolor{cellgreen!42}\textcolor{black}{.44} & \cellcolor{cellgreen!54}\textcolor{white}{.56} & \cellcolor{cellgreen!45}\textcolor{black}{.47} & \cellcolor{cellgreen!26}\textcolor{black}{.29} & \cellcolor{cellgreen!37}\textcolor{black}{.39} & \cellcolor{cellgreen!59}\textcolor{white}{.60} & \cellcolor{cellgreen!28}\textcolor{black}{.31}\\
xProvence & \cellcolor{cellgreen!32}\textcolor{black}{.35} & \cellcolor{cellgreen!52}\textcolor{black}{.54} & \cellcolor{cellgreen!43}\textcolor{black}{.46} & \cellcolor{cellgreen!23}\textcolor{black}{.26} & \cellcolor{cellgreen!30}\textcolor{black}{.33} & \cellcolor{cellgreen!36}\textcolor{black}{.38} & \cellcolor{cellgreen!23}\textcolor{black}{.26}\\
SemHL+BGE & \cellcolor{cellgreen!46}\textcolor{black}{\textbf{.48}} & \cellcolor{cellgreen!59}\textcolor{white}{.60} & \cellcolor{cellgreen!51}\textcolor{black}{.53} & \cellcolor{cellgreen!35}\textcolor{black}{.38} & \cellcolor{cellgreen!35}\textcolor{black}{.38} & \cellcolor{cellgreen!65}\textcolor{white}{.67} & \cellcolor{cellgreen!33}\textcolor{black}{.36}\\
LL2+BGE & \cellcolor{cellgreen!29}\textcolor{black}{.32} & \cellcolor{cellgreen!21}\textcolor{black}{.24} & \cellcolor{cellgreen!14}\textcolor{black}{.18} & \cellcolor{cellgreen!7}\textcolor{black}{.11} & \cellcolor{cellgreen!10}\textcolor{black}{.14} & \cellcolor{cellgreen!32}\textcolor{black}{.35} & \cellcolor{cellgreen!13}\textcolor{black}{.17}\\
\end{tabular}\end{center}
\vspace{1pt}
\begin{center}\begin{tabular}{lccccccc}
\multicolumn{8}{l}{\scriptsize\textbf{ViDoRe \textperiodcentered Industrial --- EN \textperiodcentered 283 q}}\\[2pt]
& \makecell{Extr. (125)} & \makecell{Open-\,ended (67)} & \makecell{Enum. (55)} & \makecell{Bool. (44)} & \makecell{Compare-\,contrast (29)} & \makecell{Num. (29)} & \makecell{Multi-\,hop (28)}\\[-1pt]
\midrule
\textit{Gold} & \cellcolor{cellgreen!72}\textcolor{white}{\textit{.73}} & \cellcolor{cellgreen!40}\textcolor{black}{\textit{.42}} & \cellcolor{cellgreen!41}\textcolor{black}{\textit{.43}} & \cellcolor{cellgreen!88}\textcolor{white}{\textit{.88}} & \cellcolor{cellgreen!52}\textcolor{black}{\textit{.54}} & \cellcolor{cellgreen!75}\textcolor{white}{\textit{.75}} & \cellcolor{cellgreen!33}\textcolor{black}{\textit{.35}}\\
\textbf{Comp-H} & \cellcolor{cellgreen!60}\textcolor{white}{\textbf{.61}} & \cellcolor{cellgreen!38}\textcolor{black}{\textbf{.40}} & \cellcolor{cellgreen!33}\textcolor{black}{\textbf{.36}} & \cellcolor{cellgreen!79}\textcolor{white}{\textbf{.80}} & \cellcolor{cellgreen!46}\textcolor{black}{\textbf{.48}} & \cellcolor{cellgreen!62}\textcolor{white}{\textbf{.63}} & \cellcolor{cellgreen!25}\textcolor{black}{\textbf{.28}}\\
\textbf{Comp-S} & \cellcolor{cellgreen!44}\textcolor{black}{.46} & \cellcolor{cellgreen!14}\textcolor{black}{.17} & \cellcolor{cellgreen!21}\textcolor{black}{.25} & \cellcolor{cellgreen!66}\textcolor{white}{.67} & \cellcolor{cellgreen!24}\textcolor{black}{.27} & \cellcolor{cellgreen!52}\textcolor{white}{.54} & \cellcolor{cellgreen!20}\textcolor{black}{.23}\\
Oscar & \cellcolor{cellgreen!30}\textcolor{black}{.33} & \cellcolor{cellgreen!23}\textcolor{black}{.26} & \cellcolor{cellgreen!10}\textcolor{black}{.14} & \cellcolor{cellgreen!64}\textcolor{white}{.65} & \cellcolor{cellgreen!16}\textcolor{black}{.20} & \cellcolor{cellgreen!34}\textcolor{black}{.36} & \cellcolor{cellgreen!9}\textcolor{black}{.13}\\
Zerank-2 & \cellcolor{cellgreen!42}\textcolor{black}{.45} & \cellcolor{cellgreen!28}\textcolor{black}{.31} & \cellcolor{cellgreen!28}\textcolor{black}{.31} & \cellcolor{cellgreen!59}\textcolor{white}{.61} & \cellcolor{cellgreen!17}\textcolor{black}{.20} & \cellcolor{cellgreen!38}\textcolor{black}{.40} & \cellcolor{cellgreen!11}\textcolor{black}{.15}\\
OpenProv. & \cellcolor{cellgreen!53}\textcolor{white}{.54} & \cellcolor{cellgreen!35}\textcolor{black}{.38} & \cellcolor{cellgreen!33}\textcolor{black}{.35} & \cellcolor{cellgreen!56}\textcolor{white}{.57} & \cellcolor{cellgreen!39}\textcolor{black}{.42} & \cellcolor{cellgreen!44}\textcolor{black}{.46} & \cellcolor{cellgreen!22}\textcolor{black}{.25}\\
Provence & \cellcolor{cellgreen!48}\textcolor{black}{.50} & \cellcolor{cellgreen!35}\textcolor{black}{.37} & \cellcolor{cellgreen!30}\textcolor{black}{.33} & \cellcolor{cellgreen!55}\textcolor{white}{.57} & \cellcolor{cellgreen!39}\textcolor{black}{.42} & \cellcolor{cellgreen!36}\textcolor{black}{.39} & \cellcolor{cellgreen!14}\textcolor{black}{.18}\\
xProvence & \cellcolor{cellgreen!38}\textcolor{black}{.41} & \cellcolor{cellgreen!22}\textcolor{black}{.26} & \cellcolor{cellgreen!20}\textcolor{black}{.23} & \cellcolor{cellgreen!53}\textcolor{white}{.55} & \cellcolor{cellgreen!30}\textcolor{black}{.33} & \cellcolor{cellgreen!28}\textcolor{black}{.31} & \cellcolor{cellgreen!20}\textcolor{black}{.23}\\
SemHL+BGE & \cellcolor{cellgreen!48}\textcolor{black}{.50} & \cellcolor{cellgreen!30}\textcolor{black}{.33} & \cellcolor{cellgreen!28}\textcolor{black}{.31} & \cellcolor{cellgreen!57}\textcolor{white}{.59} & \cellcolor{cellgreen!45}\textcolor{black}{.47} & \cellcolor{cellgreen!39}\textcolor{black}{.41} & \cellcolor{cellgreen!20}\textcolor{black}{.23}\\
LL2+BGE & \cellcolor{cellgreen!17}\textcolor{black}{.21} & \cellcolor{cellgreen!16}\textcolor{black}{.20} & \cellcolor{cellgreen!14}\textcolor{black}{.18} & \cellcolor{cellgreen!32}\textcolor{black}{.35} & \cellcolor{cellgreen!8}\textcolor{black}{.12} & \cellcolor{cellgreen!13}\textcolor{black}{.17} & \cellcolor{cellgreen!6}\textcolor{black}{.08}\\
\end{tabular}\end{center}
\vspace{1pt}
\begin{center}\begin{tabular}{lccccccc}
\multicolumn{8}{l}{\scriptsize\textbf{ViDoRe \textperiodcentered Pharmaceuticals --- EN \textperiodcentered 364 q}}\\[2pt]
& \makecell{Open-\,ended (144)} & \makecell{Extr. (102)} & \makecell{Compare-\,contrast (59)} & \makecell{Enum. (58)} & \makecell{Num. (50)} & \makecell{Bool. (38)} & \makecell{Multi-\,hop (35)}\\[-1pt]
\midrule
\textit{Gold} & \cellcolor{cellgreen!50}\textcolor{black}{\textit{.52}} & \cellcolor{cellgreen!67}\textcolor{white}{\textit{.68}} & \cellcolor{cellgreen!40}\textcolor{black}{\textit{.42}} & \cellcolor{cellgreen!48}\textcolor{black}{\textit{.50}} & \cellcolor{cellgreen!59}\textcolor{white}{\textit{.60}} & \cellcolor{cellgreen!84}\textcolor{white}{\textit{.84}} & \cellcolor{cellgreen!46}\textcolor{black}{\textit{.48}}\\
\textbf{Comp-H} & \cellcolor{cellgreen!43}\textcolor{black}{\textbf{.46}} & \cellcolor{cellgreen!61}\textcolor{white}{.62} & \cellcolor{cellgreen!34}\textcolor{black}{.36} & \cellcolor{cellgreen!38}\textcolor{black}{.41} & \cellcolor{cellgreen!52}\textcolor{white}{.54} & \cellcolor{cellgreen!85}\textcolor{white}{\textbf{.85}} & \cellcolor{cellgreen!30}\textcolor{black}{.33}\\
\textbf{Comp-S} & \cellcolor{cellgreen!23}\textcolor{black}{.26} & \cellcolor{cellgreen!36}\textcolor{black}{.39} & \cellcolor{cellgreen!23}\textcolor{black}{.27} & \cellcolor{cellgreen!15}\textcolor{black}{.18} & \cellcolor{cellgreen!25}\textcolor{black}{.28} & \cellcolor{cellgreen!74}\textcolor{white}{.75} & \cellcolor{cellgreen!20}\textcolor{black}{.23}\\
Oscar & \cellcolor{cellgreen!24}\textcolor{black}{.27} & \cellcolor{cellgreen!33}\textcolor{black}{.36} & \cellcolor{cellgreen!12}\textcolor{black}{.15} & \cellcolor{cellgreen!16}\textcolor{black}{.20} & \cellcolor{cellgreen!18}\textcolor{black}{.21} & \cellcolor{cellgreen!69}\textcolor{white}{.70} & \cellcolor{cellgreen!14}\textcolor{black}{.17}\\
Zerank-2 & \cellcolor{cellgreen!39}\textcolor{black}{.42} & \cellcolor{cellgreen!59}\textcolor{white}{.60} & \cellcolor{cellgreen!26}\textcolor{black}{.29} & \cellcolor{cellgreen!37}\textcolor{black}{.39} & \cellcolor{cellgreen!45}\textcolor{black}{.47} & \cellcolor{cellgreen!62}\textcolor{white}{.64} & \cellcolor{cellgreen!21}\textcolor{black}{.24}\\
OpenProv. & \cellcolor{cellgreen!41}\textcolor{black}{.43} & \cellcolor{cellgreen!53}\textcolor{white}{.55} & \cellcolor{cellgreen!32}\textcolor{black}{.35} & \cellcolor{cellgreen!36}\textcolor{black}{.39} & \cellcolor{cellgreen!27}\textcolor{black}{.30} & \cellcolor{cellgreen!62}\textcolor{white}{.64} & \cellcolor{cellgreen!25}\textcolor{black}{.28}\\
Provence & \cellcolor{cellgreen!42}\textcolor{black}{.45} & \cellcolor{cellgreen!60}\textcolor{white}{.62} & \cellcolor{cellgreen!34}\textcolor{black}{.36} & \cellcolor{cellgreen!40}\textcolor{black}{\textbf{.42}} & \cellcolor{cellgreen!48}\textcolor{black}{.50} & \cellcolor{cellgreen!66}\textcolor{white}{.68} & \cellcolor{cellgreen!21}\textcolor{black}{.24}\\
xProvence & \cellcolor{cellgreen!42}\textcolor{black}{.44} & \cellcolor{cellgreen!56}\textcolor{white}{.57} & \cellcolor{cellgreen!28}\textcolor{black}{.31} & \cellcolor{cellgreen!34}\textcolor{black}{.36} & \cellcolor{cellgreen!51}\textcolor{black}{.53} & \cellcolor{cellgreen!57}\textcolor{white}{.58} & \cellcolor{cellgreen!28}\textcolor{black}{.31}\\
SemHL+BGE & \cellcolor{cellgreen!40}\textcolor{black}{.43} & \cellcolor{cellgreen!61}\textcolor{white}{\textbf{.63}} & \cellcolor{cellgreen!35}\textcolor{black}{\textbf{.38}} & \cellcolor{cellgreen!37}\textcolor{black}{.40} & \cellcolor{cellgreen!54}\textcolor{white}{\textbf{.56}} & \cellcolor{cellgreen!73}\textcolor{white}{.74} & \cellcolor{cellgreen!30}\textcolor{black}{\textbf{.33}}\\
LL2+BGE & \cellcolor{cellgreen!10}\textcolor{black}{.13} & \cellcolor{cellgreen!12}\textcolor{black}{.16} & \cellcolor{cellgreen!6}\textcolor{black}{.10} & \cellcolor{cellgreen!6}\textcolor{black}{.09} & \cellcolor{cellgreen!6}\textcolor{black}{.10} & \cellcolor{cellgreen!32}\textcolor{black}{.35} & \cellcolor{cellgreen!6}\textcolor{black}{.07}\\
\end{tabular}\end{center}
\vspace{1pt}
\begin{center}\begin{tabular}{lccccccc}
\multicolumn{8}{l}{\scriptsize\textbf{ViDoRe \textperiodcentered Physics --- FR \textperiodcentered 302 q}}\\[2pt]
& \makecell{Open-\,ended (151)} & \makecell{Extr. (77)} & \makecell{Compare-\,contrast (66)} & \makecell{Num. (33)} & \makecell{Enum. (30)} & \makecell{Bool. (25)} & \makecell{Multi-\,hop (22)}\\[-1pt]
\midrule
\textit{Gold} & \cellcolor{cellgreen!45}\textcolor{black}{\textit{.47}} & \cellcolor{cellgreen!70}\textcolor{white}{\textit{.71}} & \cellcolor{cellgreen!51}\textcolor{black}{\textit{.53}} & \cellcolor{cellgreen!54}\textcolor{white}{\textit{.56}} & \cellcolor{cellgreen!52}\textcolor{black}{\textit{.54}} & \cellcolor{cellgreen!69}\textcolor{white}{\textit{.70}} & \cellcolor{cellgreen!35}\textcolor{black}{\textit{.38}}\\
\textbf{Comp-H} & \cellcolor{cellgreen!45}\textcolor{black}{\textbf{.48}} & \cellcolor{cellgreen!66}\textcolor{white}{\textbf{.67}} & \cellcolor{cellgreen!50}\textcolor{black}{.52} & \cellcolor{cellgreen!63}\textcolor{white}{.64} & \cellcolor{cellgreen!41}\textcolor{black}{.43} & \cellcolor{cellgreen!75}\textcolor{white}{.76} & \cellcolor{cellgreen!31}\textcolor{black}{.33}\\
\textbf{Comp-S} & \cellcolor{cellgreen!43}\textcolor{black}{.46} & \cellcolor{cellgreen!51}\textcolor{black}{.53} & \cellcolor{cellgreen!49}\textcolor{black}{.51} & \cellcolor{cellgreen!50}\textcolor{black}{.52} & \cellcolor{cellgreen!30}\textcolor{black}{.33} & \cellcolor{cellgreen!69}\textcolor{white}{.70} & \cellcolor{cellgreen!46}\textcolor{black}{\textbf{.48}}\\
Oscar & \cellcolor{cellgreen!13}\textcolor{black}{.16} & \cellcolor{cellgreen!34}\textcolor{black}{.37} & \cellcolor{cellgreen!10}\textcolor{black}{.14} & \cellcolor{cellgreen!40}\textcolor{black}{.42} & \cellcolor{cellgreen!14}\textcolor{black}{.18} & \cellcolor{cellgreen!56}\textcolor{white}{.57} & \cellcolor{cellgreen!15}\textcolor{black}{.18}\\
Zerank-2 & \cellcolor{cellgreen!43}\textcolor{black}{.45} & \cellcolor{cellgreen!62}\textcolor{white}{.64} & \cellcolor{cellgreen!53}\textcolor{white}{\textbf{.54}} & \cellcolor{cellgreen!61}\textcolor{white}{.62} & \cellcolor{cellgreen!46}\textcolor{black}{\textbf{.48}} & \cellcolor{cellgreen!77}\textcolor{white}{\textbf{.77}} & \cellcolor{cellgreen!32}\textcolor{black}{.35}\\
OpenProv. & \cellcolor{cellgreen!38}\textcolor{black}{.41} & \cellcolor{cellgreen!48}\textcolor{black}{.50} & \cellcolor{cellgreen!46}\textcolor{black}{.48} & \cellcolor{cellgreen!45}\textcolor{black}{.47} & \cellcolor{cellgreen!29}\textcolor{black}{.32} & \cellcolor{cellgreen!59}\textcolor{white}{.61} & \cellcolor{cellgreen!21}\textcolor{black}{.24}\\
Provence & \cellcolor{cellgreen!38}\textcolor{black}{.40} & \cellcolor{cellgreen!58}\textcolor{white}{.60} & \cellcolor{cellgreen!43}\textcolor{black}{.46} & \cellcolor{cellgreen!66}\textcolor{white}{\textbf{.67}} & \cellcolor{cellgreen!37}\textcolor{black}{.40} & \cellcolor{cellgreen!68}\textcolor{white}{.69} & \cellcolor{cellgreen!32}\textcolor{black}{.35}\\
xProvence & \cellcolor{cellgreen!42}\textcolor{black}{.44} & \cellcolor{cellgreen!65}\textcolor{white}{.66} & \cellcolor{cellgreen!48}\textcolor{black}{.50} & \cellcolor{cellgreen!65}\textcolor{white}{.66} & \cellcolor{cellgreen!38}\textcolor{black}{.40} & \cellcolor{cellgreen!58}\textcolor{white}{.59} & \cellcolor{cellgreen!25}\textcolor{black}{.28}\\
SemHL+BGE & \cellcolor{cellgreen!45}\textcolor{black}{.47} & \cellcolor{cellgreen!63}\textcolor{white}{.64} & \cellcolor{cellgreen!48}\textcolor{black}{.50} & \cellcolor{cellgreen!65}\textcolor{white}{.66} & \cellcolor{cellgreen!43}\textcolor{black}{.45} & \cellcolor{cellgreen!63}\textcolor{white}{.65} & \cellcolor{cellgreen!25}\textcolor{black}{.28}\\
LL2+BGE & \cellcolor{cellgreen!13}\textcolor{black}{.17} & \cellcolor{cellgreen!21}\textcolor{black}{.25} & \cellcolor{cellgreen!10}\textcolor{black}{.14} & \cellcolor{cellgreen!12}\textcolor{black}{.15} & \cellcolor{cellgreen!16}\textcolor{black}{.19} & \cellcolor{cellgreen!22}\textcolor{black}{.25} & \cellcolor{cellgreen!6}\textcolor{black}{.07}\\
\end{tabular}\end{center}
\vspace{-1em}
\caption{\textbf{Integration Test} per-query-type judge accuracy (2/2)}
\label{tab:integration-qtype2}
\end{table*}

We report performance by query type (Tables~\ref{tab:integration-qtype} and \ref{tab:integration-qtype2}) for
datasets that originally contain a query type classification (Academic Papers, EnterpriseRAG, HotpotQA, and ViDoRe).

\subsection{Qualitative Examples}\label{sec:appendix_integration_examples} 
See Figures~\ref{fig:ex-academic-soft}, \ref{fig:ex-finqa-table}, and \ref{fig:ex-hotpot} for examples.
\begin{figure}[h!]
\begin{qexample}{\texttt{Academic Papers}: soft compression regenerates the number instead of copying it}
\textbf{Q.} What is the minimum hydrostatic pressure required for hydrogen atom movement to
become observable in the trigonal polymorph of brucite at 1750\,K?\\
\textbf{Gold.} 10.1 GPa\ \ (``the onset pressure for proton mobility decreases further to
\textbf{10.1} and 13.4 GPa when the temperatures are set to 1750K and 2000K'')\\[3pt]
\begin{minipage}[t]{0.47\linewidth}
\textbf{Comp-S}, 94\% compressed (\uscore{0.00})\\[1pt]
{\footnotesize $\Rightarrow$ ``\textbf{Around 19 GPa}.''\\[2pt]
Oscar (\uscore{0.00}) reports the value is not stated and then guesses ``around 1 GPa''. Both
produce the right unit and a plausible magnitude, but not the number in the text.}
\end{minipage}\hfill
\begin{minipage}[t]{0.47\linewidth}
\textbf{Comp-H}, 99\% compressed (\gscore{1.00})\\[1pt]
{\footnotesize Keeps the answer-bearing sentence verbatim and quotes it.\\[2pt]
$\Rightarrow$ ``\textbf{10.1 GPa}''.}
\end{minipage}\\[4pt]
\textbf{Analysis.} Extractive compression discards 99\% of the retrieved tokens yet leaves the
one sentence intact, so the generator can copy the digits. The soft compressors replace each
passage with a fixed number of memory embeddings, so the value has to be reconstructed from a
lossy representation and both models emit a fabricated pressure. Semantic Highlight and the
gold-document oracle both answer 10.1 GPa.
\end{qexample}
\vspace{-0.7em}
\caption{\textbf{Integration Test Example (Academic Papers).} Exact numeric values survive
aggressive extractive pruning but not fixed-rate soft compression.}
\label{fig:ex-academic-soft}
\end{figure}

\begin{figure}[ht!]
\begin{qexample}{\texttt{FinQA}: deleting table delimiters removes the row the question needs}
\textbf{Q.} Excluding accretion, what was the ending balance of the asset retirement
liability as of September 27, 2003 (in millions)?\\
\textbf{Gold.} 6\ \ (\$7.2 ending balance $-$ \$1.2 accretion)\\[3pt]
\begin{minipage}[t]{0.47\linewidth}
\textbf{LLMLingua-2}, 77\% compressed (\uscore{0.01})\\[1pt]
{\footnotesize\texttt{retirement liability 29 2002,\$ 5., obligations,.,., retirement
liability 27 2003,\$ 7.}\\[2pt]
$\Rightarrow$ quotes this string back, answers \textbf{\$7 million}, and states that the
context gives no accretion figure to subtract.}
\end{minipage}\hfill
\begin{minipage}[t]{0.47\linewidth}
\textbf{Comp-H}, 86\% compressed (\gscore{1.00})\\[1pt]
{\footnotesize\texttt{[['...september 29 2002', '\$ 5.5'], ['additional asset retirement
obligations recognized', '0.5'], ['accretion recognized', '1.2'], ['...september 27
2003', '\$ 7.2']]}\\[2pt]
$\Rightarrow$ \$7.2 $-$ \$1.2 $=$ \textbf{\$6.0 million}.}
\end{minipage}\\[4pt]
\textbf{Analysis.} The cell delimiters collapse into \texttt{,.,.,.}, dissolving the
label--value pairs: the accretion row (\texttt{1.2}) disappears entirely and both
balances lose their decimals (\texttt{5.5}$\to$\texttt{5.},
\texttt{7.2}$\to$\texttt{7.}). The subtraction the question asks for is no longer
expressible in the compressed text, so the generator returns the truncated ending balance
instead. Comp-H keeps the table rows atomic and performs the subtraction.
\end{qexample}
\vspace{-0.7em}
\caption{\textbf{Integration Test Example (FinQA).} Two models both compress the gold document, but token-level compression detaches
numbers from their row labels, so a value the question depends on becomes unrecoverable.}
\label{fig:ex-finqa-table}
\end{figure}

\begin{figure}[ht!]
\begin{qexample}{\texttt{HotpotQA}: a distractor's number survives without its antecedent}
\textbf{Q.} ``Mis'ry's Comin' Aroun'\,'' is by an American librettist that won how many Tony
Awards?\\
\textbf{Gold.} eight Tony Awards\ \ (hop 1: the song is by Oscar Hammerstein II; hop 2:
``Hammerstein won eight Tony Awards'')\\[3pt]
\begin{minipage}[t]{0.47\linewidth}
\textbf{Comp-H}, 87\% compressed (\uscore{0.00})\\[1pt]
{\footnotesize Kept, in full: ``\emph{Mis'ry's Comin' Aroun' is a once-neglected song from the
1927 musical Show Boat by Jerome Kern and Oscar Hammerstein II.}'' $+$
``\emph{He won three Tony Awards and three Academy Awards}'' --- the latter from the page of
\textbf{Alan Jay Lerner}, a different American librettist. Hammerstein's document is pruned
from 215 tokens to \textbf{0}.\\[2pt]
$\Rightarrow$ ``...\,won \textbf{three} Tony Awards.''}
\end{minipage}\hfill
\begin{minipage}[t]{0.47\linewidth}
\textbf{Oscar (soft)}, no readable context (\gscore{0.99})\\[1pt]
{\footnotesize $\Rightarrow$ ``Mis'ry's Comin' Aroun' is by Oscar Hammerstein II, who won
\textbf{eight} Tony Awards.''}
\end{minipage}\\[4pt]
\textbf{Analysis.} Both candidates state a Tony count and neither names the song, so a
query-similarity scorer has little to separate ``\emph{Hammerstein won eight Tony
Awards\,\dots}'' from ``\emph{He won three Tony Awards\,\dots}''. Retrieval is not the
bottleneck: Hammerstein's page is among the five documents Comp-H selects, and Comp-H deletes
all 215 of its tokens. What survives is the closing sentence of Lerner's page without the
earlier sentences that name him --- a Tony count on a bare pronoun, in a context where the only
people named are Jerome Kern and Oscar Hammerstein II. The generator reads the count as
Hammerstein's, while Oscar, which encodes the same documents as memory embeddings instead of
selecting sentences, answers eight.
\end{qexample}
\vspace{-0.7em}
\caption{\textbf{Integration Test Example (HotpotQA).} Sentence pruning deletes the
answer-bearing second-hop sentence and keeps a near-identical sentence about a different
librettist, stripped of the antecedent that would identify him.}
\label{fig:ex-hotpot}
\end{figure}

\clearpage
\raggedbottom
\section{System Test}\label{sec:appendix_system_test}
\subsection{BrowseComp-Plus}\label{sec:appendix_browsecomp_result}

\subsubsection{Setup}
\paragraph{Benchmark and retrieval.}
BrowseComp-Plus~\citep{chen2025browsecompplusfairtransparentevaluation} contains
$830$ queries over a fixed corpus of $100{,}195$ documents. The first stage is
dense retrieval with Qwen3-Embedding-8B over the prebuilt
\texttt{Tevatron/browsecomp-plus-indexes}. For each search call issued by the
agent, we retrieve the top-$50$ documents, rerank and compress them, and return
the top-$5$ as agent-visible snippets.

\paragraph{Agent.}
We use \texttt{openai/gpt-oss-120b} at medium reasoning effort, served by vLLM
through an OpenAI-compatible Responses API. We disable the whole-document reader
(\texttt{get\_document}): if it is enabled, the agent can bypass the compressor
and read the raw document again, which confounds any measurement of compression
quality. We set \texttt{max\_output\_tokens}${=}10{,}000$, allow at most $100$
tool-calling rounds per query, and retry up to $3$ times on transient
Responses/Harmony tool-parser errors.

\paragraph{Compressor.}
For the system test, we try to follow the recommended / default compression thresholds of each baseline whenever available and adopt a conservative compression strategy to avoid overly aggressive token reduction that may harm answer quality. Specifically, we set $\tau=0.1$ for Comp-H, OpenProvence, Provence, and Semantic Highlight. Semantic Highlight's default is $0.5$, but since it compresses aggressively on our system test, we use $0.1$. We use $0.3$ for XProvence, $0.4$ for DSLR-Zerank, $0.5$ for LLMLingua-2, and $0.55$ for LongLLMLingua.

\paragraph{Judge.}
We follow the standard LLM-as-judge protocol of BrowseComp-Plus with one change:
we grade with Qwen3.5-35B-A3B under deterministic decoding, so that the judge is
the same as in our Unit Test and Integration Test.

\subsubsection{Effect of the Token Cap}\label{appendix:browsecomp_token_cap}

The original BrowseComp-Plus truncates each document at $512$ tokens for budget reasons. In our main
experiments, we follow the original truncation settings for the baseline, but not for evaluated compression methods, because we want to isolate the effect of
compression on long documents. In Table~\ref{tab:browsecomp_cap_vs_nocap}, we directly compare the truncated and un-truncated settings for the baseline.
The two pipelines differ only in the truncation
step. 

\noindent \textbf{No Cap}:

{\small
\begin{equation*}
\begin{aligned}
\underbrace{100\text{K corpus}}_{\text{fixed}}
&\xrightarrow{\text{Qwen3-Embed-8B}} \text{top-50}
\xrightarrow{\text{rerank \& compress}} \text{top-5} \\
&\xrightarrow{\phantom{x}} \text{agent}
\xrightarrow{\phantom{x}} \text{judge}
\end{aligned}
\end{equation*}
}

\noindent
\textbf{512 Cap}:
{\small
\begin{equation*}
\begin{aligned}
\underbrace{100\text{K corpus}}_{\text{fixed}}
&\xrightarrow{\text{Qwen3-Embed-8B}} \text{top-50}
\xrightarrow{\text{rerank \& compress}} \text{top-5} \\
&\xrightarrow{\phantom{x}} \text{per-doc 512-token cap}
\xrightarrow{\phantom{x}} \text{agent}
\xrightarrow{\phantom{x}} \text{judge}
\end{aligned}
\end{equation*}
}

\noindent
For completeness, we additionally run all compression methods under
the original $512$-token cap, which corresponds to a realistic
budget-constrained setting.

\begin{table}[ht!]
\centering
\small
\setlength{\tabcolsep}{4pt}
\begin{tabular}{llccc}
\toprule
\textbf{Method} & \textbf{Mode} & \textbf{Acc.} & \textbf{Recall} & \textbf{Search} \\
\midrule
\multirow{2}{*}{\textbf{512 Cap}}
& Retrieval-only &40.8 & 46.0   & 9.6 \\
& Retrieval+Rerank &45.9 & 50.7 & 9.5  \\
\midrule
\multirow{2}{*}{\textbf{No Cap}}
& Retrieval-only & 26.4 & 32.2 & 4.7 \\
& Retrieval+Rerank &18.6 & 28.0& 3.1 \\
\bottomrule
\end{tabular}
\vspace{-0.6em}
\caption{\textbf{System Test} results with and without document truncation to 512 tokens. Search agent is gpt-oss-120B. Without truncation, the agent issues fewer searches and reaches lower accuracy.} 
\vspace{-0.8em}
\label{tab:browsecomp_cap_vs_nocap}
\end{table}

\subsubsection{Results}

Table~\ref{tab:browsecomp_512cap_result} reports the 512-token capped setting with the
gpt-oss-20B and 120B agents. Two observations hold for both agents. First,
compression alone already improves accuracy over the uncompressed
retrieval-only baseline ($+2.2$ for 20B, $+7.7$ for 120B), so the gain is not
only an effect of a larger effective context. Second, reranking adds a much
larger gain on top of compression ($+18.1$ and $+16.0$), and the same ordering
appears in retrieval recall and in calibration error. This is consistent with
the no-cap results in the main text.

\begin{table*}[ht]
\centering
\small
\setlength{\tabcolsep}{4pt}
\resizebox{\textwidth}{!}{%
\begin{tabular}{llcccccccccc}
\toprule
\multirow{2}{*}{\textbf{Method}} & \multirow{2}{*}{\textbf{Mode}} & \multicolumn{2}{c}{\textbf{Answer Accuracy}} & \multicolumn{2}{c}{\textbf{Retrieval Recall}} & \multicolumn{2}{c}{\textbf{Calibration Err.}$\downarrow$} & \multicolumn{2}{c}{\textbf{Comp.\%}} & \multicolumn{2}{c}{\textbf{Search}} \\
\cmidrule(lr){3-4} \cmidrule(lr){5-6} \cmidrule(lr){7-8} \cmidrule(lr){9-10} \cmidrule(lr){11-12}
 & & \textbf{20B} & \textbf{120B} & \textbf{20B} & \textbf{120B} & \textbf{20B} & \textbf{120B} & \textbf{20B} & \textbf{120B} & \textbf{20B} & \textbf{120B} \\
\midrule
\multirow{2}{*}{\textbf{Baseline}} 
& Retrieval-only & 32.6 & 40.8 & 42.0 & 46.0 & 35.7 & 38.1 & --- & --- & 12.7 & 9.6 \\
& Rerank-only (BGE) &39.6 &45.9 & 46.5&  50.7& 32.2&34.7  & --- & --- & 12.4& 9.5\\
\midrule
\multirow{2}{*}{\textbf{Comp-H}}
 & Compress-only & 34.8\,\textsubscript{\textcolor{blue}{(+2.2)}} & 48.5\,\textsubscript{\textcolor{blue}{(+7.7)}} & 44.9\,\textsubscript{\textcolor{blue}{(+2.9)}} & 51.0\,\textsubscript{\textcolor{blue}{(+5.0)}} & 35.6\,\textsubscript{\textcolor{blue}{(--0.1)}} & 33.2\,\textsubscript{\textcolor{blue}{(--4.9)}} & 85.3 & 84.2 & 15.1 & 13.6 \\
 & Rerank+Compr. & \textbf{52.9}\,\textsubscript{\textcolor{blue}{(+18.1)}} & \textbf{64.5}\,\textsubscript{\textcolor{blue}{(+16.0)}} & \textbf{61.2}\,\textsubscript{\textcolor{blue}{(+16.3)}} & \textbf{66.3}\,\textsubscript{\textcolor{blue}{(+15.3)}} & \textbf{27.4}\,\textsubscript{\textcolor{blue}{(--8.2)}} & \textbf{26.6}\,\textsubscript{\textcolor{blue}{(--6.6)}} & 75.6 & 75.6 & 13.5 & 11.0 \\
\midrule
\multirow{2}{*}{DSLR-Zerank} 
 & Compress-only & 29.6 & 34.5 & 44.6 & 46.3 & 33.5 & 37.2 & 81.8 & 79.9 & 19.1 & 14.2 \\
 & Rerank+Compr. & 41.2 & 49.5 & 57.8 & 59.8 & 31.5 & 30.8 & 70.7 & 70.7 & 18.4 & 12.6 \\
\midrule
\multirow{2}{*}{OpenProvence}
 & Compress-only & 31.3 & 46.0 & 44.0 & 49.8 & 34.5 & 35.1 & 77.6 & 75.6 & 14.7 & 13.0 \\
 & Rerank+Compr. & 48.3 & 58.5 & 57.0 & 61.4 & 29.4 & 31.5 & 63.3 & 62.8 & 14.5 & 10.9 \\
\midrule
\multirow{2}{*}{Provence}
 & Compress-only & 27.5 & 41.8 & 41.4 & 49.1 & 36.5 & 36.7 & 72.8 & 72.0 & 15.3 & 12.9 \\
 & Rerank+Compr. & 38.3 & 52.9 & 54.9 & 61.0 & 34.5 & 30.8 & 71.7 & 70.7 & 14.9 & 12.3 \\
\midrule
\multirow{2}{*}{XProvence} 
 & Compress-only & 26.3 & 29.2 & 44.2 & 44.6 & 29.3 & 37.9 & 84.4 & 83.3 & 18.7 & 13.8 \\
 & Rerank+Compr. & 24.2& 31.1 &39.0 & 42.9 & 33.3 & 36.6 & 82.8 & 80.7 & 18.2 & 14.1 \\
\midrule
\multirow{2}{*}{Semantic HL}
 & Compress-only & 31.0 & 43.0 & 44.2 & 49.4 & 35.9 & 36.0 & 81.0 & 81.3 & 15.6 & 14.2 \\
 & Rerank+Compr. & 41.0 & 51.6 & 50.0 & 54.0 & 28.0 & 32.4 & 74.6 & 75.0 & 14.9 & 12.6 \\
\midrule
\multirow{2}{*}{LLMLingua-2} 
& Compress-only & 26.5 & 40.0 & 39.0 & 45.9 & 40.6 & 37.4 & 6.4 & 6.6 & 12.1 & 9.7 \\
 & Rerank+Compr. & 35.1 & 44.0 &44.7 &49.2 & 33.9 & 34.1 &1.94 &  2.2 & 11.8 & 9.3\\
\midrule
\multirow{2}{*}{LongLLMLingua}
 & Compress-only & 15.2 & 25.1 & 35.4 & 41.3 & 47.9 & 48.2 & 4.0 & 4.1 & 12.2 & 9.9 \\
 & Rerank+Compr. & 20.7 & 25.8 & 41.5 & 43.6 & 44.0 & 48.9 & 5.3 & 5.7 & 12.1 & 9.3 \\
\bottomrule
\end{tabular}}
\vspace{-0.6em}
\caption{\textbf{Full BrowseComp-Plus results} under a \textbf{512-token context cap}, with the
gpt-oss-20B and 120B agents. Subscripts for Comp-H show the gain of
Compress-only over the retrieval-only baseline and of Rerank+Compress over
Compress-only. Calibration Error is lower-is-better ($\downarrow$). Comp.\% is
computed with respect to the 512-token context cap and is measured using the
same Qwen3 tokenizer as the context cap.}
\label{tab:browsecomp_512cap_result}
\vspace{-0.8em}
\end{table*}

\subsubsection{Qualitative Example}
\definecolor{good}{RGB}{27,138,101}
\definecolor{bad}{RGB}{176,58,58}
\ifdefined\qexample\else
  \newtcolorbox{qexample}[1]{enhanced, colback=black!2, colframe=black!30,
    boxrule=0.4pt, arc=2pt, left=5pt, right=5pt, top=4pt, bottom=4pt,
    fonttitle=\bfseries\small, title=#1}
\fi
\providecommand{\uscore}[1]{\textcolor{bad}{\textbf{#1}}}
\providecommand{\gscore}[1]{\textcolor{good}{\textbf{#1}}}
\providecommand{\ok}{\textcolor{good}{\ding{51}}}
\providecommand{\bad}{\textcolor{bad}{\ding{55}}}

\begin{figure*}[t]
\begin{qexample}{BrowseComp-Plus \#922: compression makes the evidence salient; reranking verifies the timeline}
\textbf{Q.}
A letter was written in the first half of the twentieth century from a ruler of one country to another
and delivered by an official nicknamed after a body part. Written about six months after the author came
to power, it mentions that the country had regained a region. What region?

\textbf{Gold.} \textbf{North Transylvania}.\\[4pt]
\begin{minipage}[t]{0.325\textwidth}
\raggedright\setlength{\parindent}{0pt}
\textbf{Baseline (Retrieval-only)}~\bad\\
{\footnotesize
no compr.\ $\bullet$ 27.9k ctx tok \\$\bullet$ recall 0.33 $\bullet$
\uscore{Alsace-Lorraine}}\\[3pt]
{\footnotesize
\textbf{Key observation.}
On the tenth search the correct FRUS page is retrieved, and the answer sentence
(\emph{``regaining North Transylvania''}) already appears at token 384 of the
512-token chunk.\\[2pt]
\textbf{Failure.}
The sentence is buried inside dense diplomatic prose.
The agent continues following the nickname clue, invents a German official
nicknamed ``the Ear'', and answers from parametric knowledge instead of the
retrieved evidence. Rerank-only makes the same mistake.}
\end{minipage}\hfill
\begin{minipage}[t]{0.325\textwidth}
\raggedright\setlength{\parindent}{0pt}
\textbf{Comp-H (Compress-only)}~\ok\\
{\footnotesize
99.4\% comp.\ (1.60M$\rightarrow$10.1k) $\bullet$ \\recall 0.67 $\bullet$
\gscore{North Transylvania}}\\[3pt]
{\footnotesize
\textbf{Key observation.}
The top-5 pages are read in full and compressed from 29.8k tokens to 730,
of which the answer page keeps 334---enough to leave the answer sentence
immediately salient.\\[2pt]
\textbf{Outcome.}
The agent identifies the correct region on its first search.
However, it incorrectly dates King Michael's accession, spends later searches
checking the nickname clue, and never verifies the six-month constraint or cites
the supporting page.}
\end{minipage}\hfill
\begin{minipage}[t]{0.325\textwidth}
\raggedright\setlength{\parindent}{0pt}
\textbf{Comp-H (Rerank+Compr.)}~\ok\\
{\footnotesize
99.2\% comp.\ (1.94M$\rightarrow$14.6k)\\ $\bullet$ recall 1.0 $\bullet$
\gscore{Northern Transylvania}}\\[3pt]
{\footnotesize
\textbf{Key observation.}
Reranking additionally surfaces the passage describing the 23 Aug.\ 1944 coup
and retrieves the second gold page on Northern Transylvania.\\[2pt]
\textbf{Outcome.}
The agent verifies the coup date, the six-month interval, and the regained
region before answering; only the nickname clue remains a guess.
It finishes in ten searches with perfect recall, five citations,
and the highest confidence.}
\end{minipage}
\end{qexample}
\vspace{-1.2em}
\caption{
\textbf{Agent trajectory on BrowseComp-Plus (gpt-oss-120B).}
Compression makes the answer sentence prominent enough to identify the correct
region, while reranking contributes complementary evidence that verifies the
timeline and pins the region to both gold pages.}
\label{fig:ex-browsecomp-traj}
\end{figure*}
Figure~\ref{fig:ex-browsecomp-traj} shows one BrowseComp-Plus query with the
trajectories of three configurations: retrieval-only baseline, Compress-only,
and Rerank+Compress.

\subsection{OfficeQA}\label{sec:appendix_officeqa_result}

\subsubsection{Setup}

OfficeQA~\citep{officeqa,opsahlong2026officeqaproenterprisebenchmark} is a
document-grounded numeric QA benchmark over U.S.\ Treasury Bulletins, with a
page-level corpus of ${\sim}83$K pages; we use the \texttt{full} version
($246$ queries).
Retrieval uses \texttt{Embed-v4}: we load precomputed corpus embeddings, encode
each agent query at inference time, and search a flat inner-product index over
the ${\sim}83$K pages. As in
Appendix~\ref{sec:appendix_browsecomp_result}, each search call retrieves the
top-$50$ pages, reranks and compresses them, and returns the top-$5$ as
agent-visible snippets. Since OfficeQA answers are numeric, the main text reports numeric-tolerance accuracy following \citet{officeqa,opsahlong2026officeqaproenterprisebenchmark}. In this section, we additionally report judge accuracy using an LLM-as-a-judge protocol with \texttt{Qwen/Qwen3.5-35B-A3B} as the judge model, consistent with BrowseComp-Plus.


\subsubsection{Results}\label{app:officeqa_judge}

Table~\ref{tab:officeqa_judge_nocap_result} reports judge accuracy with the
gpt-oss-120B agent. The absolute numbers are lower than on BrowseComp-Plus, and compression
alone is not sufficient here. Under retriever order every system stays below the
retrieval-only baseline ($22.36$), including ours ($16.67$). With reranking,
Comp-H is the only system above the baseline ($23.17$) and
DSLR-Zerank matches it ($22.36$). One likely reason is the numeric nature of
OfficeQA: an answer often depends on a single cell of a long table, so pruning a
page is riskier than in BrowseComp-Plus. The two token-level pruners are much weaker; LongLLMLingua almost
never produces a correct answer ($1.63$) even though it keeps more than half of
the tokens.

\begin{table}[ht!]
\centering
\footnotesize
\setlength{\tabcolsep}{3pt}
\resizebox{\columnwidth}{!}{%
\begin{tabular}{llccc}
\toprule
\textbf{Method} & \textbf{Mode} & \textbf{Acc.} & \textbf{Recall} & \textbf{Comp.\%} \\
\midrule
\multirow{2}{*}{Baseline}
 & Retrieval-only & 22.36 & 39.73 & --- \\
 & Rerank-only (BGE) & \textbf{30.08} & 48.53 & --- \\
\midrule
\multirow{2}{*}{\textbf{Comp-H}}
 & Compress-only & 16.67\,\textsubscript{\textcolor{red}{(--5.69)}} & 40.17\,\textsubscript{\textcolor{blue}{(+0.44)}} & 83.77 \\
 & Rerank+Compr. & 23.17\,\textsubscript{\textcolor{blue}{(+6.50)}} & \textbf{59.76}\,\textsubscript{\textcolor{blue}{(+19.59)}} & 84.62 \\
\midrule
\multirow{2}{*}{DSLR-Zerank} 
 & Compress-only & 14.23 & 39.78 & 49.49 \\
 & Rerank+Compr. & 22.36 & 46.67 & 37.80 \\
\midrule
\multirow{2}{*}{OpenProvence}
 & Compress-only & 16.67 & 39.49 & 78.34 \\
 & Rerank+Compr. & 17.89 & 49.92 & 79.24 \\
\midrule
\multirow{2}{*}{Provence}
 & Compress-only & 15.04 & 39.69 & 80.08 \\
 & Rerank+Compr. & 16.26 & 43.70 & 81.14 \\
\midrule
\multirow{2}{*}{XProvence} 
 & Compress-only & 7.32 & 41.51 & 94.31 \\
 & Rerank+Compr. & 7.72 & 51.00 & 93.51 \\
\midrule
\multirow{2}{*}{Semantic HL}
 & Compress-only & 19.92 & 38.78 & 44.42 \\
 & Rerank+Compr.$^{\dagger}$ & 21.14 & 46.58 & 43.14 \\
\midrule
\multirow{2}{*}{LLMLingua-2}
 & Compress-only & 13.82 & 38.79 & 36.76 \\
 & Rerank+Compr.$^{\dagger}$ & 13.41 & 46.93 & 36.56 \\
\midrule
\multirow{2}{*}{LongLLMLingua}
 & Compress-only & 0.81 & 37.63 & 42.45 \\
 & Rerank+Compr. & 1.63 & 41.92 & 42.50 \\
\bottomrule
\end{tabular}}
\vspace{-0.6em}
\caption{\textbf{OfficeQA} with the gpt-oss-120B agent, graded by the LLM
judge. Subscripts show the change of Compress-only against the retrieval-only
baseline and of Rerank+Compress against Compress-only. Comp.\% is the
percentage of retrieved tokens removed before generation.
$^{\dagger}$Semantic HL and LLMLingua-2 have no internal reranker, so their
Rerank+Compress row uses an external BGE reranker.}
\label{tab:officeqa_judge_nocap_result}
\vspace{-0.8em}
\end{table}

\subsubsection{Qualitative Example}\label{sec:appendix_officeqa_ex}
\providecommand{\ok}{\textcolor{good}{\ding{51}}}
\providecommand{\bad}{\textcolor{bad}{\ding{55}}}
\providecommand{\sq}[1]{\textit{\textcolor{black!70}{``#1''}}}
\begin{figure*}[t]
\begin{qexample}{\texttt{multi\_source} OfficeQA \#168: numeric aggregation needs reranking to drive recall}
\textbf{Q.} In the 1939 U.S.\ Treasury Bulletins, locate the four monthly grand totals of
\emph{aggregate international flows of liquid banking funds} for May--August 1939 and sum
them.

\textbf{Gold.} \textbf{566{,}840} (the sum of the four monthly \emph{Net Capital Movement
between the United States and Foreign Countries} grand totals).\\[4pt]

\begin{minipage}[t]{0.325\textwidth}
\raggedright
\textbf{Comp-H (Rerank+Compr.)}~\ok\\
{\footnotesize comp.\ 82\%\;\textbullet\;recall 1.00\;\textbullet\;answer \gscore{566{,}840}}\\[3pt]
{\footnotesize
\textbf{Subqueries $\rightarrow$ table surfaced}\\
1.~\sq{...liquid banking funds ... May 1939 grand total} $\rightarrow$ \emph{Net Capital Movement} (through May)\\
2.~\sq{... June 1939 ... grand total} $\rightarrow$ \emph{Net Capital Movement} (through June)\\
3.~\sq{Net Capital Movement July 1939} $\rightarrow$ \emph{Net Capital Movement} (through July \emph{and} August)\\[3pt]
\textbf{Outcome.} Reranking surfaces the same correct series every hop, so all four monthly
grand totals are kept (in compact compressed tables) and summed exactly.}
\end{minipage}\hfill
\begin{minipage}[t]{0.325\textwidth}
\raggedright
\textbf{DSLR (Rerank+Compr.)}~\bad\\
{\footnotesize comp.\ 6\%\;\textbullet\;recall 0.25\;\textbullet\;answer \uscore{462{,}311}}\\[3pt]
{\footnotesize
\textbf{Subqueries $\rightarrow$ table surfaced}\\
1.~\sq{... liquid banking funds May 1939} $\rightarrow$ \emph{Short-Term Foreign Assets} (May)\\
3--6.~\sq{... "Short-Term Foreign Liabilities" ... Total June/August} $\rightarrow$ \emph{Short-Term Foreign Liabilities} (June/Aug)\\
7.~\sq{July 1939 Short-Term Foreign Liabilities} $\rightarrow$ same wrong series\\[3pt]
\textbf{Where it breaks.} The reranker locks onto a lexically similar but \emph{wrong} table
family (short-term liabilities/assets) and never returns the \emph{Net Capital Movement}
grand totals. It barely compresses yet sums the wrong rows.}
\end{minipage}\hfill
\begin{minipage}[t]{0.325\textwidth}
\raggedright
\textbf{Comp-H (Compress-only)}~\bad\\
{\footnotesize comp.\ 68\%\;\textbullet\;recall 0.50\;\textbullet\;answer \uscore{413{,}610}}\\[3pt]
{\footnotesize
\textbf{Subqueries $\rightarrow$ table surfaced}\\
1,3.~\sq{... liquid banking funds 1939} $\rightarrow$ \emph{Net Capital Movement} (May, June)\\
  4--14.~Literal \sq{"Total June/July"} / \sq{"Grand total"} searches and doc-ID look-ups return
  section-divider or empty pages, pruned to a title line\\
15.~\sq{International flows liquid banking 1939} $\rightarrow$ back to May/June\\[3pt]
\textbf{Where it breaks.} With the retriever's ranking (no reranking) the July/August
grand-total pages never surface; the agent spends 15 searches, half of them on empty pages,
and covers only two of four months.}
\end{minipage}
\end{qexample}
\vspace{-1.2em}
\caption{\textbf{Agentic trajectory on a multi-source OfficeQA query (gpt-oss-120B, \#168).}
The task needs one grand-total figure from each of four monthly bulletins. \textbf{Comp-H
(Rerank+Compr.)} reranks every search hop onto the correct \emph{Net Capital Movement}
series, so all four figures are retained even at ${\sim}82\%$ compression and the sum is
exact. \textbf{DSLR (Rerank+Compr.)} reranks to a near-duplicate but wrong table family and, despite
compressing only ${\sim}6\%$, sums the wrong rows. \textbf{Comp-H (Compress-only)} keeps the right
format but, lacking reranking, the retriever never surfaces the July/August pages, so the
agent burns searches on header-only pages and covers half the months.}
\label{fig:ex-officeqa-traj}
\end{figure*}
Figure~\ref{fig:ex-officeqa-traj} shows one OfficeQA query. With reranking, our
model keeps the numeric evidence needed for the answer while dropping most of
the irrelevant text.

\subsubsection{End-to-End Token Accounting in Agentic Systems}
\label{app:token-accounting}

Compression ratio measures only the fraction of retrieved tokens removed before they reach the
agent, not the total cost of an agentic trajectory. We therefore measure consumption end to end by
reporting search calls, retrieved tokens, reasoning tokens, and total tokens. Total tokens include both input and output tokens. Reasoning tokens are reported separately as a
breakdown of output tokens and should not be added to Total.

\begin{table*}[t] 
\centering 
\small \begin{tabular}{lrrrrrrr} \toprule & \multicolumn{1}{c}{Comp.} & \multicolumn{1}{c}{Search} & \multicolumn{1}{c}{Retrieved} & \multicolumn{1}{c}{Reason.} & \multicolumn{1}{c}{Total} & \multicolumn{1}{c}{$\Delta$Total} & \multicolumn{1}{c}{Acc.}\\ & \multicolumn{1}{c}{(\%)} & \multicolumn{1}{c}{calls} & \multicolumn{1}{c}{tok.} & \multicolumn{1}{c}{tok.} & \multicolumn{1}{c}{tok.} & \multicolumn{1}{c}{(\%)} & \multicolumn{1}{c}{(\%)}\\ \midrule Rerank-only (BGE) & 0.0 & 3.78 & 38.7K & 2{,}184 & 136.9K & -- & \textbf{31.0}\\ \midrule Comp-H & 84.4 & 6.52 & 10.2K & 2{,}213 & \textbf{77.3K} & $\mathbf{-43.5}$ & 23.4\\ DSLR-Zerank & 37.9 & 4.87 & 28.9K & 2{,}263 & 135.9K & $-$0.7 & 22.6\\ Semantic HL & 43.7 & 4.65 & 26.1K & 2{,}275 & 118.5K & $-$13.5 & 21.8\\ OpenProvence & 79.1 & 6.97 & 13.3K & 2{,}408 & 92.8K & $-$32.2 & 18.4\\ Provence & 80.8 & 7.96 & 11.9K & 2{,}427 & 95.3K & $-$30.4 & 16.7\\ LLMLingua-2 & 36.6 & 5.09 & 32.6K & 2{,}628 & 148.5K & $+$8.5 & 13.8\\ XProvence & 93.5 & 12.00 & 6.5K & 2{,}315 & 83.6K & $-$39.0 & 7.9\\ LongLLMLingua & 42.4 & 7.83 & 36.6K & 1{,}835 & 204.3K & $+$49.2 & 1.7\\ \bottomrule \end{tabular} \caption{\textbf{Per-query token consumption on OfficeQA.} Averaged over the 239 queries completed by every configuration; seven excluded queries
are all context-overflow failures of the rerank-only reference. $\Delta$Total relative to rerank-only.} 
\label{tab:officeqa-token-accounting} 
\end{table*}

\paragraph{OfficeQA.}
As in Table \ref{tab:officeqa-token-accounting}, Comp-H removes 84.4\% of retrieved tokens but reduces total consumption by 43.5\%, without substantially increasing generation cost: reasoning tokens increase by only 1.3\%
and output tokens by 3.9\%, while the input tokens contributed by retrieved context decrease by
71.0\%. The gap between per-search compression and end-to-end
savings is larger for other methods. XProvence removes 93.5\% of retrieved tokens but reduces total
consumption by 39.0\% while nearly tripling the number of searches, and DSLR-Zerank removes 37.9\%
while reducing total consumption by only 0.7\%. Two methods increase overall cost despite compression: LLMLingua-2 by 8.5\%, accompanied by a 20.3\% increase in reasoning tokens, and LongLLMLingua by 49.2\%, driven by additional search rounds. Rerank-only baseline remains the most accurate
configuration on this benchmark (31.0\%), while Comp-H provides the largest reduction in end-to-end token consumption among the compression methods.

\begin{table*}[t] 
\centering \small \begin{tabular}{lrrrrrrrr} \toprule & \multicolumn{1}{c}{Complete} & \multicolumn{1}{c}{Tok Reduc.} & \multicolumn{1}{c}{Search} & \multicolumn{1}{c}{Retrieved} & \multicolumn{1}{c}{Reason.} & \multicolumn{1}{c}{Total} & \multicolumn{1}{c}{$\Delta$Total} & \multicolumn{1}{c}{Acc.}\\ & \multicolumn{1}{c}{(\%)} & \multicolumn{1}{c}{(\%)} & \multicolumn{1}{c}{calls} & \multicolumn{1}{c}{tok.} & \multicolumn{1}{c}{tok.} & \multicolumn{1}{c}{tok.} & \multicolumn{1}{c}{(\%)} & \multicolumn{1}{c}{(\%)}\\ \midrule Rerank-only (cap512) & 100.0 & 95.6 & 9.39 & 22.2K & 1{,}854 & 168.3K & -- & 47.2\\ \midrule Comp-H & 99.4 & 98.8 & 10.57 & 10.4K & 1{,}945 & \textbf{103.7K} & $\mathbf{-38.4}$ & \textbf{64.9}\\ OpenProvence & 98.9 & 97.4 & 10.11 & 17.8K & 1{,}922 & 143.9K & $-$14.5 & 57.2\\ Provence & 99.6 & 95.7 & 11.57 & 13.0K & 2{,}108 & 125.0K & $-$25.7 & 53.7\\ Semantic HL & 95.1 & 97.7 & 11.37 & 18.0K & 2{,}112 & 164.6K & $-$2.2 & 52.5\\ DSLR-Zerank & 90.1 & 97.5 & 11.40 & 22.4K & 2{,}110 & 196.9K & $+$17.0 & 49.9\\ LongLLMLingua & 94.7 & 45.4 & 7.94 & 44.8K & 1{,}872 & 251.8K & $+$49.6 & 30.0\\ \midrule \multicolumn{9}{l}{\emph{Below 90\% completion; reported on their own completed subsets, so accuracies are optimistically biased.}}\\ LLMLingua-2 & 37.8 & 46.4 & 4.68 & 73.3K & 1{,}336 & 364.5K & $+$116.6 & 61.1\\ XProvence & 78.6 & 95.5 & 11.92 & 41.5K & 2{,}345 & 781.6K & $+$364.4 & 29.4\\ \bottomrule \end{tabular} \caption{\textbf{Per-query token consumption on BrowseComp-Plus.} Rows above the final block are averaged over the 670 queries completed by every configuration reaching $\geq$90\% completion, with $\Delta$Total relative to rerank-only. Completion is the fraction of the 830 queries that did not fail due to context overflow. Compression ratio is measured against the original retrieved context before the 512-token cap. The rerank-only baseline caps documents at 512 tokens while compressors are uncapped; therefore, token savings are conservative, and the Token Reduction column should be interpreted with this cap difference in mind.}
\label{tab:browsecomp-token-accounting} 
\end{table*}

\paragraph{BrowseComp-Plus.}
As shown in Table~\ref{tab:browsecomp-token-accounting}, compression is needed not only to save tokens but also to process these long-context queries successfully. Without a
document cap the uncompressed pipelines exhaust the context window: retrieval-only fails on 457 of
830 queries and rerank-only on 622, in every case because the accumulated context exceeds the
model's limit (Table \ref{tab:browsecomp_cap_vs_nocap}). This is why baselines cap
documents at 512 tokens. Comp-H, uncapped, fails on 5 of 830. Because the cap already removes
95.6\% of retrieved tokens from the reference, per-search compression ratios are less informative on
this benchmark and we compare end-to-end consumption instead. Comp-H obtains the best trade-off,
reducing total token usage by 38.4\% while raising accuracy from 47.2\% to 64.9\% and leaving
reasoning tokens nearly unchanged ($+4.9\%$). DSLR-Zerank and LongLLMLingua increase total
consumption despite compressing the retrieved context, as do LLMLingua-2 and XProvence in the
low-completion block.






\section{Model Training}

\subsection{Data Sources}\label{app:train_data}

Table~\ref{tab:train_data} provides an overview of the training datasets used.

\subsection{Data Preparation}\label{app:train_proc}
We re-use the already processed and filtered train splits for NQ, MS MARCO, GooAQ, JaQuAD, JSQuAD and DuReader from \citet{yuichi-tateno-2025-open-provence} and \citet{zilliz2026semantic_highlight}.
For all PDF-based datasets, we parse documents from PDF or image to a markdown representation if the original source does not provide it.
All datasets from both pools are then identically annotated with sequence-level relevance annotations, generated answers and rerank teacher scores.
For sequence-level relevance annotations, we follow  \citet{chirkova2025provenceefficientrobustcontext} and prompt the LLM to output indices of relevant sequences given a query and a document with numbered sequences.
As a final processing step, we apply three data filters:
1) If the positive document of a sample has no spans labelled as relevant, it is discarded entirely. 2) If an individual document has more than 80\% of its text labelled as relevant, the document is discarded. 3) If a document is empty or has otherwise failed LLM span annotation, it is also discarded.

\subsection{Training Details}\label{app:train_details}
Soft compression is trained end-to-end supervised on the generated target answers.
Following \citet{louis2026oscar}, we fine-tune the generator with LoRA \citep{hu2022lora}.
Hard compression is trained with the compressor only on a combined cross-entropy and margin-MSE \citep{hofstatter2020improving} loss for token relevance labeling and reranking, respectively.
We use 5 hard negatives per query and a maximum sequence length of 2,048 tokens per query-document pair for training.
For soft compression, we use a fixed compression ratio of 16 by compressing each chunk of 128 tokens to 8 memory tokens.
All models are implemented and trained with HuggingFace Transformers \citep{wolf2020transformers}.
A full training takes around 18 hours on two nodes of 8$\times$NVIDIA H100.

\subsection{Training Ablation}\label{app:train_abl}

\begin{table}[h!]
\centering
\small
\setlength{\tabcolsep}{3pt}
\label{tab:hard-compression-training-ablation}
\begin{tabular}{@{}lrrrr@{}}
\toprule
\textbf{Model} & \textbf{\shortstack{Evidence\\Recall}} & \textbf{\shortstack{Comp.\\\%}} & \textbf{NDCG@5} & \textbf{\shortstack{Judge\\Score}} \\
\midrule
XProvence & 40.88 & 58.90 & 66.22 & 0.48 \\
Comp-H ($-$ PDF) & 53.03 & 82.90 & 78.11 & 0.54 \\
Comp-H ($-$ model) & 58.44 & 76.44 & 73.10 & 0.58 \\
Comp-H (final) & 68.87 & 73.35 & 78.41 & 0.63 \\
\bottomrule
\end{tabular}
\caption{Hard compression model training ablation. All models evaluated with threshold 0.2.}
\label{tab:training_ablation}
\end{table}

In Table~\ref{tab:training_ablation}, we analyze the effect of the training choices we made for our hard compression model.
We compare against XProvence \citep{mohamed2026XProvencezerocostmultilingualcontext} as the reference multilingual hard compression and reranking model.
XProvence uses \texttt{bge-reranker-v2-m3} \citep{chen2024bge} as its base model and trains primarily on multilingual MS MARCO \citep{bajaj2016ms} data.
In the first ablation \textit{Comp-H ($-$ PDF)}, we remove all data from our new PDF document pool and keep only the web and Wikipedia training sets, but with our own teacher labels and the BidirLM base model.
In the second ablation \textit{Comp-H ($-$ model)}, we keep both the Web+Wikipedia and PDF training pools, but switch back to the same \texttt{bge-reranker-v2-m3} base model XProvence uses.
All models are evaluated with threshold 0.2.

We can see that both the choice of base model ($0.54$) and the addition of the PDF training pool ($0.58$) independently show significant lift over the baseline reference ($0.48$) as measured by the judge score.
But only the combination of both results in the final performance ($0.63$).
Notably, the ablation with PDF training data removed learns the highest compression ratio at the same score threshold.
This could be attributable to the fact that Web+Wikipedia training queries typically can be answered with shorter spans than those in the PDF training pool.

\begin{table*}[t]
\centering
\small
\setlength{\tabcolsep}{3pt}
\label{tab:training-data-overview}
\begin{tabularx}{\linewidth}{@{}l r l X l@{}}
\toprule
\textbf{Dataset} & \textbf{Queries} & \textbf{Language} & \textbf{Domain} & \textbf{Source} \\
\midrule
\multicolumn{4}{@{}l}{\emph{Web+Wikipedia pool}} \\
\midrule
Natural Questions & 10,168 & English & Open-domain questions; Wikipedia & \citet{kwiatkowski-etal-2019-natural} \\
MS MARCO & 216,963 & English & Web search & \citet{bajaj2016ms} \\
GooAQ & 109,104 & English & Google-search open QA & \citet{khashabi2021gooaq} \\
DuReader & 46,330 & Chinese & Baidu-search reading comprehension & \citet{he2018dureader} \\
JaQuAD & 13,874 & Japanese & Wikipedia reading comprehension & \citet{so2022jaquad} \\
JSQuAD & 25,772 & Japanese & Wikipedia reading comprehension & \citet{kurihara-etal-2022-jglue} \\
MIRACL & 26,320 & 18 languages & Wikipedia retrieval & \citet{zhang-etal-2023-miracl} \\
mMARCO (multilingual) & 402,128 & 13 languages & Machine-translated web search & \citet{bonifacio2021mmarco} \\
\midrule
\multicolumn{5}{@{}l}{\emph{PDF pool}} \\
\midrule
ColPali PDFs & 43,221 & English & PDFs with synthetic LLM queries & \citet{faysse2025colpali} \\
TAT-DQA & 12,392 & English & Financial reports & \citet{zhu2022tatdqa} \\
VisRAG--Manuallib & 113,668 & English & Product manuals; synthetic LLM queries & \citet{yu2024visrag} \\
VisRAG--NeurIPS & 27,932 & English & NeurIPS papers; synthetic LLM queries & \citet{yu2024visrag} \\
VisRAG--Textbooks & 52,416 & English & College textbooks; synthetic LLM queries & \citet{yu2024visrag} \\
VDR (de) & 51,663 & German & Visual document retrieval & \citet{llamaindex2024vdr} \\
VDR (en) & 46,093 & English & Visual document retrieval & \citet{llamaindex2024vdr} \\
VDR (es) & 49,902 & Spanish & Visual document retrieval & \citet{llamaindex2024vdr} \\
VDR (fr) & 49,481 & French & Visual document retrieval & \citet{llamaindex2024vdr} \\
\midrule
\textbf{Total} & \textbf{1,297,427} & --- & --- & --- \\
\bottomrule
\end{tabularx}

\vspace{2pt}

\caption{Training data overview for our hard and soft compression models. Query counts are after filtering is applied.}
\label{tab:train_data}
\end{table*}

\section{Efficiency: Latency, Memory, and Generator Compute}\label{app:efficiency}

\paragraph{Setup.} Unlike our full experiments, which share one generator
endpoint, each system here gets a dedicated two-GPU node (NVIDIA B200) so that
queueing from other jobs does not enter the latency measurements: GPU~0 serves a
local gpt-oss-120b (\texttt{max\_model\_len}${=}131072$, reasoning effort
\texttt{medium}), GPU~1 hosts retrieval and compression.
For the efficiency evaluation, we use the first 30 top-level queries from each
System Test benchmark. Model loading, warmup, and
LLM-as-a-judge scoring are excluded. Each system contributes a single end-to-end run under the vLLM Responses API
defaults (temperature${=}1.0$, top-$p{=}1.0$, top-$k{=}0$, no request-level
seed), so controlled comparisons rely on replay. For replay we
freeze, per benchmark, those 30 queries together with their exact top-50
retrieved documents, and every system re-runs this identical input outside the
agent loop: one warmup call per query, then three timed repeats synchronized with
\texttt{torch.cuda.synchronize()}.
Compressor models are called with an identical chunk batch size of 32 across methods for comparability.

\paragraph{Metrics.} \emph{Agent run} (s/query): per-query medians of end-to-end
wall-clock (E2E), summed generator latency (LLM), and summed time inside
search-tool calls (Tool). Directly measured retrieval latency is small and stable across systems
(0.08--0.3\,s/query), so most excess Tool time reflects reranking and
compression. \emph{Replay} (s/call): P50
and P95 over the 30 per-query medians, i.e.\ the cost of processing one
50-document candidate set at batch size~32. \emph{Memory} (GiB): peak CUDA memory allocated during replay, in a process that
loads only the compressor, with peak statistics reset after one warmup call. It
covers weights plus peak working memory.
\emph{Prefill} (TFLOPs/query): $2 \times (\text{active parameters per token})
\times (\text{uncached input tokens})$, from the API-reported token accounting.
Since the uncached count depends on the serving stack's caching policy, prefill
is comparable across systems rather than absolute.

\begin{table*}[t]
\centering
\small
\begin{tabular}{llcrrrrrrr}
    \toprule
    System & Mode & $\tau$ & \multicolumn{3}{c}{Agent run (s/query)} & \multicolumn{2}{c}{Replay (s/call)} & Mem & Prefill \\
    \cmidrule(lr){4-6} \cmidrule(lr){7-8}
     & & & E2E & LLM & Tool & P50 & P95 & (GiB) & (TFLOPs/q) \\
    \midrule
    \multicolumn{10}{l}{\emph{OfficeQA}} \\
\midrule
Baseline & -- & -- & 4.05 & 3.66 & 0.12 & -- & -- & -- & 154.980 \\
Comp-H & C & 0.1 & 3.27 & 2.88 & 0.37 & 0.15 & 0.29 & 2.23 & 36.224 \\
SemanticHL & C & 0.1 & 2.77 & 2.57 & 0.24 & 0.12 & 0.17 & 3.69 & 74.103 \\
LLMLingua-2 & C & 0.5 & 3.40 & 3.16 & 0.29 & 0.15 & 0.27 & 2.63 & 76.816 \\
\cmidrule{1-10}
Comp-H & R & 0.1 & 7.32 & 5.08 & 1.89 & 1.41 & 2.26 & 2.85 & 153.033 \\
Comp-H & RC & 0.1 & 5.29 & 2.61 & 2.47 & 1.59 & 2.50 & 2.85 & 45.035 \\
DSLR-Zerank & RC & 0.4 & 11.46 & 3.67 & 8.02 & 7.90 & 11.20 & 74.89 & 127.709 \\
OpenProvence & RC & 0.1 & 3.13 & 2.20 & 0.90 & 0.57 & 0.89 & 4.88 & 48.053 \\
Provence & RC & 0.1 & 3.44 & 2.03 & 1.19 & 0.98 & 1.80 & 5.96 & 40.041 \\
XProvence & RC & 0.1 & 6.95 & 3.30 & 3.46 & 2.84 & 6.44 & 29.11 & 126.220 \\
LongLLMLingua & RC & 0.55 & 12.93 & 2.00 & 11.08 & 8.63 & 12.83 & 67.25 & 96.791 \\
    \specialrule{0.08em}{0.8ex}{0.6ex}
    \multicolumn{10}{l}{\emph{BrowseComp-Plus}} \\
\midrule
Baseline$^\dag$ & -- & -- & 4.40 & 4.08 & 0.20 & -- & -- & -- & 326.425 \\
Comp-H & C & 0.1 & 2.97 & 1.94 & 0.82 & 0.18 & 1.38 & 3.02 & 9.127 \\
SemanticHL & C & 0.1 & 2.43 & 2.07 & 0.39 & 0.14 & 0.70 & 5.79 & 24.085 \\
LLMLingua-2$^\dag$ & C & 0.5 & 3.41 & 2.98 & 0.66 & 0.20 & 1.32 & 2.81 & 158.203 \\
\cmidrule{1-10}
Comp-H$^\dag$ & R & 0.1 & 13.02 & 6.09 & 5.75 & 2.91 & 7.47 & 3.02 & 448.935 \\
Comp-H & RC & 0.1 & 5.78 & 1.95 & 3.73 & 2.92 & 7.63 & 3.02 & 20.264 \\
DSLR-Zerank$^\dag$ & RC & 0.4 & 22.77 & 2.41 & 18.67 & 32.16 & 85.01 & 42.76 & 132.225 \\
OpenProvence & RC & 0.1 & 4.87 & 2.26 & 2.16 & 1.36 & 3.30 & 4.89 & 18.063 \\
Provence & RC & 0.1 & 4.61 & 1.96 & 2.41 & 2.82 & 6.88 & 5.96 & 19.541 \\
XProvence$^\dag$ & RC & 0.1 & 21.13 & 5.58 & 15.81 & 12.13 & 21.83 & 29.12 & 413.888 \\
LongLLMLingua & RC & 0.55 & 10.95 & 2.53 & 7.61 & 6.37 & 8.60 & 67.24 & 59.042 \\
\bottomrule
    \end{tabular}

\caption{Efficiency on the first 30 top-level queries from each System Test
benchmark, using one agent run per system.
Modes are R=Reranking, C=Compression, RC=Reranking+Compression.
Agent-run columns are per-query medians, and prefill is a per-query
arithmetic mean; replay columns and peak memory come from controlled replay on an
identical frozen top-50 workload. $\dag$~marks systems that did not complete all
30 BrowseComp-Plus queries because the 131K context window was exceeded: Baseline (28/30 completed), Comp-H R (25/30),
DSLR-Zerank RC (29/30), LLMLingua-2 (29/30), and XProvence (28/30). Their agent-run and prefill values are computed only over completed
queries. For compressor rows marked with $\dagger$, controlled replay still
completed all 30 frozen workloads with three repeats each, so replay latency and
peak memory are unaffected by the agent-side overflow.
}  
\label{tab:efficiency-30}
\end{table*}

\paragraph{Summary.} Table~\ref{tab:efficiency-30} shows that compression-only can
remove most of the generator's uncached prefill while simultaneously reducing end-to-end latency on both OfficeQA and BrowseComp-Plus. Latency and prefill do not rank systems
identically---the fastest compressor per call is not the one that removes the most
generator compute---so we report both.
Peak compressor memory separates systems
further, growing with document length for the sentence-level models. 
As expected, reranking generally induces additional latency to the tool calls and thus the full pipeline.
Additionally enabling compression partially offsets the reranking-induced latency increases, resulting in end-to-end latencies close to the baseline for some of the methods.
Importantly, compression models here are not individually optimized, so these values do not indicate best possible latencies for every method.


\end{document}